\documentclass[twoside,english]{elsarticle}
\usepackage[T1]{fontenc}
\usepackage[latin9]{inputenc}
\usepackage{geometry}
\usepackage{array}
\usepackage{float}
\usepackage{booktabs}
\usepackage{multirow}
\usepackage{amsmath}
\usepackage{amssymb}
\usepackage{graphicx}
\graphicspath{{Figures/}}
\usepackage{svg}
\usepackage{pdfpages}
\usepackage{wasysym}
\usepackage{enumerate}
\usepackage{amsthm,amsmath,amssymb}
\usepackage{mathrsfs}
\usepackage{indentfirst}

\makeatletter

\providecommand{\tabularnewline}{\\}

\journal{Computer Methods in Applied Mechanics and Engineering}
\usepackage{algorithm,algpseudocode}
\PassOptionsToPackage{sort&compress}{natbib}
\usepackage{calc}

\@ifundefined{showcaptionsetup}{}{%
 \PassOptionsToPackage{caption=false}{subfig}}
\usepackage{subfig}
\makeatother
\usepackage{babel}
\begin{document}

\begin{frontmatter}{}

\title{An open-source ABAQUS implementation of the scaled boundary finite
element method to study interfacial problems using polyhedral meshes}

\author[unsw]{Shukai~Ya}

\author[unsw]{Sascha~Eisenträger}

\author[unsw]{Chongmin~Song}

\author[dlut]{Jianbo~Li\corref{cor1}}

\ead{jianboli@dlut.edu.cn}

\cortext[cor1]{Corresponding author}

\address[unsw]{School of Civil and Environmental Engineering, University of New
South Wales, Sydney 2052, Australia}

\address[dlut]{Institute of Earthquake Engineering, State Key Laboratory of Coastal
and Offshore Engineering, Dalian University of Technology, Dalian
116024, China}
\begin{abstract}
The scaled boundary finite element method (SBFEM) is capable of generating
polyhedral elements with an arbitrary number of surfaces. This salient
feature significantly alleviates the meshing burden being a bottleneck
in the analysis pipeline in the standard finite element method (FEM).
In this paper, we implement polyhedral elements based on the SBFEM
into the commercial finite element software ABAQUS. To this end, user
elements are provided through the user subroutine UEL. Detailed explanations
regarding the data structures and implementational aspects of the
procedures are given. The focus of the current implementation is on
interfacial problems and therefore, element-based surfaces are created
on polyhedral user elements to establish interactions. This is achieved
by an overlay of standard finite elements with negligible stiffness,
provided in the ABAQUS element library, with polyhedral user elements.
By means of several numerical examples, the advantages of polyhedral
elements regarding the treatment of non-matching interfaces and automatic
mesh generation are clearly demonstrated. Thus, the performance of
ABAQUS for problems involving interfaces is augmented based on the
availability of polyhedral meshes. Due to the implementation of polyhedral
user elements, ABAQUS can directly handle complex geometries given
in the form of digital images or stereolithography (STL) files. In
order to facilitate the use of the proposed approach, the code of
the UEL is published open-source and can be downloaded from https://github.com/ShukaiYa/SBFEM-UEL.
\end{abstract}
\begin{keyword}
Scaled boundary finite element method; ABAQUS UEL; Polyhedral element;
Interfacial problems
\end{keyword}

\end{frontmatter}{}

\section{Introduction\label{sec:Introduction}}

The finite element method (FEM) has been widely used in engineering
and science for solving partial differential equations arising in
many different areas of application. To this end, the FEM discretizes
a problem domain into a number of subdomains of simple shape, so-called
finite elements, and assembles them to derive numerical solutions.
Geometrically simple elements are commonly utilized in the standard
FEM, e.g., tetrahedral and hexahedral elements in three-dimensional
applications. Many commercial finite element software packages are
available, such as ANSYS, ABAQUS, MARC, etc. which makes the FEM easily
accessible for engineers and researchers alike. Nonetheless, there
are some aspects that the FEM is unable to handle in a satisfactory
manner: a) stress singularities occurring in fracture mechanics problems
\citep{fawkes1979} and b) the representation of the unbounded domains
\citep{medina1983}. Additionally, mesh generation is still laborious
and time-consuming and therefore, often constitutes a severe bottleneck
in the analysis pipeline. Considering special data formats, such as
digital images and stereolithography (STL), which are nowadays widely
adopted by industry, this problem is prominent. To alleviate the aforementioned
issues, various alternatives have been proposed by researchers, to
name a few, the boundary element method (BEM) \citep{banerjee1981},
the scaled boundary finite element method (SBFEM) \citep{Song1997},
and the extended finite element method (X-FEM) \citep{belytschko1999}. 

The SBFEM is in its core nature a semi-analytical method which was
first proposed by Wolf and Song \citep{Song1997,wolf1996finite} for
modeling wave propagation in unbounded domains. However, over the
last two decades, the SBFEM has been extended to various types of
analyses including wave propagation in bounded domains \citep{bazyar2008},
fracture mechanics \citep{Saputra2015,song2018review,coelho2020},
acoustics \citep{Lehmann2006,birk2016}, contact mechanics \citep{Xing2018,xing2019},
seepage \citep{Liu2018}, elasto-plasticity \citep{ooi2014}, damage
analysis \citep{zhang2018}, adaptive analysis \citep{Hirshikesh2019,Zhang2020},
among many others \citep{natarajan2015,eisentrager2020,liu2016,Ye2018}.
Due to its versatility and applicability to a wide class of problems,
the SBFEM can be seen as a general numerical method to solve PDEs.

One of the main advantages of SBFEM over standard FEM is that it offers
greater flexibility regarding the spatial discretization, as polytope
elements of an arbitrary number of vertices, edges and surfaces can
be straightforwardly constructed. This advantage significantly reduces
the meshing burden encountered in the standard FEM. Firstly, the SBFEM
is highly complementary to the hierarchical tree-based mesh generation
technique, which is a robust mesh generation technology that can automatically
discretize complicated geometries with highly irregular boundaries
\citep{Yerry1984,Shephard1991,greaves1999}. In recent years, such
polytope elements have been employed in SBFEM analyses for both two-
\citep{Ooi2015} and three-dimensional \citep{saputra2017,Liu2017}
problems. The geometry can be either stored as digital images obtained
from computed tomography (CT) scans or as STL files widely used in
3D printing. Secondly, the polytope elements allow meshes with rapid
transitions in element size, which is beneficial for the modeling
of complex domains \citep{Huang2015}. To obtain analysis-ready meshes
for complex structures, such as particle reinforced composites, human
bones, or other micro-structured materials, the geometric details
of the structure require small element sizes at the boundary, while
in the interior larger elements are appropriate. Hence, we require
transition elements allowing adjacent polytope elements with different
sizes, effectively reducing the number of elements and also the computational
resources required. Finally, owing to the higher degree of flexibility
of polytope elements, the initially non-matching interface meshes
can be straightforwardly converted to matching ones by splitting interface
edges (2D) or surfaces (3D) \citep{Zhang2019}. The matching discretization
on the interface benefits the analyses involving multiple domains
by facilitating the application of interface constraints. Common areas
of application include domain decomposition \citep{Zhang2019}, contact
\citep{Xing2018,xing2019}, and acoustic-structure interaction \citep{liu2019}
problems. 

Despite the aforementioned advantages of the SBFEM, the popularity
of this method is still restricted to a small part of the scientific
community. One important reason is seen in the more involved theoretical
derivation of the SBFEM compared to the FEM. The FEM is practically
taught in any engineering program at universities all over the world,
concise mathematical theories on its properties exist, and powerful
commercial software packages are readily available, while the SBFEM
has been so far only implemented for research purposes. With the implementation
of the SBFEM into a commercial software package, experienced analysts
will have easy access to SBFEM and can exploit its advantages. It
is our intent to facilitate the acceptance of the method itself and
spark its further development to other classes of problems. Considering
scholars interested in starting research on the SBFEM, the availability
of the implementation will provide a user-friendly and robust platform
and comprehensive access to non-linear solvers.

The possibility of implementing other numerical methods which are
related to the FEM into commercial software packages provides a whole
new set of possibilities and can be realized through programming user
element subroutines. Many commercial FEM codes support this functionality,
e.g., $\mathrm{USERELEM}$ in ANSYS \citep{Ansys2013}, $\mathrm{USELEM}$
in MARC \citep{marc2005}, and $\mathrm{UEL}$ in ABAQUS \citep{ABA2016}.
In recent years, many non-standard numerical methods have been implemented
in ABAQUS. Before the X-FEM was available as an additional module
of commercial software packages, it was implemented in the form of
a user element \citep{shi2008x,mcnary2009,Giner2009}. Elguedj et
al. \citep{elguedj2012} introduced the user element implementation
of NURBS based on isogeometric analysis. The cell-based smoothed finite
element method (CSFEM) was implemented as user elements in ABAQUS
by Cui et al. \citep{Cui2020} and Kumbhar et al. \citep{Kumbhar2020}
for alleviating mesh distortion issues seen in standard finite elements.
Yang et al. \citep{Yang2020} implemented the SBFEM in ABAQUS and
showed the benefits of using polytope (polygon/polyhedron) elements
in linear elastic stress analyses.

In the article at hand, linear elastic polyhedral element straightforwardly
constructed by the SBFEM is implemented into ABAQUS through $\mathrm{UEL}$.
To be consistent with ABAQUS' element library, the surfaces of polyhedral
element are discretized only as triangular and quadrilateral shapes,
and only low-order (linear and quadratic) finite elements are used
for the surface tessellation in this paper. In this way, the SBFEM
and FEM can be easily coupled since the discretization of element
surfaces is identical. In addition, another ABAQUS' user subroutine,
$\mathrm{UEXTERNALDB}$ to manage external databases, is involved
in the current implementation. Because of the arbitrariness of polyhedron
in terms of topology, the nodal connectivity of the polyhedral user
element should be recovered by a supplementary input file. The $\mathrm{UEXTERNALDB}$
is used to open the supplementary input file and store the polyhedral
mesh information in a memory-efficient way.

The main goal of this paper is to illustrate the advantages of the
SBFEM user element with a special focus on three-dimensional interfacial
problems. For this type of applications, some salient features of
ABAQUS can be utilized, such as i) the comprehensive approaches for
the contact modeling, ii) enriched library of cohesive elements for
the modeling of interfacial behaviors, and iii) powerful non-linear
solution techniques. It is also our belief that ABAQUS can benefit
from the salient features of the SBFEM in different aspects mentioned
in the previous paragraphs. In addition to a more flexible meshing
paradigm provided by the SBFEM, the performance of ABAQUS can be enhanced
for further specific applications. ABAQUS models containing non-matching
interface meshes, e.g., fail to pass the patch test both in domain
decomposition \citep{Zhang2019} and contact \citep{Xing2018,xing2019}
problems, especially when the interfaces are curved. Using polytope
elements, the theoretical results are recovered again, as we can easily
generate matching surfaces for a wide variety of cases.

The remaining part of this paper is organized as follows: A brief
review of the theoretical derivations of the SBFEM is presented in
Section~\ref{sec:Theory}. Section~\ref{sec:Implementation} introduces
the implementation of the SBFEM user element in ABAQUS, including
its workflow, ABAQUS input file formats, data structures for storing
the mesh information, a detailed explanation of the implementation
and the definition of element-based surfaces needed for contact modeling.
The numerical examples include academic benchmark tests and more complex
structures clearly demonstrating the main advantages of the SBFEM
(see Section~\ref{sec:Numerical-examples}). Finally, important conclusions
are provided in Section~\ref{sec:Conclusion}.

\section{Theory of the scaled boundary finite element method\label{sec:Theory}}

This section briefly summarizes the construction of polyhedral elements
using SBFEM. Only the key concepts and equations related to the implementation
of the SBFEM as a user element are presented. Readers interested in
detailed discussions and derivations of this method are referred to
the monograph by Song \citep{Song2018}. 

\subsection{Polyhedral elements in the SBFEM}

\begin{figure}
\noindent
 \hfill{}
 \subfloat[Volume and scaling center\label{fig:Volume-and-scaling}]
 {\noindent \centering{}\includegraphics[scale=0.8]{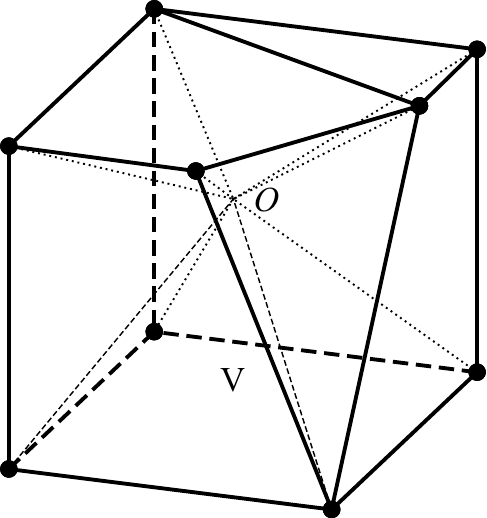}}
 \hfill{}
 \subfloat[Surface descretization\label{fig:Surface-descretization}]
 {\noindent \centering{}\includegraphics[scale=0.8]{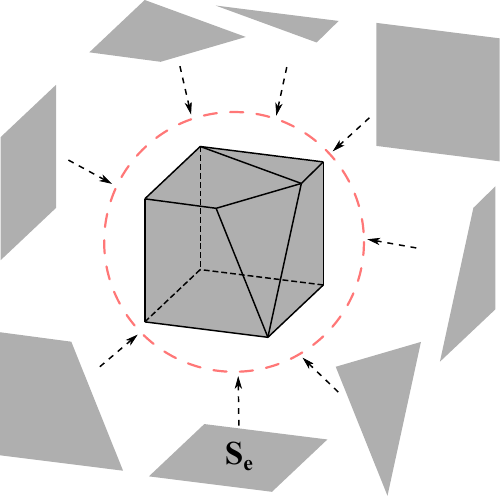}}
 \hfill{}
\noindent \caption{Schematic of a polyhedral mesh\label{fig:Schematic-polyhedron}}
\end{figure}

Polyhedral elements offer a great deal of flexibility in terms of
mesh generation and mesh transitioning, especially for geometrically
complex structures and adaptive analyses. One of the salient features
of SBFEM is that we can straightforwardly derive such elements. In
general, a polyhedral mesh can be created by discretizing the boundary
into surface elements, while the volume of the domain is not discretized
itself \citep{Song1997}. A typical example of a polyhedral element
employing the SBFEM is illustrated in Fig.~\ref{fig:Schematic-polyhedron}.
A so-called scaling center (denoted as ``$\mathbf{\mathit{O}}$''
in Fig.~\ref{fig:Volume-and-scaling}) is selected within the volume,
such that every point on the boundary is directly visible, i.e., the
domain has to fulfill the requirements of star-convexity. The boundary
of the polyhedral domain is discretized using surface elements, which
are commonly of triangular and quadrilateral shapes as depicted in
Fig.~\ref{fig:Surface-descretization}. The volume of the polyhedral
mesh is represented by scaling the surface elements towards the scaling
center. Only the solution on surface elements is interpolated whereas
the solution inside the polyhedral element will be obtained analytically,
leading to a semi-analytical procedure.

It is worthwhile to mention that in the framework of the SBFEM, often
standard low-order finite elements are used for the surface tessellation,
although high-order elements~\citep{ooi2016} and arbitrary faceted
polyhedra~\citep{ooi2020dual} are possible. Besides, because it
is well-known that quadrilateral elements generally perform better
than triangular elements due to the additional higher-order terms
in the approximation space, an approach to avoid the use of triangular
elements in the surface discretization is proposed~\citep{Duczek2020}.
To be consistent with ABAQUS' element library, we only add linear
and quadratic triangular (T3 and T6) and quadrilateral (Q4 and Q8)
elements in the current implementation. In the numerical examples
presented in Section~\ref{sec:Numerical-examples}, only linear elements
are used.

\subsection{Geometry transformation}

\begin{figure}
\noindent \begin{centering}
\hfill{}\subfloat[Scaling of the surface element using radial coordinate $\xi$ ($\xi=0$
at the scaling center and $\xi=1$ on the boundary) \label{fig:Radial-coordinate}]{\noindent \centering{}\hspace{0.05\textwidth}\includegraphics[bb=0bp 0bp 112bp 173bp,scale=0.8]{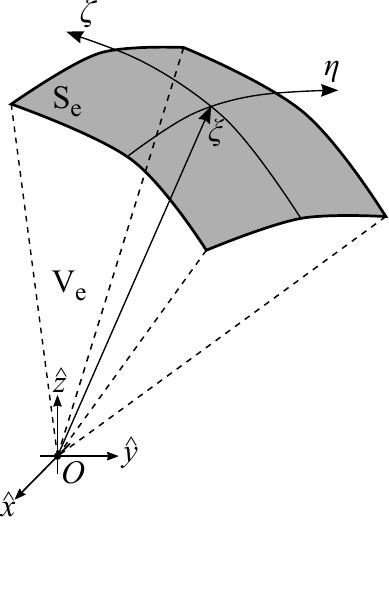}\hspace{0.05\textwidth}}\hfill{}\subfloat[Triangular and quadrilateral reference elements in natural coordinates
$\eta$, $\zeta$\label{fig:isoparametric}]{\noindent \centering{}\includegraphics[scale=0.8]{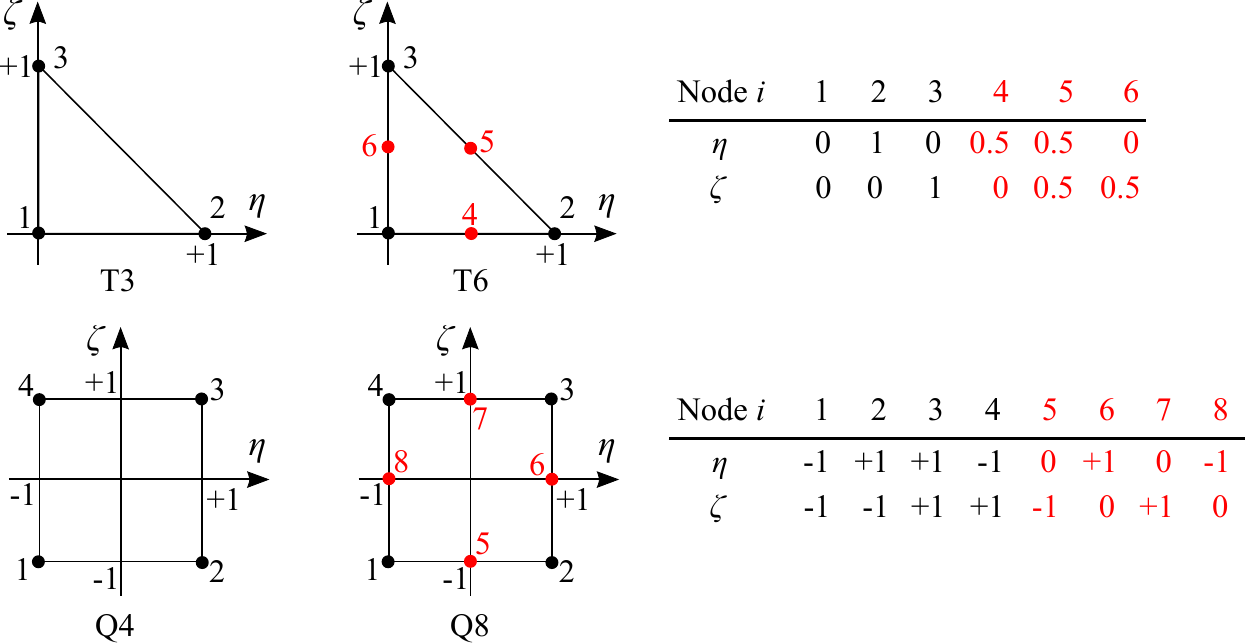}}\hfill{}
\par\end{centering}
\noindent \centering{}\caption{Three-dimensional scaled boundary coordinate system\label{fig:Three-dimensional-scaled-boundar}}
\end{figure}

The geometry transformation in the SBFEM is formulated by combining
the boundary discretization with a scaling operation of the boundary.
The scaling procedure is illustrated in Fig.~\ref{fig:Radial-coordinate}
for a quadrilateral surface element. A volume sector $\mathrm{V}_{\mathrm{e}}$
is represented by continuously scaling the surface element $\mathrm{S}_{\mathrm{e}}$
to the scaling center. A three-dimensional scaled boundary coordinate
system is established in each volume sector, with the circumferential
coordinates $\eta$, $\zeta$ on the surface element and the radial
coordinate $\xi$.

The surface elements on the boundary are isoparametric elements known
from the standard FEM~\citep{zienkiewicz1994}. The most commonly
used triangular and quadrilateral elements are depicted in Fig.~\ref{fig:isoparametric}
in its natural coordinates $\eta$, $\zeta$. The nodes marked in
red are only used for quadratic elements. The surface element's shape
functions are denoted by $\mathrm{\mathbf{N}}(\eta,\zeta)$. The radial
coordinate $\mathit{\xi}$ is dimensionless, emanating from the scaling
center and pointing towards the boundary. At the scaling center $\xi=0$
and on the boundary $\xi=1$ hold.

A point ($\hat{x},\hat{y},\hat{z}$) within the volume sector $\mathrm{V}_{\mathrm{e}}$
can be defined by the scaled boundary coordinates ($\eta,\zeta,\xi$)
as: 
\begin{subequations}
\label{eq:scaledcoordinate}
\begin{equation}
\hat{x}(\xi,\eta,\zeta)=\xi\hat{x}(\eta,\zeta)=\xi\mathbf{N}(\eta,\zeta)\hat{\mathbf{x}},
\end{equation}
\begin{equation}
\hat{y}(\xi,\eta,\zeta)=\xi\hat{y}(\eta,\zeta)=\xi\mathbf{N}(\eta,\zeta)\hat{\mathbf{y}},
\end{equation}
\begin{equation}
\hat{z}(\xi,\eta,\zeta)=\xi\hat{z}(\eta,\zeta)=\xi\mathbf{N}(\eta,\zeta)\hat{\mathbf{z}},
\end{equation}
\end{subequations}
 where $\hat{\mathbf{x}}$, $\hat{\mathbf{y}}$ and $\hat{\mathbf{z}}$
are the nodal coordinate vectors of the surface element $\mathrm{S}_{\mathrm{e}}$
in Cartesian coordinates. Note that all the coordinates in Eq.~(\ref{eq:scaledcoordinate})
are the relative coordinates with respect to the scaling center, i.e.,
the scaling center is taken as the origin of the local coordinate
system, and the hat symbol over variables indicates that they are
related to this specific local coordinate systems.

The partial derivatives with respect to the scaled boundary coordinates
can be related to the partial derivatives in Cartesian coordinates
by the following equation: 
\begin{equation}
\left\{ \begin{array}{c}
\frac{\partial}{\partial\xi}\\
\frac{\partial}{\partial\eta}\\
\frac{\partial}{\partial\zeta}
\end{array}\right\} =\left[\begin{array}{ccc}
1 & 0 & 0\\
0 & \xi & 0\\
0 & 0 & \xi
\end{array}\right]\mathbf{J}_{\mathrm{b}}(\eta,\zeta)\left\{ \begin{array}{c}
\frac{\partial}{\partial x}\\
\frac{\partial}{\partial y}\\
\frac{\partial}{\partial z}
\end{array}\right\} ,
\end{equation}
where $\mathbf{J}_{\mathrm{b}}(\eta,\zeta)$ denotes the Jacobian
matrix on the boundary ($\xi=1$), which is defined as 
\begin{equation}
\mathbf{J}_{\mathrm{b}}(\eta,\zeta)=\left[\begin{array}{ccc}
\hat{x}(\eta,\zeta) & \hat{y}(\eta,\zeta) & \hat{z}(\eta,\zeta)\\
\hat{x}(\eta,\zeta)_{,\eta} & \hat{y}(\eta,\zeta)_{,\eta} & \hat{y}(\eta,\zeta)_{,\eta}\\
\hat{x}(\eta,\zeta)_{,\zeta} & \hat{y}(\eta,\zeta)_{,\zeta} & \hat{z}(\eta,\zeta)_{,\zeta}
\end{array}\right].\label{eq:Jacobian}
\end{equation}
The determinant of $\mathbf{J}_{\mathrm{b}}(\eta,\zeta)$ is given
as
\begin{equation}
\left|\mathbf{J}_{\mathrm{b}}\right|=\hat{x}(\hat{y}_{,\eta}\hat{z}_{,\zeta}-\hat{z}_{,\eta}\hat{y}_{,\zeta})+\hat{y}(\hat{z}_{,\eta}\hat{x}_{,\zeta}-\hat{x}_{,\eta}\hat{z}_{,\zeta})+\hat{z}(\hat{x}_{,\eta}\hat{y}_{,\zeta}-\hat{y}_{,\eta}\hat{x}_{,\zeta}),
\end{equation}
where the argument ($\eta,\zeta$) has been dropped for the sake of
clarity. 

\subsubsection{Displacement field}

As a semi-analytical procedure, the SBFEM introduces unknown nodal
displacement functions $\mathbf{u}^{\mathrm{e}}(\xi)$ for the representation
of the displacement field of a volume sector $\mathrm{V}_{\mathrm{e}}$in
radial direction, i.e., those functions express the displacements
along the radial lines connecting the scaling center and boundary
nodes on \textit{e}-th surface element $\mathrm{S}_{\mathrm{e}}$,
and they are usually determined analytically. For the sake of completeness,
we remark that in Refs. \citep{ArticleChen2016,ArticleArioli2019}
a numerical approximation in radial direction has been successfully
established.

The displacements at a point ($\xi,\eta,\zeta$) within the volume
sector $\mathrm{V}_{\mathrm{e}}$ are interpolated based on $\mathbf{u}^{\mathrm{e}}(\xi)$,
thus the displacement field $\mathrm{\mathbf{u}}(\xi,\eta,\zeta)$
is expressed as 
\begin{equation}
\mathrm{\mathbf{u}}(\xi,\eta,\zeta)=\mathbf{N}_{\mathrm{u}}(\eta,\zeta)\mathbf{u}^{\mathrm{e}}(\xi),\label{eq:displacement-field}
\end{equation}
where the shape function matrix $\mathbf{N}_{\mathrm{u}}(\eta,\zeta)$
is written as
\begin{equation}
\mathbf{N}_{\mathrm{u}}(\eta,\zeta)=\left[N_{1}(\eta,\zeta)\mathbf{I},N_{2}(\eta,\zeta)\mathbf{I},\cdots,N_{n}(\eta,\zeta)\mathbf{I}\right],
\end{equation}
with a $3\times3$ identity matrix $\mathbf{I}$.

The nodal displacement functions $\mathbf{u}(\xi)$ of the entire
element are assembled by 
\begin{equation}
\mathbf{u}(\xi)=\underset{e}{\mathrm{\mathbb{A}}}\mathop{\mathbf{u}^{\mathrm{e}}(\xi)},
\end{equation}
where the symbol $\underset{e}{\mathrm{\mathbb{A}}}$ indicates the
standard finite element assembly procedure.

\subsubsection{Strain field}

In scaled boundary coordinates, the strain field $\boldsymbol{\boldsymbol{\varepsilon}}(\xi,\eta,\zeta)$
is obtained from the displacement field $\mathrm{\mathbf{u}}(\xi,\eta,\zeta)$
by applying the linear differential operator $\mathbf{L}$ 
\begin{equation}
\boldsymbol{\boldsymbol{\varepsilon}}(\xi,\eta,\zeta)=\mathbf{L}\mathrm{\mathbf{u}}(\xi,\eta,\zeta),\label{eq:strain-displacement}
\end{equation}
which is defined by 
\begin{equation}
\mathbf{L}=\mathbf{b}_{1}(\eta,\zeta)\frac{\partial}{\partial\xi}+\frac{1}{\xi}\left(\mathbf{b}_{2}(\eta,\zeta)\frac{\partial}{\partial\eta}+\mathbf{b}_{3}(\eta,\zeta)\frac{\partial}{\partial\zeta}\right),
\end{equation}
where the matrices $\mathbf{b}_{1}(\eta,\zeta)$, $\mathbf{b}_{2}(\eta,\zeta)$,
$\mathbf{b}_{3}(\eta,\zeta)$ can be found in Ref.~\citep{Song2018},
their entries are taken directly from the inverse of the Jacobian
matrix $\mathbf{J}_{\mathrm{b}}$. Substituting Eq.~(\ref{eq:displacement-field})
into Eq.~(\ref{eq:strain-displacement}), the relationship between
the strain field and the nodal displacement functions can be obtained
as 
\begin{equation}
\boldsymbol{\boldsymbol{\varepsilon}}(\xi,\eta,\zeta)=\mathbf{B}_{1}(\eta,\zeta)\mathrm{\mathbf{u}}(\xi)_{,\xi}+\frac{1}{\text{\ensuremath{\xi}}}\mathbf{B}_{2}(\eta,\zeta)\mathrm{\mathbf{u}}(\xi),\label{eq:strain-nodal}
\end{equation}
where 
\begin{subequations}
\label{eq:B1B2}
\begin{equation}
\mathbf{B}_{1}(\eta,\zeta)=\mathbf{b}_{1}(\eta,\zeta)\mathbf{N}_{\mathrm{u}}(\eta,\zeta),
\end{equation}
\begin{equation}
\mathbf{B}_{2}(\eta,\zeta)=\mathbf{b}_{2}(\eta,\zeta)\mathbf{N}_{\mathrm{u}}(\eta,\zeta)_{,\eta}+\mathbf{b}_{3}(\eta,\zeta)\mathbf{N}_{\mathrm{u}}(\eta,\zeta)_{,\zeta}.
\end{equation}
\end{subequations}

\subsubsection{Stress field}

For linear elastic problems, the stress field $\boldsymbol{\sigma}(\xi,\eta,\zeta)$
is related to the strain field by 
\begin{equation}
\boldsymbol{\sigma}(\xi,\eta,\zeta)=\mathbf{D}\boldsymbol{\boldsymbol{\varepsilon}}(\xi,\eta,\zeta),\label{eq:stress-strain}
\end{equation}
where $\mathbf{D}$ is the elasticity matrix, which is expressed as
follows for linear isotropic material:

\begin{equation}
\mathbf{D}=\frac{E}{(1+\nu)(1-2\nu)}\left[\begin{array}{cccccc}
1-\nu & \nu & \nu & 0 & 0 & 0\\
 & 1-\nu & \nu & 0 & 0 & 0\\
 &  & 1-\nu & 0 & 0 & 0\\
 &  &  & \frac{1-2\nu}{2} & 0 & 0\\
 &  &  &  & \frac{1-2\nu}{2} & 0\\
 &  & \mathrm{Symmetric} &  &  & \frac{1-2\nu}{2}
\end{array}\right].\label{eq:Dmatrix}
\end{equation}
Substituting Eq.~(\ref{eq:strain-nodal}) into Eq.~(\ref{eq:stress-strain}),
the relationship between the stress field and the nodal displacement
functions can be expressed as 
\[
\boldsymbol{\sigma}(\xi,\eta,\zeta)=\mathbf{D}\left(\mathbf{B}_{1}(\eta,\zeta)\mathrm{\mathbf{u}}(\xi)_{,\xi}+\frac{1}{\text{\ensuremath{\xi}}}\mathbf{B}_{2}(\eta,\zeta)\mathrm{\mathbf{u}}(\xi)\right).
\]

\subsection{Scaled boundary finite element equation}

By applying Galerkin's method, the scaled boundary finite element
equation in terms of nodal displacement functions is derived as 
\begin{equation}
\mathbf{E}_{0}\xi^{2}\mathbf{u}(\xi)_{,\xi\xi}+\left(2\mathbf{E}_{0}-\mathbf{E}_{1}+\mathbf{E}_{1}^{\mathrm{T}}\right)\text{\ensuremath{\xi}}\mathbf{u}(\xi)_{,\xi}+\left(\mathbf{E}_{1}^{\mathrm{T}}-\mathbf{E}_{2}\right)\mathbf{u}(\xi)+\mathbf{P}(\xi)-\xi^{2}\mathbf{M}_{0}\ddot{\mathbf{u}}(\xi)=\mathbf{0}.\label{eq:second-order}
\end{equation}
The coefficient matrices $\mathbf{E}_{0}$, $\mathbf{E}_{1}$, $\mathbf{E}_{2}$
and $\mathbf{M}_{0}$ of the entire element are assembled from the
coefficient matrices $\mathbf{E}_{0}^{\mathrm{e}}$, $\mathbf{E}_{1}^{\mathrm{e}}$,
$\mathbf{E}_{2}^{\mathrm{e}}$ and $\mathbf{M}_{0}^{\mathrm{e}}$
belonging to each surface element. The coefficient matrices $\mathbf{E}_{0}^{\mathrm{e}}$,
$\mathbf{E}_{1}^{\mathrm{e}}$, $\mathbf{E}_{2}^{\mathrm{e}}$ and
$\mathbf{M}_{0}^{\mathrm{e}}$ of a surface element $\mathrm{S}_{\mathrm{e}}$
are given as 
\begin{subequations}
\label{eq:coefficient_matrices}
\begin{equation}
\mathbf{E}_{0}^{\mathrm{e}}=\int_{\mathrm{S}_{\mathrm{e}}}\mathbf{B}_{1}^{\mathrm{T}}\mathbf{D}\mathbf{B}_{1}\left|\mathbf{J}_{\mathrm{b}}\right|\mathrm{d}\eta\mathrm{d}\zeta,
\end{equation}
\begin{equation}
\mathbf{E}_{1}^{\mathrm{e}}=\int_{\mathrm{S}_{\mathrm{e}}}\mathbf{B}_{2}^{\mathrm{T}}\mathbf{D}\mathbf{B}_{1}\left|\mathbf{J}_{\mathrm{b}}\right|\mathrm{d}\eta\mathrm{d}\zeta,
\end{equation}
\begin{equation}
\mathbf{E}_{2}^{\mathrm{e}}=\int_{\mathrm{S}_{\mathrm{e}}}\mathbf{B}_{2}^{\mathrm{T}}\mathbf{D}\mathbf{B}_{2}\left|\mathbf{J}_{\mathrm{b}}\right|\mathrm{d}\eta\mathrm{d}\zeta,
\end{equation}
\begin{equation}
\mathbf{M}_{0}^{\mathrm{e}}=\int_{\mathrm{S}_{\mathrm{e}}}\mathbf{N}_{\mathrm{u}}^{\mathrm{T}}\rho\mathbf{N}_{\mathrm{u}}\left|\mathbf{J}_{\mathrm{b}}\right|\mathrm{d}\eta\mathrm{d}\zeta,
\end{equation}
\end{subequations}
 where $\rho$ is the mass density. The load function vector $\mathbf{P}(\xi)$
may include contributions due to surface traction, body forces, and
initial stresses; readers are referred to Ref.~\citep{Song2018}
for details.

The system of non-homogeneous second-order ordinary differential equations
(ODEs) in Eq.~(\ref{eq:second-order}) can be transferred into a
system of homogeneous first-order ODEs by doubling the number of unknowns
and neglecting the inertial and external loads. The variable 
\begin{equation}
\mathbf{X}(\xi)=\left[\begin{array}{c}
\xi^{0.5}\mathrm{\mathbf{u}}(\xi)\\
\xi^{-0.5}\mathrm{\mathbf{q}}(\xi)
\end{array}\right]\label{eq:X}
\end{equation}
is introduced, where $\mathrm{\mathbf{q}}(\xi)$ denotes the vector
of internal nodal force functions on a surface with constant $\xi$
and can be formulated as~\citep{Song2018}
\begin{equation}
\mathbf{q}(\xi)=\xi\left(\mathbf{E}_{0}\xi\mathrm{\mathbf{u}}(\xi)_{,\xi}+\mathbf{E}_{1}^{\mathrm{T}}\mathrm{\mathbf{u}}(\xi)\right).\label{eq:Q}
\end{equation}
Substituting Eqs.~(\ref{eq:X})~and~(\ref{eq:Q}) into Eq.~(\ref{eq:second-order}),
the scaled boundary finite element equation is reformulated as 
\begin{equation}
\xi\mathbf{X}(\xi)_{,\xi}=\mathbf{Z}_{\mathrm{p}}\mathbf{X}(\xi),\label{eq:Euler-Cauchy}
\end{equation}
where the coefficient matrix $\mathbf{\mathbf{Z}_{\mathrm{p}}}$ is
a Hamiltonian matrix and defined as
\begin{equation}
\mathbf{Z}_{\mathrm{p}}=\left[\begin{array}{cc}
-\mathbf{E}_{0}^{-1}\mathbf{E}_{1}^{\mathrm{T}}+0.5\mathbf{I} & \mathbf{E}_{0}^{-1}\\
\mathbf{E}_{2}-\mathbf{E}_{1}\mathbf{E}_{0}^{-1}\mathbf{E}_{1}^{\mathrm{T}} & \mathbf{E}_{1}\mathbf{E}_{0}^{-1}-0.5\mathbf{I}
\end{array}\right].\label{eq:ConstructionZp}
\end{equation}

\subsection{Solution of the scaled boundary finite element equation}

The scaled boundary finite element equation constitutes a system of
Euler-Cauchy equations, see Eq.~(\ref{eq:Euler-Cauchy}), which can
be solved by applying eigenvalue or block-diagonal Schur decompositions~\citep{Song2004}.
The simpler eigenvalue method is utilized for the numerical implementation
of our work, and therefore, a brief summary is presented in this subsection.

The eigenvalue problem of $\mathbf{Z}_{\mathrm{p}}$ can be expressed
in matrix form as 
\begin{equation}
\mathbf{Z}_{\mathrm{p}}\mathbf{\Phi}=\mathbf{\Phi}\mathbf{\Lambda},\label{eq:eigenform}
\end{equation}
where $\mathbf{\Phi}$ denotes the eigenvector matrix, and $\mathbf{\Lambda}$
is the diagonal matrix of eigenvalues. The general solution of $\mathbf{X}(\xi)$
is expressed as: 
\begin{equation}
\mathbf{X}(\xi)=\mathbf{\Phi}\xi^{\mathbf{\text{\ensuremath{\Lambda}}}}\mathbf{C},\label{eq:general solution}
\end{equation}
where $\mathbf{C}$ is the vector of integration constants which depends
on the boundary conditions.

The general solution of $\mathbf{X}(\xi)$, given in Eq.~(\ref{eq:general solution}),
can be sorted by the real parts of the eigenvalues in descending order.
It is well-known that if $\lambda$ is an eigenvalue of a Hamiltonian
matrix, $-\lambda$ is also an eigenvalue of this particular matrix
\citep{van1984}. Thus, Eq.~(\ref{eq:general solution}) can be rewritten
as 
\begin{equation}
\mathbf{X}(\xi)=\mathbf{\left[\begin{array}{cc}
\mathbf{\Phi}_{\mathrm{u1}} & \mathbf{\Phi}_{\mathrm{u2}}\\
\mathbf{\Phi}_{\mathrm{q1}} & \mathbf{\Phi}_{\mathrm{q2}}
\end{array}\right]}\left[\begin{array}{cc}
\xi^{\mathbf{\Lambda}^{+}} & 0\\
0 & \xi^{\mathbf{\Lambda}^{-}}
\end{array}\right]\mathbf{\left[\begin{array}{c}
\mathbf{C}_{\mathrm{1}}\\
\mathbf{C}_{\mathrm{2}}
\end{array}\right]},\label{eq:sorted_solution}
\end{equation}
where $\mathbf{\Phi}_{\mathrm{u1}}$, $\mathbf{\Phi}_{\mathrm{u2}}$,
$\mathbf{\Phi}_{\mathrm{q1}}$, and $\mathbf{\Phi}_{\mathrm{q2}}$
are the sub-matrices of $\mathbf{\Phi}$, and $\mathbf{\Lambda}^{+}$
and $\mathbf{\Lambda}^{-}$ are the diagonal matrices of eigenvalues
with positive and negative real parts, respectively. $\mathbf{C}_{\mathrm{1}}$
and $\mathbf{C}_{\mathrm{2}}$ denote the vectors of integration constants
related to $\mathbf{\Lambda}^{+}$ and $\mathbf{\Lambda}^{-}$, respectively.
For bounded domains, the displacements have to remain finite inside
the element and therefore, the integration constants contained in
$\mathbf{C}_{2}$ have to be equal to zero. Otherwise the power function
$\xi^{\mathbf{\Lambda}^{-}}$ would go to infinity near the scaling
center ($\xi=0$). Consequently, the general solution given by Eq.~(\ref{eq:sorted_solution})
can be simplified as 
\begin{equation}
\mathbf{X}(\xi)=\mathbf{\left[\begin{array}{c}
\mathbf{\mathbf{\Phi}}_{\mathrm{u1}}\\
\mathbf{\mathbf{\Phi}}_{\mathrm{q1}}
\end{array}\right]\xi^{\mathbf{\Lambda}^{+}}\mathbf{C}_{\mathrm{1}}}.\label{eq:simply_solution}
\end{equation}
Using the definition of $\mathbf{X}(\xi)$ in Eq.~(\ref{eq:X}) together
with the general solution provided in Eq.~(\ref{eq:simply_solution}),
the unknown nodal displacement functions $\mathrm{\mathbf{u}}(\xi)$
and nodal force functions $\mathrm{\mathbf{q}}(\xi)$ can be expressed
as 
\begin{subequations}
\begin{equation}
\mathbf{u}(\xi)=\mathbf{\mathbf{\Phi}}_{\mathrm{u1}}\xi^{\mathbf{\Lambda}^{+}-0.5\mathbf{I}}\mathbf{C}_{1},
\end{equation}
\begin{equation}
\mathbf{q}(\xi)=\mathbf{\mathbf{\Phi}}_{\mathrm{q1}}\xi^{\mathbf{\Lambda}^{+}+0.5\mathbf{I}}\mathbf{C}_{1}.
\end{equation}
\end{subequations}

The relationship between the nodal displacement functions and the
nodal force functions can be obtained by eliminating the integration
constants as 
\begin{equation}
\mathbf{q}(\xi)=\mathbf{\mathbf{\Phi}}_{\mathrm{q1}}\mathbf{\mathbf{\Phi}}_{\mathrm{u1}}^{-1}\xi\mathbf{u}(\xi).
\end{equation}
On the boundary ($\xi=1$), the nodal force vector $\mathbf{F}=\mathbf{q}(\xi=1)$
and the nodal displacement vector $\mathbf{d}=\mathbf{u}(\xi=1)$
apply. The relationship between nodal force and displacement vectors
is $\mathbf{F=Kd}$, therefore, the stiffness matrix of the subdomain
is expressed as 
\begin{equation}
\mathbf{K}=\mathbf{\mathbf{\Phi}}_{\mathrm{q1}}\mathbf{\mathbf{\Phi}}_{\mathrm{u1}}^{-1}.\label{eq:stiffmatrix}
\end{equation}

The mass matrix of a volume element is formulated as~\citep{Song2018}
\begin{equation}
\mathbf{M}=\mathbf{\mathbf{\Phi}}_{\mathrm{u1}}^{-\mathrm{T}}\varint_{0}^{1}\xi^{\mathbf{\Lambda}^{+}}\mathbf{m}_{0}\xi^{\mathbf{\Lambda}^{+}}\xi\mathrm{d}\xi\mathbf{\mathbf{\Phi}}_{\mathrm{u1}}^{-1},\label{eq:massmatrix}
\end{equation}
where the coefficient matrix $\mathbf{m}_{0}$ is given by 
\begin{equation}
\mathbf{m}_{0}=\mathbf{\mathbf{\Phi}}_{\mathrm{u1}}^{\mathrm{T}}\mathbf{M}_{0}\mathbf{\mathbf{\Phi}}_{\mathrm{u1}}.
\end{equation}
Transferring the integration part in Eq.~(\ref{eq:massmatrix}) into
a matrix form, the expression of the mass matrix can be rewritten
as 
\begin{equation}
\mathbf{M}=\mathbf{\mathbf{\Phi}}_{\mathrm{u1}}^{-\mathrm{T}}\mathbf{m}\mathbf{\mathbf{\Phi}}_{\mathrm{u1}}^{-1},
\end{equation}
where 
\begin{equation}
\mathbf{m}=\varint_{0}^{1}\xi^{\mathbf{\Lambda}^{+}}\mathbf{m}_{0}\xi^{\mathbf{\Lambda}^{+}}\xi\mathrm{d}\xi.
\end{equation}
Each entry of $\mathbf{m}$ can be evaluated in an analytical fashion,
resulting in 
\begin{equation}
m_{ij}=\frac{m_{0ij}}{\lambda_{ii}^{+}+\lambda_{jj}^{+}+2},\label{eq:massmatrix_2}
\end{equation}
 where $m_{0ij}$ is the entry of $\mathbf{m}_{0}$ and $\lambda_{ii}^{+}$
and $\lambda_{jj}^{+}$ are particular entries of $\mathbf{\Lambda}^{+}$.

\section{Implementation of the SBFEM user element\label{sec:Implementation}}

The implementation of SBFEM user elements involves two user subroutines.
One is required to manage external databases ($\mathrm{UEXTERNALDB}$)
provided by the user, while the other defines the specific user element
($\mathrm{UEL}$). They are written in FORTRAN \citep{ABA2016} and
incorporated into one single source code ({*}.for). The subroutine
$\mathrm{UEXTERNALDB}$ imports the polyhedral mesh data from a supplementary
input file and stores the mesh information in several user-defined
COMMON blocks. The subroutine $\mathrm{UEL}$ performs the computations
of the matrices and load vectors of SBFEM user elements.

In this paper, both the pre- and post-processing are executed outside
of ABAQUS. The pre-processing, performed by an in-house code, prepares
two files for the simulation, an ABAQUS input file ({*}.inp) and a
text file ({*}.txt) for describing the topology of the defined user
elements. There are also commercial software packages available for
generating polyhedral meshes and can be used together with our source
code, to name a few, STAR-CCM+ \citep{cd2017star}, ANSYS Fluent \citep{ANSYSFluent12},
and VoroCrust \citep{abdelkader2019vorocrust}. As the visualization
of user elements is not directly supported by ABAQUS CAE, we output
the nodal displacements through ABAQUS and write the result to external
files which can be visualized by ParaView~\citep{squillacote2007paraview}.
ParaView is a popular visualization tool implemented on the Visualization
ToolKit (VTK). The VTK is a software system that supports a variety
of visualization algorithms including scalar, vector, tensor, which
are compatible with the analysis results (nodal displacements, stresses,
etc.). 

At the end, four files, i.e., the source code containing $\mathrm{UEXTERNALDB}$
and $\mathrm{UEL}$, an ABAQUS input file, a text file storing the
mesh data, and a user-defined include file ({*}.inc) specifying the
dimensions of arrays in the COMMON blocks, should be stored in the
working directory in order to run an analysis in ABAQUS featuring
the proposed user element. The actual analysis is carried out by executing
the following command:
\noindent \begin{center}
\texttt{abaqus job}\textit{=<input file name> }\texttt{user}\textit{=<fortran
file name>.}
\par\end{center}

In this section, detailed explanations concerning the implementation
are given. For a better understanding of the analysis process in ABAQUS
involving user subroutines, a workflow considering the general static
analysis step is provided. The input file formats, including the ABAQUS
input file and the supplementary text file, are demonstrated explicitly
by means of an example, followed by the data structures for storing
polyhedral elements. The structure and detailed descriptions of the
$\mathrm{UEL}$ are presented as well. In addition, a method for defining
element-based surfaces for SBFEM user elements is proposed. These
surfaces are needed to establish interactions for interfacial problems.

\subsection{General workflow of ABAQUS analysis involving user subroutines\label{subsec:General-workflow}}

\begin{figure}
\noindent \begin{centering}
\includegraphics[scale=0.8]{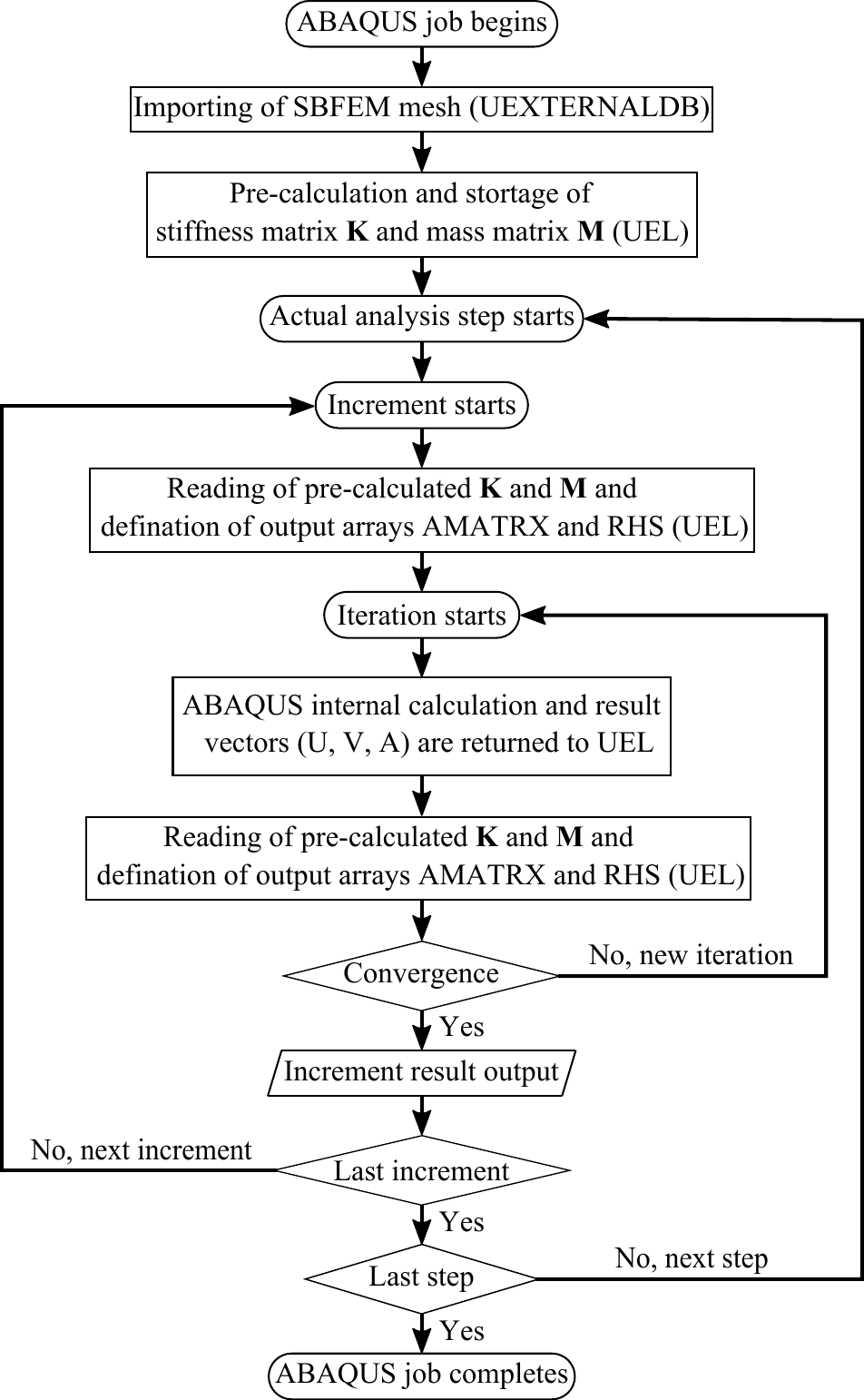}
\par\end{centering}
\noindent \centering{}\caption{General workflow for ABAQUS/Standard analysis involving SBFEM user
subroutines\label{fig:ABAQUS/Standard-logic}}
\end{figure}

The general workflow of an ABAQUS analysis involving SBFEM user subroutines
is depicted in Fig.~\ref{fig:ABAQUS/Standard-logic}. Only the simulation
pipeline is included in the flowchart, neglecting both pre- and post-processing
procedures. The simulation procedure is similar to a standard ABAQUS
analysis except for the fact that two user subroutines $\mathrm{UEXTERNALDB}$
and $\mathrm{UEL}$ are called during run-time. Before the actual
analysis starts, $\mathrm{UEXTERNALDB}$ is called for importing the
polyhedral mesh, and $\mathrm{UEL}$ is called for the pre-calculation
and storage of the stiffness and mass matrices. In each increment
of the actual analysis, the pre-calculated stiffness and mass matrices
are read by $\mathrm{UEL}$, and directly used for the definition
of required output arrays AMATRX and RHS. During the iteration procedure,
the $\mathrm{UEL}$ is called again to check the convergence of the
solution. 

In the present implementation,  the SBFEM user element is linear elastic.
To avoid repetitive calculation of the stiffness matrices of the user
element in a non-linear analysis, a pre-calculation analysis step
(general static) is created before the actual load steps. In such
a step, there is only one increment with no external loading. It converges
after one iteration. The stiffness and mass matrices are calculated
and stored  to provide the required output arrays of $\mathrm{UEL}$
in the iterative analysis of actual load steps. The implementation
is presented in Section \ref{subsec:User-element}.

\subsection{Input file formats}

\begin{figure}
\noindent \begin{centering}
\hfill{}\subfloat[Node set]{\noindent \centering{}\includegraphics[scale=0.8]{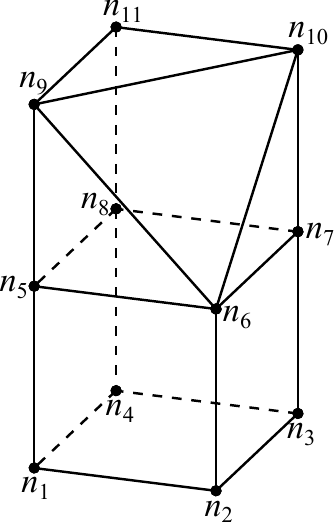}}\hfill{}\subfloat[Surface set]{\noindent \centering{}\includegraphics[scale=0.8]{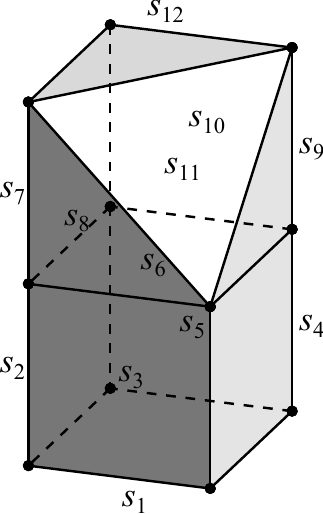}}\hfill{}\subfloat[Element set]{\noindent \centering{}\includegraphics[scale=0.8]{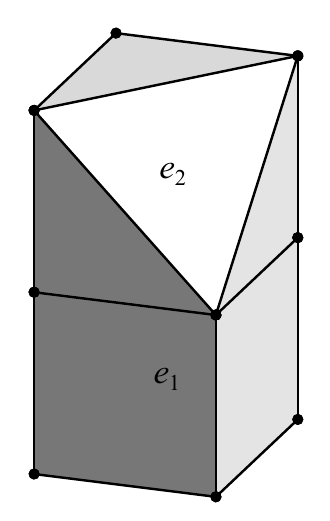}}\hfill{}
\par\end{centering}
\noindent \centering{}\caption{A polyhedron mesh with two elements\label{fig:A-polyhedron-mesh}}
\end{figure}

The input formats are illustrated by means of a simple example consisting
of two polyhedral elements, as depicted in Fig.~\ref{fig:A-polyhedron-mesh}.
The two polyhedral elements are comprised of 11 nodes and 12 surface
elements, respectively. Element 1 consists of six square-shaped surface
elements with eight unique nodes, while Element 2 features seven surface
elements (four triangles, three quadrilaterals) with seven nodes.
The ABAQUS input file involving these SBFEM user elements is depicted
in Fig.~\ref{fig:inp_example}. The text file being read by the user
subroutine $\mathrm{UEXTERNALDB}$ is shown in Fig.~\ref{fig:txt_example}.
For the sake of simplicity, only the first three nodes and surfaces
are included in the listing. 

\subsubsection{ABAQUS input file format}

The ABAQUS input file consists of the following information: nodal
coordinates, SBFEM user elements definition, material properties,
boundary conditions, and the analysis type. Compared to a standard
ABAQUS analysis using default element types, only the formats for
defining user elements and their material properties are different.
\begin{figure}
\noindent \begin{centering}
\includegraphics[scale=0.8]{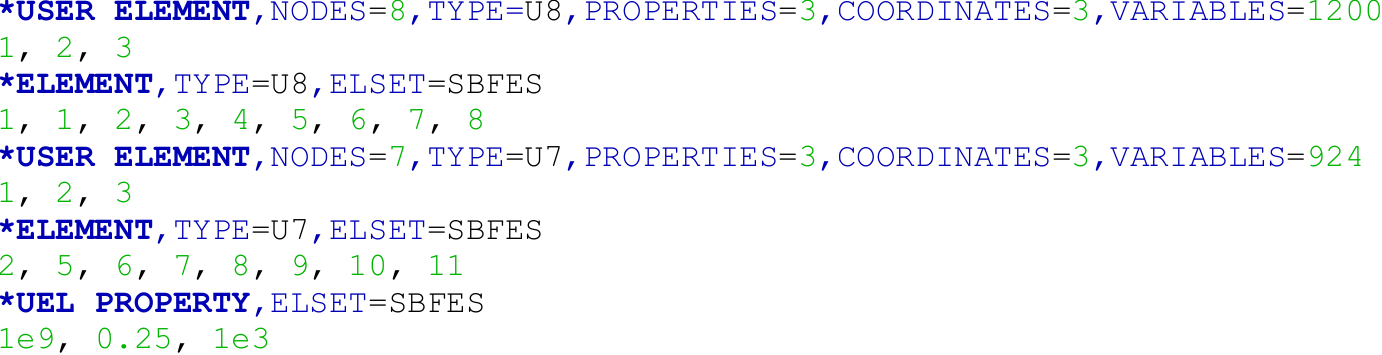}
\par\end{centering}
\noindent \centering{}\caption{An example of ABAQUS input format involving SBFEM user elements\label{fig:inp_example}}
\end{figure}

Figure \ref{fig:inp_example} lists the definition used for the SBFEM
user elements depicted in Fig.~\ref{fig:A-polyhedron-mesh}. With
the keyword $\mathrm{\mathit{\mathrm{*USER\ ELEMENT}}}$, a new type
of user element is declared. Then additional information regarding
this element type, such as the number of nodes ($\mathrm{NODES}$),
the type name ($\mathrm{TYPE}$), the number of property values ($\mathrm{PROPERTIES}$),
the number of degrees of freedom (DOFs) per node ($\mathrm{COORDINATES}$),
and the number of solution-dependent variables ($\mathrm{VARIABLES}$)
are given. In the example, two user element types U8 and U7 are defined.
In the next line, the DOFs per node are specified. In our examples,
we use three-dimensional elements and therefore, each node has three
translational DOFs. In ABAQUS, 1 denotes a displacement DOF in $x$-direction,
while 2 and 3 represent the corresponding displacement DOFs in $y$-
and $z$-directions, respectively. The definition of individual elements
is similar to the standard format starting with the keyword $\mathrm{*ELEMENT}$.
Here, only the element type that usually belongs to the standard ABAQUS
element library is replaced by our user element type. The following
line creates a single element by listing the element number and its
nodes. The element property definition starts with the keyword $\mathrm{*UEL\ PROPERTY}$,
then specifies the element set ($\mathrm{ELSET}$) it is assigned
to. In the example, the element set is SBFES which includes Element
1 and Element 2 defined in the user element types U8 and U7, respectively.
The property values are provided in the following line representing
Young's modulus, Poisson's ratio and mass density. In our implementation,
the solution-dependent variables contains static residual of the previous
increment, static residual of the current increment, stiffness matrix,
and mass matrix. Therefore for a user element with $n$ DOFs, the
value of $\mathrm{VARIABLES}$ should be $n+n+n^{2}+n^{2}$.

Among the various definitions, two points deserve special mentions: 
\begin{enumerate}
\item The SBFEM user element type is essentially determined by the number
of nodes. For a two- or three-dimensional problem, the number of element
nodes determines the number of DOFs of each element and further the
dimensions of the output arrays of the $\mathrm{UEL}$. The SBFEM
user elements having the same number of nodes can be classified as
the same type, regardless of the element topology.
\item There is no requirement for the node ordering for one SBFEM user element,
i.e., it can be arbitrary. For an ABAQUS built-in element, the node
ordering should follow specific rules because it defines the topology
of this element type. However, it is impossible to set a universal
principle of describing the topology of the polyhedral element based
only on the node ordering. The element topology will be described
in a supplementary input text file introduced in Section~\ref{subsec:Text-file-format}.
The node ordering of the SBFEM user element only affects the location
vector for the assembly of the global stiffness matrix and force vector.
\end{enumerate}

\subsubsection{Text file format\label{subsec:Text-file-format}}

As mentioned above, the ABAQUS input file only lists the node identifiers
of each element, which is insufficient to describe the topology of
polyhedral elements. Therefore, it is compulsory to provide additional
information for the SBFEM user elements through a supplementary input
file. There are four types of information in the supplementary file,
shown in Fig.~\ref{fig:txt_example}, i.e., (i) nodal coordinates,
(ii) surface information, (iii) element information, and (iv) scaling
center coordinates. At the beginning of each type of information,
one integer is declared to identify the number of the specific entity
such as nodes, surfaces, and elements. For each surface, one integer
indicating the number of nodes is declared followed by the nodal connectivity
of this surface, either clockwise or counter-clockwise. For the definition
of a polyhedron, the number of surfaces is stated first, and then
the identifiers of the surfaces enclosing the polyhedron are listed.
The sign of surface ID is positive if the normal vector of the surface
is pointing outward of the cell, otherwise it is negative. 
\begin{figure}
\noindent \begin{centering}
\includegraphics[scale=0.8]{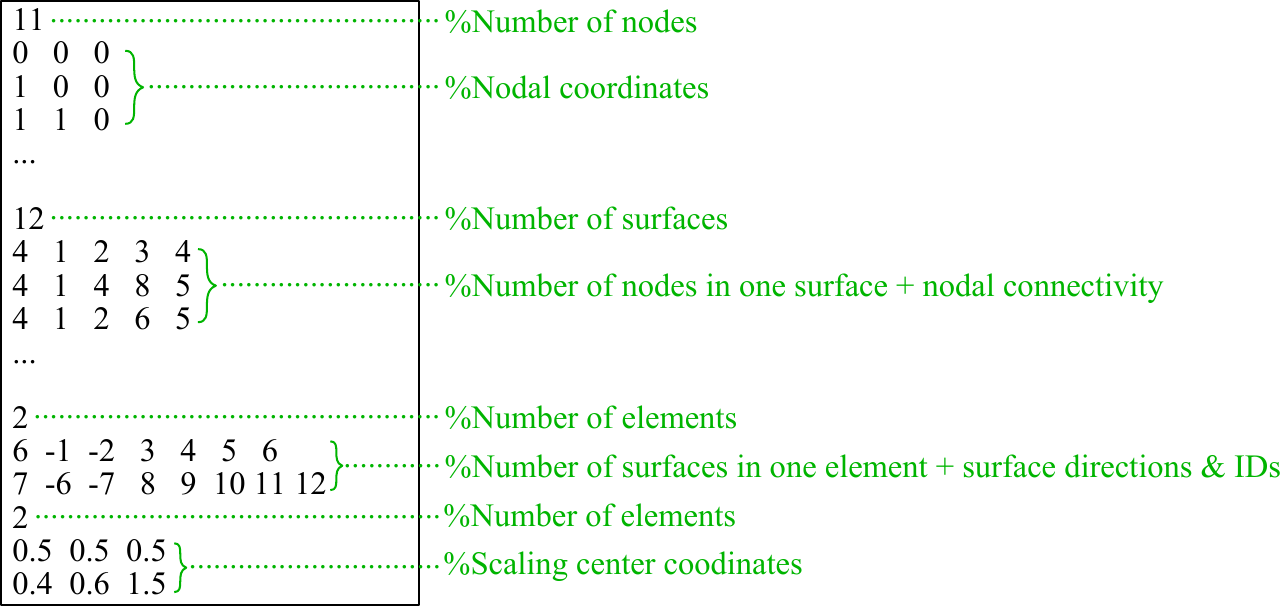}
\par\end{centering}
\noindent \centering{}\caption{An example of text input format describing SBFEM polyhedral mesh\label{fig:txt_example}}
\end{figure}

\subsection{Data structure of the mesh information}

The user subroutine $\mathrm{UEXTERNALDB}$ is the interface subroutine
which allows communication between other programs and user subroutines
within ABAQUS/Standard. In our implementation, it only plays a role
at the beginning of the analysis. $\mathrm{UEXTERNALDB}$ reads the
text file describing the SBFEM discretization and stores the mesh
data into user-defined COMMON blocks. The mesh information allows
the $\mathrm{UEL}$ to construct the element topology for defining
the stiffness matrix $\mathbf{K}$, the mass matrix $\mathbf{M}$,
etc.

Three COMMON blocks are created when the subroutine $\mathrm{UEXTERNALDB}$
is called. They are named as NDINFO, SFINFO and ELINFO, for storing
the node, surface and element information, respectively. The explicit
data structure of each block for the previous example is shown in
Table~\ref{tab:Data-structure-stored}. 
\begin{table}
\noindent \begin{centering}
\caption{Data structure stored in the COMMON blocks\label{tab:Data-structure-stored}}
\subfloat[Node information block NDINFO]{\noindent \begin{centering}
\par\end{centering}
\noindent \centering{}\hspace{0.5cm}%
\begin{tabular}{>{\centering}m{0.75cm}>{\centering}m{0.75cm}>{\centering}m{0.75cm}}
\hline 
\multicolumn{3}{c}{NDCRD}\tabularnewline
\hline 
0.0 & 0.0 & 0.0\tabularnewline
1.0 & 0.0 & 0.0\tabularnewline
1.0 & 1.0 & 0.0\tabularnewline
 & $\cdots$ & \tabularnewline
\hline 
\end{tabular}\hspace{0.5cm}}\hspace{0.5cm}\subfloat[Surface information block SFINFO]{\noindent \centering{}%
\begin{tabular}{>{\centering}m{0.375cm}>{\centering}m{0.375cm}>{\centering}m{0.375cm}>{\centering}m{0.375cm}>{\centering}m{0.375cm}>{\centering}m{0.375cm}>{\centering}m{0.375cm}>{\centering}m{0.375cm}>{\centering}m{0.375cm}>{\centering}m{0.375cm}>{\centering}m{0.375cm}>{\centering}m{0.375cm}}
\hline 
\multicolumn{12}{c}{SFCONN}\tabularnewline
\hline 
1 & 2 & 3 & 4 & 1 & 4 & 8 & 5 & 1 & 2 & 6 & $\cdots$\tabularnewline
\hline 
 &  &  &  &  &  &  &  &  &  &  & \tabularnewline
\hline 
\multicolumn{12}{c}{SFINDEX}\tabularnewline
\hline 
4 & 8 & 12 & 16 & 20 & 24 & 28 & 31 & 34 & 38 & 41 & 44\tabularnewline
\hline 
\end{tabular}}
\par\end{centering}
\noindent \centering{}\subfloat[Element information block ELINFO]{\noindent \centering{}%
\begin{tabular}{>{\centering}m{0.375cm}>{\centering}m{0.375cm}>{\centering}m{0.375cm}>{\centering}m{0.375cm}>{\centering}m{0.375cm}>{\centering}m{0.375cm}>{\centering}m{0.375cm}>{\centering}m{0.375cm}>{\centering}m{0.375cm}>{\centering}m{0.375cm}>{\centering}m{0.375cm}>{\centering}m{0.375cm}>{\centering}m{0.375cm}}
\hline 
\multicolumn{13}{c}{ELCONN}\tabularnewline
\hline 
-1 & -2 & 3 & 4 & 5 & 6 & -6 & 7 & 8 & 9 & 10 & 11 & 12\tabularnewline
\hline 
 &  &  &  &  &  &  &  &  &  &  &  & \tabularnewline
\cline{3-5} \cline{4-5} \cline{5-5} \cline{8-12} \cline{9-12} \cline{10-12} \cline{11-12} \cline{12-12} 
 &  & \multicolumn{3}{c}{ELINDEX} &  &  & \multicolumn{5}{c}{CTCRD} & \tabularnewline
\cline{3-5} \cline{4-5} \cline{5-5} \cline{8-12} \cline{9-12} \cline{10-12} \cline{11-12} \cline{12-12} 
 &  & 6 &  & 13 &  &  & 0.5 &  & 0.5 &  & 0.5 & \tabularnewline
\cline{3-5} \cline{4-5} \cline{5-5} 
 &  &  &  &  &  &  & 0.4 &  & 0.6 &  & 1.5 & \tabularnewline
\cline{8-12} \cline{9-12} \cline{10-12} \cline{11-12} \cline{12-12} 
\end{tabular}}
\end{table}

The COMMON block NDINFO contains a matrix named NDCRD with dimension
$n_{\mathrm{nd}}\times3$ to store nodal coordinates, in which $n_{\mathrm{nd}}$
denotes the number of nodes. 

The COMMON block SFINFO contains two vectors named SFCONN and SFINDEX,
respectively. The vector SFCONN stores the nodal connectivity of each
surface in a continuous fashion. An indexing vector SFINDEX is stored
to recover the individual surface elements. It has $n_{\mathrm{sf}}$
entries, where $n_{\mathrm{sf}}$ represents the number of surfaces.
Each entry of SFINDEX indicates the final storage position in SFCONN
for each surface. The number of nodes for one surface can be calculated
by the difference between the positioning entries of itself and the
former surface. The first entry of SFINDEX is the number of nodes
for the first surface as well.

The COMMON block ELINFO contains a vector named ELCONN, a vector named
ELINDEX, and a matrix named CTCRD with dimension $n_{\mathrm{el}}\times3$,
where $n_{\mathrm{el}}$ is the number of elements. Similar to SFCONN,
ELCONN stores the surfaces enclosing each element continuously, by
the order of element ID. The indexing vector ELINDEX has $n_{\mathrm{el}}$
entries, which identify the final storage position of each element
in ELCONN. The matrix CTCRD stores the scaling center coordinates
of each element. For a subdomain without cracking, the scaling center
may be calculated as the average coordinates of all nodes in each
element. However, for cracked subdomain used in fracture mechanics,
the scaling center should be specified as the crack tip~\citep{Saputra2015}.
To make the implementation more versatile, we purposely decided to
provide the scaling center as additional input parameter.

The proposed scheme for storing polyhedral meshes is comparably memory-efficient.
Only the matrices for storing coordinates, i.e., NDCRD and CTCRD,
are provided as float data type while the components of other vectors
are all integers. To minimize the memory requirements, we opted to
store the surface and element information in vectors instead of matrices.
Note that the number of nodes of each surface is not uniform and the
polyhedral elements have arbitrary surfaces, which may induce zero
entries in matrices. This issue can be avoided by using vectors to
store the surface and mesh information.

\subsection{User element\label{subsec:User-element}}

The user subroutine $\mathrm{UEL}$ is used to define a special element
type not available in ABAQUS. It must provide all element level calculations,
such as the definition of the stiffness and mass matrices as well
as the residual force vector. To this end, ABAQUS provides some input
variables to the $\mathrm{UEL}$, such as the current element number
$\mathrm{JELEM}$, nodal coordinates $\mathrm{COORDS}$, material
properties $\mathrm{PROPS}$, current nodal displacement vector U,
and current solution procedure type $\mathrm{LFLAGS}$. Those variables
are defined in the ABAQUS input file or generated automatically according
to the simulation procedure. The most important arrays to output by
the $\mathrm{UEL}$ are $\mathrm{AMATRX}$ and $\mathrm{RHS}$. For
static analysis, $\mathrm{AMATRX}$ and $\mathrm{RHS}$ are the current
stiffness matrix and residual force vector, respectively. In other
analysis types, these generic arrays have different meanings. Detailed
definitions of $\mathrm{AMATRX}$ and $\mathrm{RHS}$ for different
solution procedures are out of the scope of the current section. The
reader is referred to the ABAQUS user subroutine reference manual
\citep{ABA2016} for a comprehensive discussion.

The flowchart of the $\mathrm{UEL}$ subroutine implementing SBFEM
is shown in Fig. \ref{fig:Flowchart-of-UEL}. In the following, the
functions and specific implementations of each process are explained
in detail: 
\begin{figure}[t]
\noindent \begin{centering}
\includegraphics[scale=0.8]{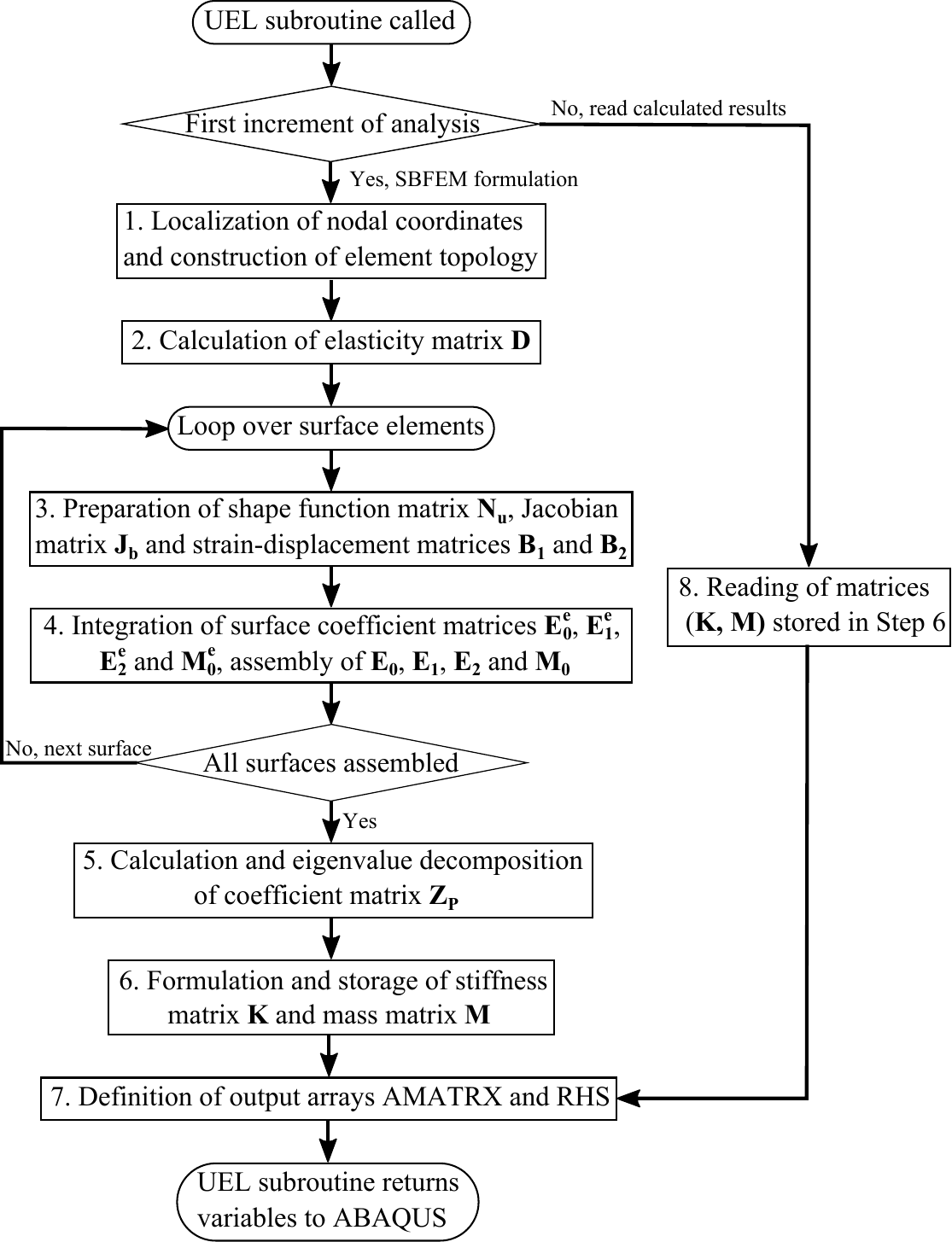}
\par\end{centering}
\noindent \centering{}\caption{Flowchart of user subroutine UEL for SBFEM\label{fig:Flowchart-of-UEL}}
\end{figure}

\begin{enumerate}
\item Localization of nodal coordinates and construction of element topology:
In this step, the global nodal coordinates are transformed into relative
nodal coordinates with respect to the scaling center, and the element
topology is constructed using local nodal identifiers. Two matrices
named as RCRD and SELE are generated for storing the local nodal coordinates
and the element topology, respectively. The global nodal coordinates
($x,y,z$) of the nodes belonging to the current element are passed
into $\mathrm{UEL}$ as input matrix $\mathrm{COORDS}$. They are
transformed into relative nodal coordinates ($\hat{x},\hat{y},\hat{z}$)
with respect to the scaling center, which are used in Eq.~(\ref{eq:scaledcoordinate}).
The sequence of the nodal coordinates conforms to the node ordering
of each element given in the ABAQUS input file. The element topology
can be extracted from the COMMON blocks when the element number $\mathrm{JELEM}$
is known. Note that the connectivity is given in terms of the global
nodal identifiers. To meet the global assembly requirement, the calculation
of $\mathrm{AMATRX}$ and $\mathrm{RHS}$ should match the node ordering
given in the ABAQUS input file. Therefore, the element topology should
be described using local nodal identifiers. Matching the nodal coordinates
stored in $\mathrm{COORDS}$ and NDINFO, the correspondence between
local and global nodal identifiers is obtained. Then, the element
topology can be re-constructed by local nodal identifiers. During
the construction of the element topology, the nodal connectivity of
negative surfaces should be reordered for the mapping consideration.
The RCRD and SELE for Element 2 (depicted in Fig. \ref{fig:A-polyhedron-mesh})
are shown explicitly in Table \ref{tab:Data-structures-of-localization},
in which $n_{\mathrm{nd}}^{\mathrm{s}}$ represents the node number
of each surface. 
\begin{table}
\noindent \begin{centering}
\caption{Data structures of relative nodal coordinates and element topology\label{tab:Data-structures-of-localization}}
\par\end{centering}
\noindent \centering{}\subfloat[Node ID correspondence]{\noindent \centering{}\hspace{1.1cm}%
\begin{tabular}{>{\centering}p{0.75cm}>{\centering}p{0.75cm}}
\hline 
Global & Local\tabularnewline
\hline 
5 & 1\tabularnewline
6 & 2\tabularnewline
7 & 3\tabularnewline
8 & 4\tabularnewline
9 & 5\tabularnewline
10 & 6\tabularnewline
11 & 7\tabularnewline
\hline 
\end{tabular}\hspace{1.1cm}}\hspace{0.4cm}\subfloat[RCRD]{\noindent \centering{}%
\begin{tabular}{>{\raggedleft}p{0.6cm}>{\raggedleft}p{0.6cm}>{\raggedleft}p{0.6cm}}
\hline 
$\hat{x}$ & $\hat{y}$ & $\hat{z}$\tabularnewline
\hline 
-0.4 & -0.6 & -0.5\tabularnewline
0.6 & -0.6 & -0.5\tabularnewline
0.6 & 0.4 & -0.5\tabularnewline
-0.4 & 0.4 & -0.5\tabularnewline
-0.4 & -0.6 & 0.5\tabularnewline
0.6 & 0.4 & 0.5\tabularnewline
-0.4 & 0.4 & 0.5\tabularnewline
\hline 
\end{tabular}}\hspace{1.5cm}\subfloat[SELE]{\noindent \centering{}%
\begin{tabular}{>{\centering}p{0.6cm}>{\centering}p{0.6cm}>{\centering}p{0.6cm}>{\centering}p{0.6cm}>{\centering}p{0.6cm}}
\hline 
$n_{\mathrm{nd}}^{\mathrm{s}}$ & \multicolumn{4}{c}{Local nodal connectivity}\tabularnewline
\hline 
4 & 4 & 3 & 2 & 1\tabularnewline
4 & 5 & 7 & 4 & 1\tabularnewline
3 & 1 & 2 & 5 & 0\tabularnewline
3 & 2 & 3 & 6 & 0\tabularnewline
4 & 3 & 4 & 7 & 6\tabularnewline
3 & 2 & 6 & 5 & 0\tabularnewline
3 & 5 & 6 & 7 & 0\tabularnewline
\hline 
\end{tabular}}
\end{table}
\item Calculation of the elasticity matrix $\mathbf{D}$: The material properties
such as Young's modulus, Poisson's ratio and mass density are stored
in the input array named PROPS. Using Young's modulus and Poisson's
ratio, the elasticity matrix $\mathbf{D}$ can be formulated as in
Eq.~(\ref{eq:Dmatrix}).
\item Preparation of matrices $\mathbf{N_{\mathrm{u}}}$, $\mathbf{J}_{\mathrm{b}}$,
$\mathbf{B}_{1}$ and $\mathbf{B}_{2}$: The shape function matrix
$\mathbf{N_{\mathrm{u}}}$ depends on the surface element type, which
can be either a triangular or quadrilateral element, with linear or
quadratic shape functions. The surface element type has been defined
in the matrix SELE generated in the previous step. Extracting the
nodal coordinate vectors $\mathbf{x}$, $\mathbf{y}$, and $\mathbf{z}$
of the surface from the matrix RCRD, together with the shape function
matrix $\mathbf{N}_{\mathrm{u}}$, the Jacobian matrix $\mathbf{J}_{\mathrm{b}}$
is formulated following Eqs.~(\ref{eq:scaledcoordinate})~and~(\ref{eq:Jacobian}).
After computing the matrices $\mathbf{b}_{1}$, $\mathbf{b}_{2}$
and $\mathbf{b}_{3}$ based on $\mathbf{J}_{\mathrm{b}}^{-1}$, the
matrices $\mathbf{B}_{1}$ and $\mathbf{B}_{2}$ are calculated by
Eq.~(\ref{eq:B1B2}).
\item Integration of $\mathbf{E}_{0}^{\mathrm{e}}$, $\mathbf{E}_{1}^{\mathrm{e}}$,
$\mathbf{E}_{2}^{\mathrm{e}}$ and $\mathbf{M}_{0}^{\mathrm{e}}$:
The integrals to determine the coefficient matrices of the individual
surface elements, see Eq.~(\ref{eq:coefficient_matrices}), are solved
using numerical quadrature rules, in our case, a standard Gaussian
quadrature rule. Then the coefficient matrices $\mathbf{E}_{0}$,
$\mathbf{E}_{1}$, $\mathbf{E}_{2}$ and $\mathbf{M}_{0}$ of the
entire element are assembled from the coefficient matrices $\mathbf{E}_{0}^{\mathrm{e}}$,
$\mathbf{E}_{1}^{\mathrm{e}}$, $\mathbf{E}_{2}^{\mathrm{e}}$ and
$\mathbf{M}_{0}^{\mathrm{e}}$ of all surface elements.
\item Calculation and eigenvalue decomposition of $\mathbf{Z}_{\mathrm{p}}$:
The expression of the coefficient matrix $\mathbf{Z}_{\mathrm{p}}$
is provided in Eq.~(\ref{eq:ConstructionZp}), which can be computed
based on $\mathbf{E}_{0}$, $\mathbf{E}_{1}$, $\mathbf{E}_{2}$ and
$\mathbf{E}_{0}^{-1}$. For computing the inverse of $\mathbf{E}_{0}$,
two subroutines DGETRF and DGETRS from the Linear Algebra PACKage
(LAPACK)~\citep{Anderson1999} are called. DGETRF prepares the LU
factorization of $\mathbf{E}_{0}$ for DGETRS to compute $\mathbf{E}_{0}^{-1}$.
The eigenvalue decomposition of $\mathbf{Z}_{\mathrm{p}}$ is implemented
by calling the subroutine DGEEV of LAPACK. DGEEV provides a matrix
and two vectors as output parameters. The matrix alternately stores
the real and imaginary parts of the right eigenvectors in its columns.
The two vectors store the real and imaginary parts of eigenvalues,
respectively. They are essentially the eigenvector matrix and eigenvalues
in Eq.~(\ref{eq:eigenform}), but in the DGEEV output format. The
real parts of eigenvalues are sorted by the descending order through
bubble sort algorithm \citep{astrachan2003bubble}. The imaginary
parts of eigenvalues and eigenvectors are sorted also following the
sorted real parts of the eigenvalues. Combining real parts and imaginary
parts, the eigenvector matrix $\mathbf{\Phi}$ and the diagonal matrix
of eigenvalues $\mathbf{\Lambda}$ given in Eq.~(\ref{eq:sorted_solution})
are obtained. They are complex matrices and stored with double complex
data type. The sub-matrices $\mathbf{\Phi}_{\mathrm{u1}}$, $\mathbf{\Phi}_{\mathrm{q1}}$
and $\mathbf{\Lambda}^{+}$ can be extracted easily from $\mathbf{\Phi}$
and $\mathbf{\Lambda}$. 
\item Formulation and storage of $\mathbf{K}$ and $\mathbf{M}$: The stiffness
matrix $\mathbf{K}$ can be calculated by Eq.~(\ref{eq:stiffmatrix}),
in which the calculation of the inverse of sub-matrices $\mathbf{\Phi}_{\mathrm{u1}}$
can be done analogously to the inversion of $\mathbf{E}_{0}$ discussed
in Step 5. The mass matrix $\mathbf{M}$ is computed step by step
through implementing the Eqs.~(\ref{eq:massmatrix})--(\ref{eq:massmatrix_2}).
At the end of the first increment of an analysis, the stiffness and
mass matrices are stored in the array $\mathrm{SVARS}$ of the $\mathrm{UEL}$
for later use. 
\item Definition of $\mathrm{AMATRX}$ and $\mathrm{RHS}$: Generally speaking,
for all types of solution procedure related to solid mechanics, we
can defined the output arrays $\mathrm{AMATRX}$ and $\mathrm{RHS}$
based on the existing calculated arrays and input arrays. $\mathrm{AMATRX}$
is always defined as stiffness matrix $\mathbf{K}$, mass matrix $\mathbf{M}$,
damping matrix $\mathbf{C}$ or a combination thereof obeying specific
rules. The damping matrix $\mathbf{C}$ of a linear user element can
be defined by Rayleigh damping~\citep{ABA2016}
\begin{equation}
\mathbf{C}=\alpha\mathbf{M}+\beta\mathbf{K},
\end{equation}
where $\alpha$ and $\beta$ are the user-specified Rayleigh damping
factors. $\mathrm{RHS}$ is always defined through the internal force
vector or its combination with the inertia force vector obeying specific
rules. The internal force vector can be obtained by 
\begin{equation}
\mathrm{\mathbf{F}_{S}=\mathbf{KU}},
\end{equation}
where $\mathbf{U}$ is the current nodal displacement vector, one
input array of the $\mathrm{UEL}$. For linear static problem, $\mathrm{RHS}$
is formulated as $-\mathbf{F}_{\mathrm{S}}$. The inertia force vector
can be calculated as
\begin{equation}
\mathrm{\mathbf{F}_{I}=\mathbf{MA}}.
\end{equation}
$\mathbf{A}$ is the current nodal acceleration vector, and it is
also one input array of the $\mathrm{UEL}$. The inertia force vector
$\mathbf{F}_{\mathrm{I}}$ contributes to $\mathrm{RHS}$ in dynamic
problems.
\item Reading of matrices ($\mathbf{K}$, $\mathbf{M}$) stored in Step
6: For the increments except for the first one, the stiffness and
mass matrices are retrieved directly from the array $\mathrm{SVARS}$.
\end{enumerate}

\subsection{Element-based surface definition\label{subsec:Element-based-surface-definition}}

The element-based surface definition for the SBFEM user element is
important to take advantage of different features inherent to ABAQUS.
By defining an element-based surface, distributed loads such as pressure
and surface traction can be assigned to that surface. More importantly,
for interfacial problems involving interactions such as tie constraints
or contact definitions, it is a prerequisite to define surfaces to
establish these interactions.

Currently, element-based surfaces can be created on solid, continuum
shell, and cohesive elements in ABAQUS. On user elements, only node-based
surfaces can be created, and they act only as slave surfaces and always
use node-to-surface schemes in contact analysis, which is a severe
shortcoming as the node-to-surface contact may fail the contact patch
test even for a flat interface~\citep{xing2019}.

By overlaying standard elements with nearly zero stiffness on the
SBFEM user elements, it is possible to define element-based surfaces
for the SBFEM user elements. In the numerical examples, presented
in the following section, the Young's modulus of the overlaid elements
is chosen in accordance with the value of the adjacent user element
and scaled accordingly, i.e., $E_{\mathrm{ov}}=10^{-16}E_{\mathrm{UEL}}$.
This methodology has originally been proposed in the ABAQUS manual~\citep{ABA2016}
to visualize the user elements and is modified in this contribution
to define element-based surfaces. The modification is related to the
fact that instead of overlaying the entire polyhedral element with
the standard finite elements only the volume sector needed for the
surface definition are included. The volume sectors of the three-dimensional
SBFEM user elements are either of tetrahedral or pyramidal shape depending
on the corresponding surface element type, i.e., either triangular
or quadrilateral. Those shapes can be easily overlaid by standard
elements from ABAQUS' element library. As the scaling center fulfills
the star-convexity requirement, i.e., the entire boundary is directly
visible from this point, the process of creating this overlay mesh
is robust for all types of polyhedrons. Hence, the element-based surfaces
can be directly created on the overlaid standard elements as they
share the DOFs with the corresponding SBFEM user elements. Note that
this methodology does not affect the solution quality as the overlay
elements contribute a negligible stiffness to the global system. Considering
the shape functions on the surface of the SBFEM user elements it must
be stressed again that they coincide with those of the elements from
ABAQUS' element library.

\begin{figure}
\hfill{}\subfloat[Overlaying operation\label{fig:Linear-overlaying}]{\centering{}\includegraphics[scale=0.8]{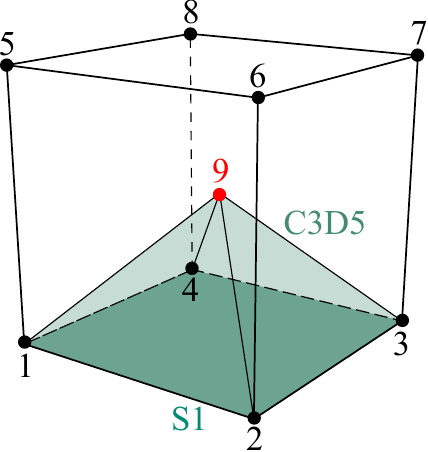}}\hfill{}\subfloat[Input format\label{fig:Input-format-overlay}]{\centering{}\includegraphics[scale=0.8]{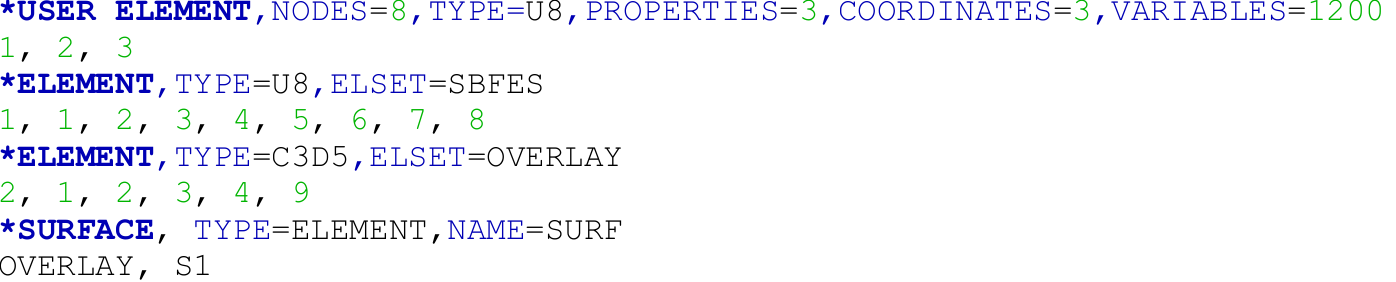}}\hfill{}

\caption{Overlaying standard element on SBFEM user element for defining surface\label{fig:Overlaying-standard-element}}
\end{figure}

The operation of overlaying standard elements on the volume sectors
of the SBFEM user element is illustrated in Fig.~\ref{fig:Overlaying-standard-element}.
This particular example is based on the SBFEM user element depicted
in Fig.~\ref{fig:Linear-overlaying} which consists of six surfaces
and eight nodes. Note that the scaling center is marked in red (Node
9). Here, a quadrilateral element-based surface is created by introducing
a pyramidal overlay element that is superimposed on the volume sector
created be surface element S1 and the scaling center. This procedure
is straightforward as ABAQUS offers pyramidal finite elements such
as the C3D5 element. The ABAQUS input file format for defining this
element-based surface is given in Fig.~\ref{fig:Input-format-overlay}.
It is important to keep in mind that in our implementation, the scaling
center is not an actual node of the element and consequently, must
be defined in the ABAQUS input file. The surface identifier S1 indicates
the surface formed by the first four nodes of the pyramidal element~\citep{ABA2016}.
If a triangular element-based surface is to be created, the corresponding
volume sector must be superimposed by a standard tetrahedral element,
e.g., the C3D4 element. 

\section{Numerical examples\label{sec:Numerical-examples}}

The objective of this section is to verify the implementation and
demonstrate the advantages of including SBFEM user elements. This
is achieved by means of various numerical examples. In the first subsection,
benchmark tests are performed to check the correct implementation.
The second subsection presents examples considering the treatment
of non-matching meshes. The third subsection includes examples that
are related to automatically generated meshes based on image and STL
data. In these cases, robust and efficient octree mesh generation
algorithms are employed. The computational times reported are measured
on a DELL workstation with an Intel Xeon E5-2637 CPU running Windows
10 Pro, 64 bit.

\subsection{Benchmark tests}

This subsection is designated to verify the implementation of SBFEM
user elements by performing static, modal, and transient analyses
on simple geometries. A patch test is performed on a quadrangular
prism under uniaxial tension in order to check the implementation
of the stiffness matrix. Modal and transient analyses are performed
on a cantilever beam to verify the implementation of the mass matrix.

\subsubsection{Patch test}

\begin{figure}
\noindent \begin{centering}
\includegraphics[scale=0.8]{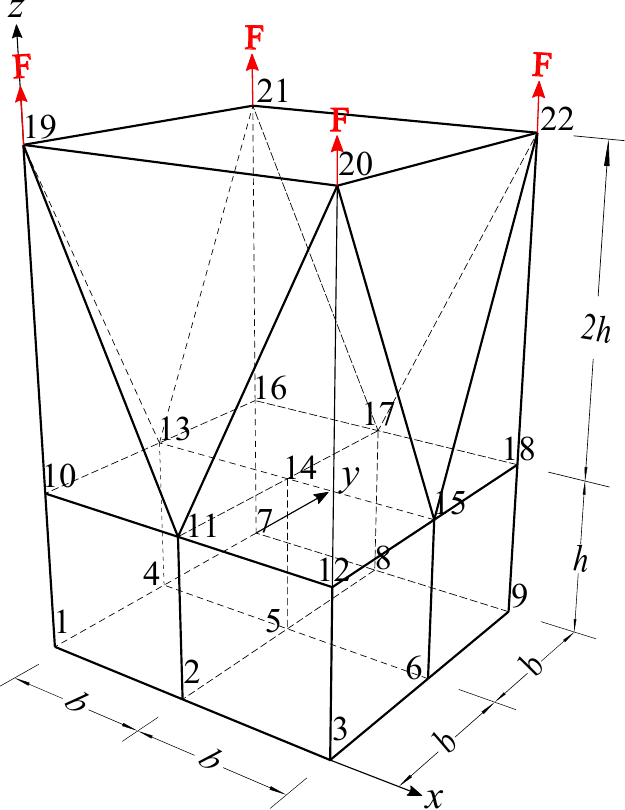}
\par\end{centering}
\noindent \centering{}\caption{Modeling of a quadrangular prism\label{fig:Modeling-prism}}
\end{figure}

A quadrangular prism is discretized by five polyhedral elements as
shown in Fig.~\ref{fig:Modeling-prism}. The dimensions are indicated
in Fig.~\ref{fig:Modeling-prism} with $b=h=1\,\mathrm{m}$. The
material constants are Young's modulus $E=10\,\mathrm{GPa}$ and Poisson's
ratio $\nu=0.25$. The left surface (nodes with $x=0$) of the quadrangular
prism is constrained in $x$-direction, the front surface (nodes with
$y=0$) is constrained in $y$-direction, and the bottom surface (nodes
with $z=0$) is constrained in $z$-direction. A vertical force $F=1000\,\mathrm{kN}$
is applied at Nodes 19, 20, 21 and 22 on the top surface.

\begin{figure}
\noindent \begin{centering}
\hfill{}\subfloat[Displacement contour $u_{z}$ (Unit: m)]{\noindent \centering{}\includegraphics[scale=0.04]{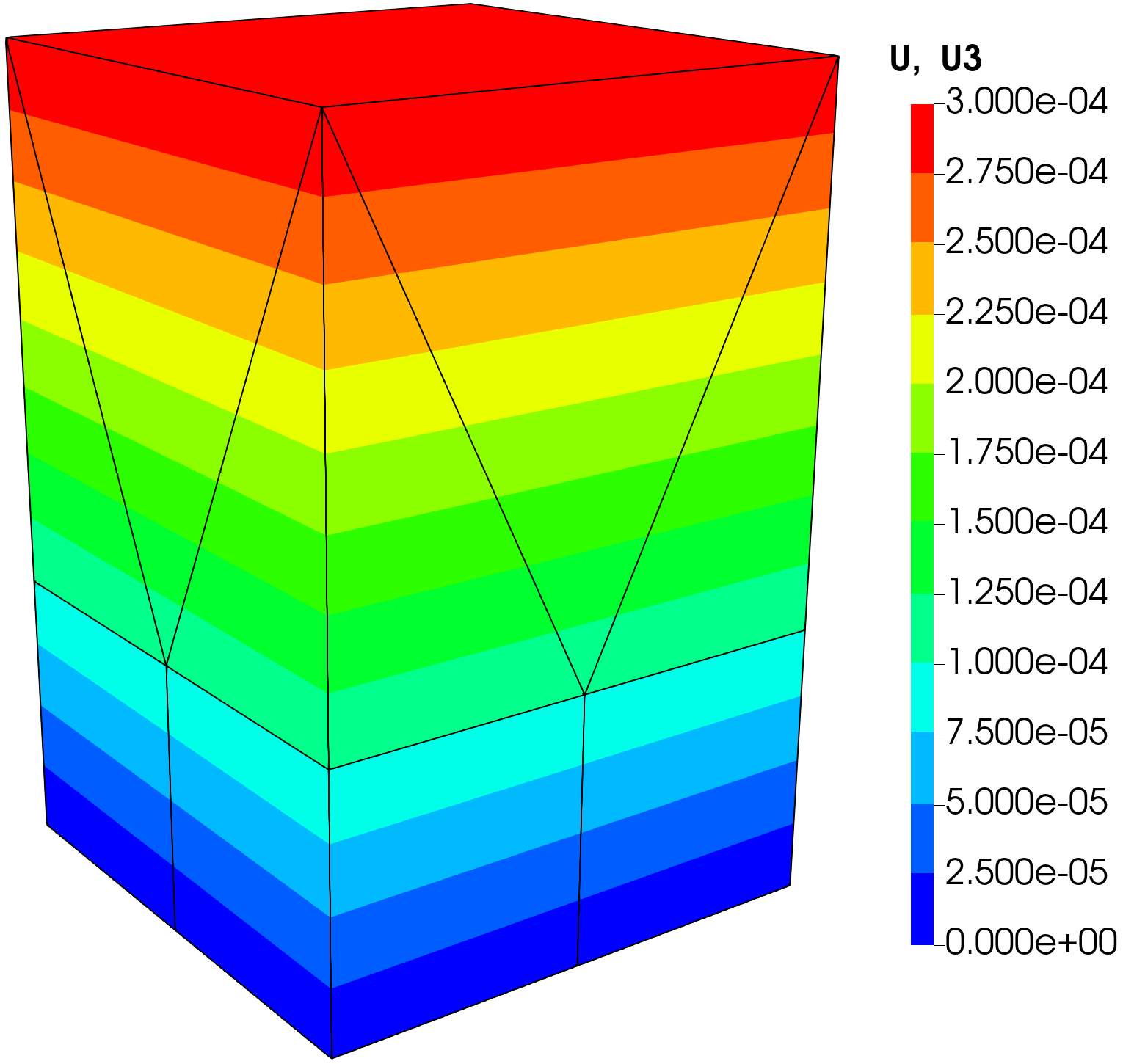}}\hfill{}\subfloat[Stress contour $\sigma_{z}$ (Unit: Pa)]{\noindent \centering{}\includegraphics[scale=0.04]{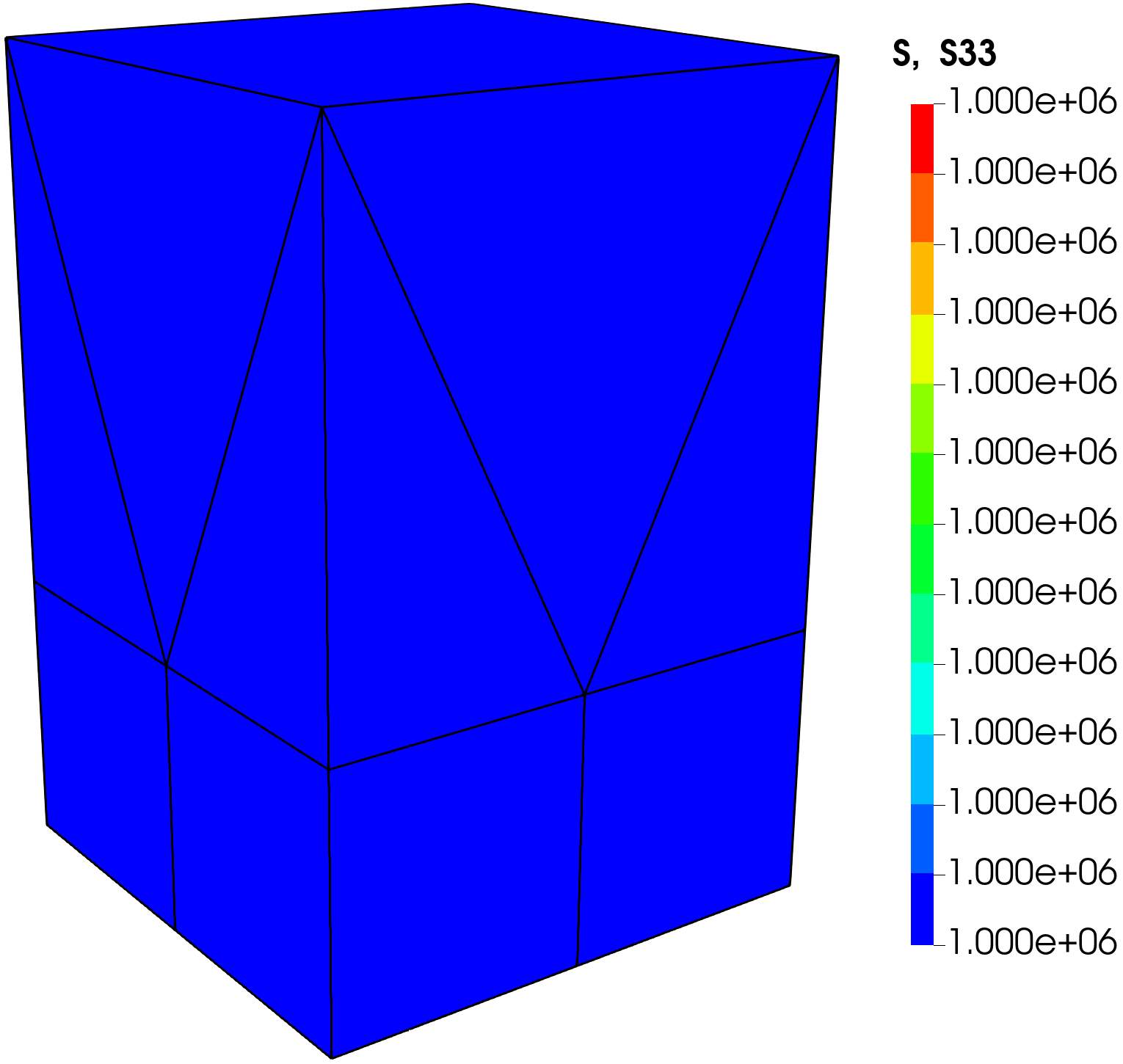}}\hfill{}
\par\end{centering}
\noindent \centering{}\caption{Contours of a quadrangular prism under uniaxial tension\label{fig:Contours-of-a-prism}}
\end{figure}

The theoretical solutions of displacement $u_{z}$ and stress $\sigma_{z}$
of the uniaxial tension problem are expressed as: 
\begin{subequations}
\label{eq:analytical_patch}
\begin{equation}
\sigma_{z}=\frac{F}{b^{2}}=1\,\mathrm{MPa},
\end{equation}
\begin{equation}
u_{z}=\left(\frac{\sigma_{z}}{E}-\frac{\nu\sigma_{x}}{E}-\frac{\nu\sigma_{y}}{E}\right)z=\frac{F}{Eb^{2}}z=0.0001z.
\end{equation}
\end{subequations}
 The corresponding contours are depicted in Fig.~\ref{fig:Contours-of-a-prism}.
The maximum nodal $L^{2}$ error norms of displacement $u_{z}$ and
stress $\sigma_{z}$, which are calculated based on the analytical
solutions in Eq.~(\ref{eq:analytical_patch}), are listed in Table
\ref{tab:patch_error}. The displacement and stress results are considered
to be accurate up to the machine precision. 
 
\begin{table}
\centering{}\caption{Maximum nodal $L^{2}$ error norm of a quadrangular prism under uniaxial
tension\label{tab:patch_error}}
\begin{tabular}{>{\centering}p{3cm}>{\centering}p{2cm}>{\centering}p{2cm}}
\toprule 
 & $u_{z}$ & $\sigma_{z}$\tabularnewline
\midrule 
$L^{2}$ error norm & 1.199e-14 & 1.695e-14\tabularnewline
\bottomrule
\end{tabular}
\end{table}

\subsubsection{Modal analysis of a cantilever beam\label{subsec:Modal-analysis}}

\begin{figure}
\noindent \begin{centering}
\hfill{}\subfloat[Geometry and boundary condition\label{fig:Geometry-and-boundary_3d_modal}]{\noindent \centering{}\hspace{0.025\textwidth}\includegraphics[scale=0.8]{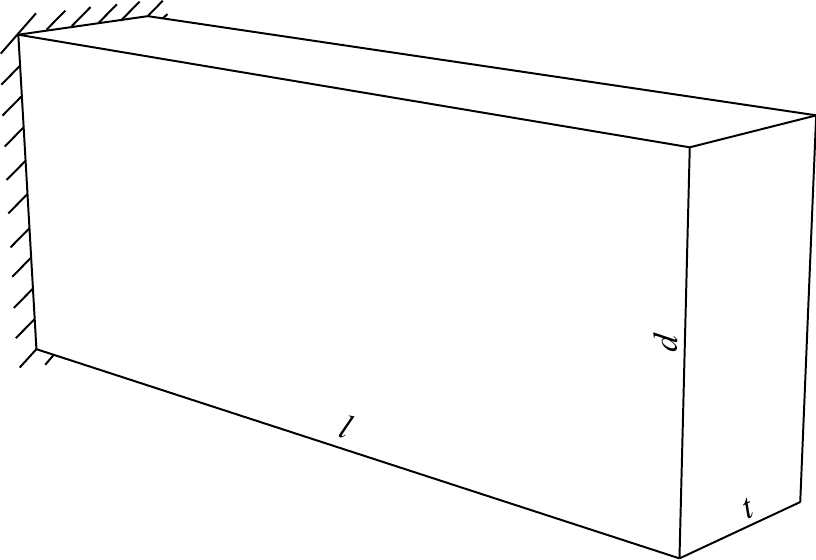}\hspace{0.025\textwidth}}\hfill{}\subfloat[Mesh\label{fig:Mesh_3D_beam}]{\noindent \centering{}\hspace{0.025\textwidth}\includegraphics[scale=0.8]{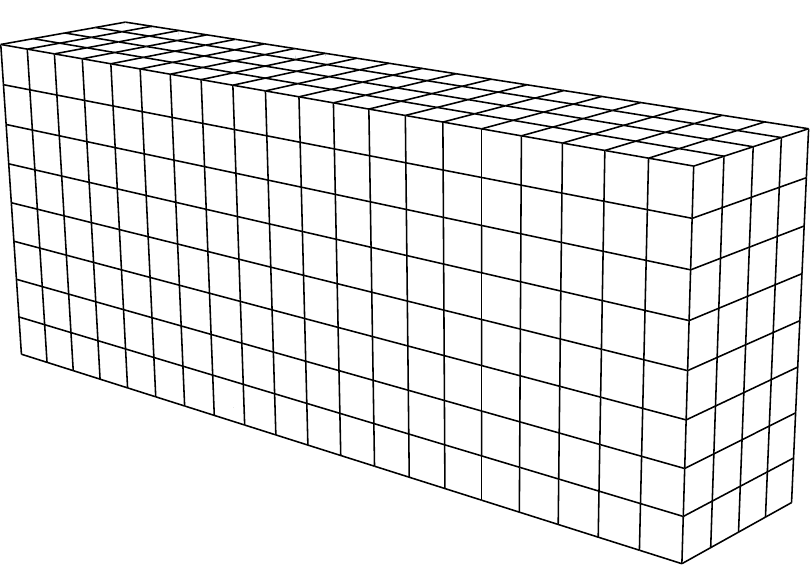}\hspace{0.025\textwidth}}\hfill{}
\par\end{centering}
\noindent \centering{}\caption{Model of a cantilever beam}
\end{figure}

The second example is a three-dimensional cantilever beam with a depth
of $d=0.4\,\mathrm{m}$, a thickness of $t=0.2\,\mathrm{m}$, and
a length of $l=1\,\mathrm{m}$, where one end of the beam is clamped,
as shown in Fig.~\ref{fig:Geometry-and-boundary_3d_modal}. The material
properties are Young's modulus $E=1\,\mathrm{MPa}$, Poisson's ratio
$\nu=0.25$, and the mass density $\rho=2000\,\mathrm{kg/m^{3}}$.

Three-dimensional SBFEM user elements and 8-node linear brick elements
(ABAQUS: C3D8) are assigned to the same mesh to compare their performances.
In the following, we refer to the mesh assigned with ABAQUS library's
elements as ABAQUS model and the mesh assigned with SBFEM user elements
as SBFEM model. Four meshes of the cantilever beam with different
element sizes, i.e., $0.1\,\mathrm{m}$, $0.05\,\mathrm{m}$, $0.025\,\mathrm{m}$
and $0.0125\,\mathrm{m}$, are generated and one is shown in Fig.~\ref{fig:Mesh_3D_beam}
(mesh size: $0.05\,\mathrm{m}$). The reference model has a mesh size
of $0.004\,\mathrm{m}$, consisting of 1,250,000 C3D8 elements and
1,292,901 nodes. 
\begin{figure}
\noindent \begin{centering}
\hfill{}\subfloat[First 10 eigenvalues\label{fig:First-10-eigenvalues}]{\noindent \centering{}\includegraphics[scale=0.8]{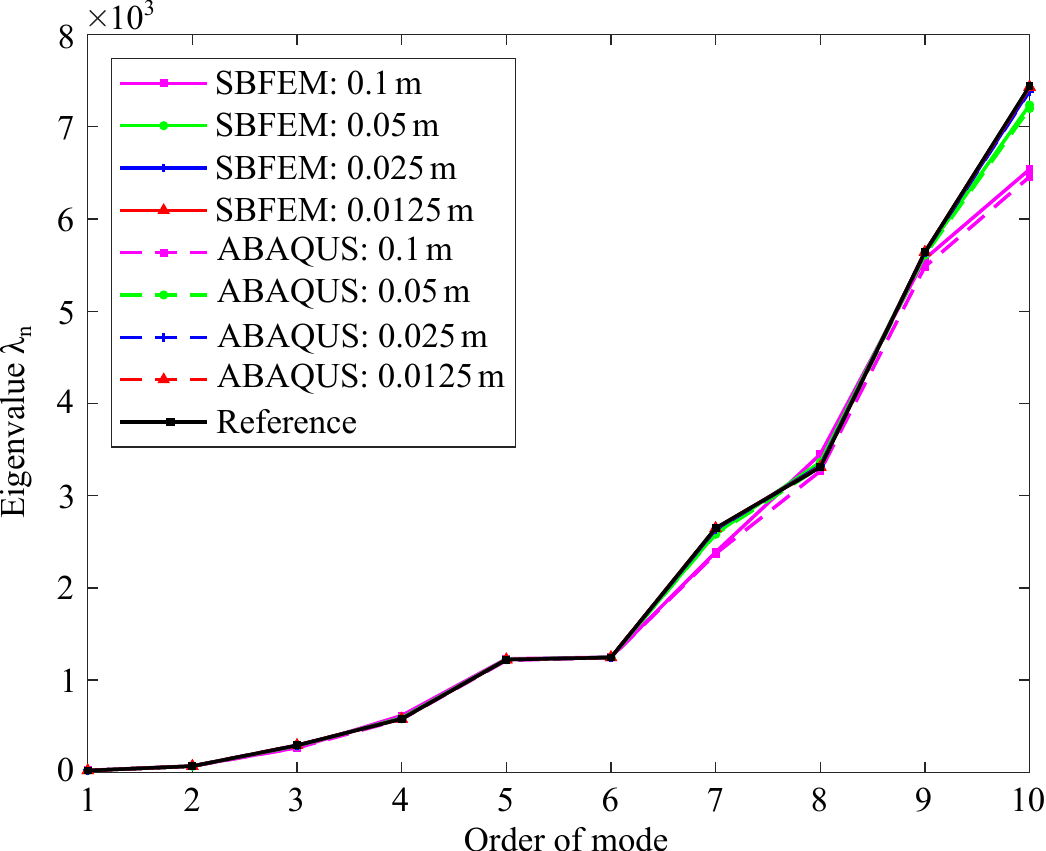}}\hfill{}\subfloat[Convergence of eigenvalue $\lambda_{10}$ \label{fig:Convergence-of-eigenvalue}]{\noindent \centering{}\includegraphics[scale=0.8]{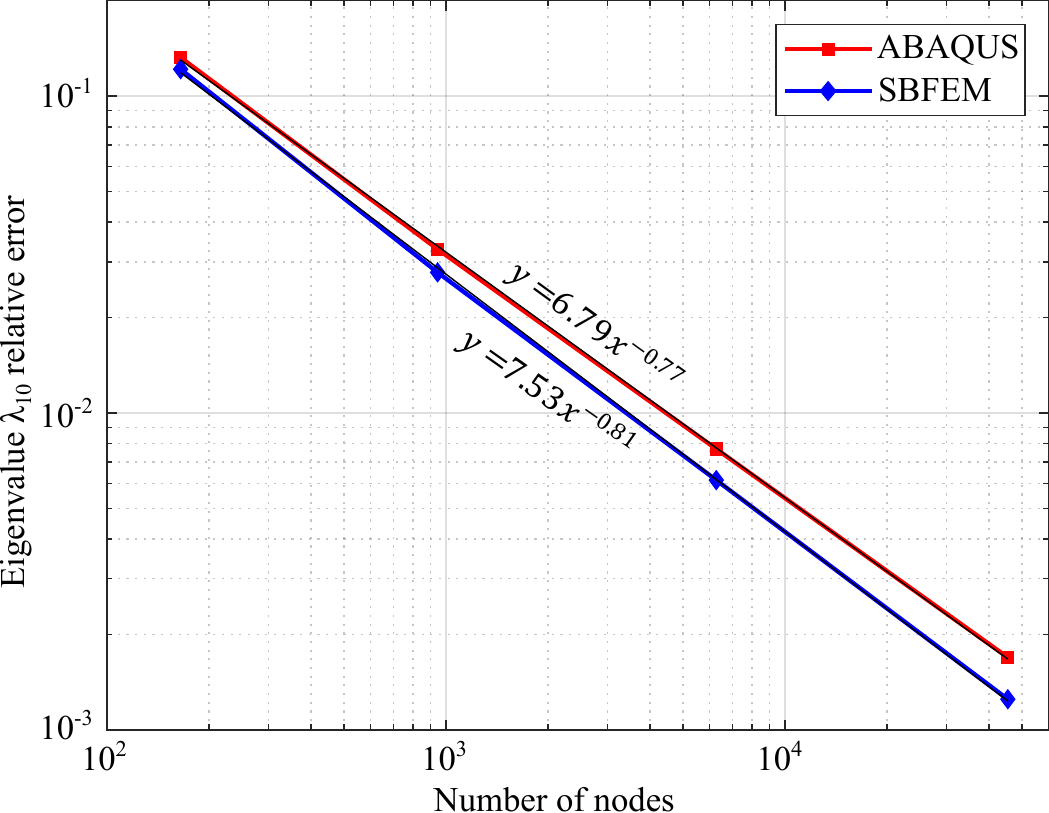}}\hfill{}
\par\end{centering}
\caption{Modal analysis comparison of a cantilever beam\label{fig:Modal-analysis-comparison}}
\end{figure}

For all models, an eigenvalue analysis is performed using the default
settings of ABAQUS' solver~\citep{ABA2016}. The first ten eigenvalues
for the different models are plotted in Fig.~\ref{fig:First-10-eigenvalues}.
As can be observed, the first ten eigenvalues converge algebraically
for both element types. The convergence in the relative error norm
of eigenvalue $\lambda_{10}$ is also studied. The relative error
of eigenvalue $\lambda_{10}$ is determined by $E_{\mathrm{rel}}=(\lambda_{10}-\lambda_{10}^{\mathrm{ref}})$/$\lambda_{10}^{\mathrm{ref}}$,
where $\lambda_{10}^{\mathrm{ref}}$ is the 10th eigenvalue obtained
from the reference model. The relative errors versus the number of
nodes for the two types of models are plotted in Fig.~\ref{fig:Convergence-of-eigenvalue}.
The results are in good agreement, while the SBFEM models achieve
slightly more accurate results.

\subsubsection{Transient analysis of a cantilever beam\label{subsec:Transient-analysis}}

\begin{figure}
\hfill{}\subfloat[Boundary and load conditions\label{fig:Boundary_Load_3D_beam}]{\centering{}\includegraphics[scale=0.8]{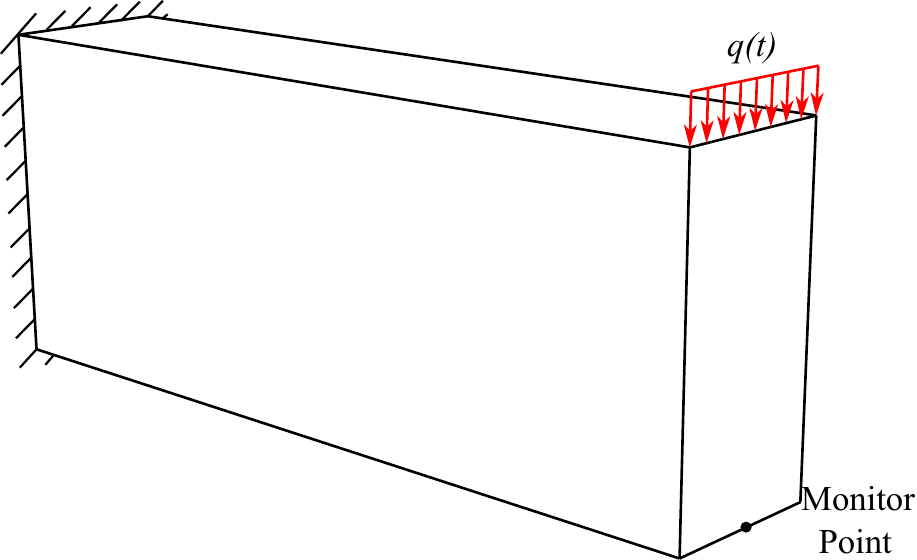}}\hfill{}\subfloat[Load history\label{fig:Load-history_3D_beam}]{\begin{centering}
\includegraphics[scale=0.8]{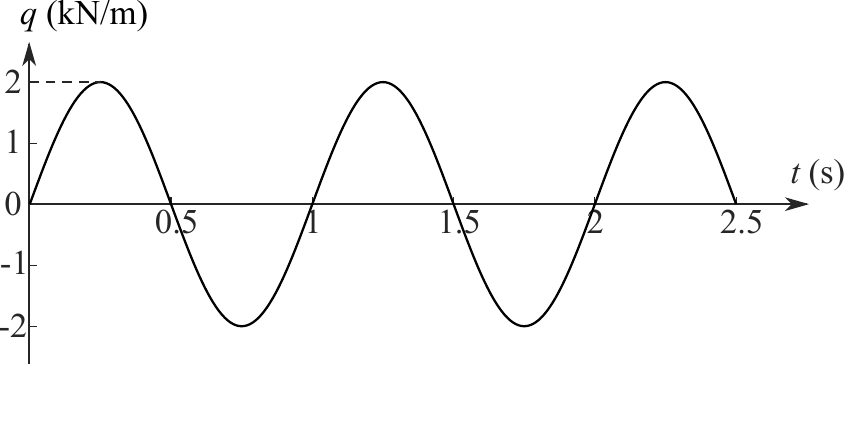}
\par\end{centering}

}\hfill{}

\caption{A cantilever beam subjected to dynamic loading\label{fig:Dynamic_3D_beam}}
\end{figure}

A transient analysis is performed on the three-dimensional cantilever
beam structure introduced in the previous section. The boundary conditions
and load history are shown in Fig.~\ref{fig:Dynamic_3D_beam}. A
distributed line load $q(t)=2\sin(2\pi t)$ (Unit: $\mathrm{kN/m}$)
acting on the top edge of the free end varies harmonically in time
for $2.5\,\mathrm{s}$, as depicted in Fig.~\ref{fig:Load-history_3D_beam}.
As a monitor point, we choose the middle point on the lower edge of
the free end of the cantilever beam as indicated in Fig.~\ref{fig:Boundary_Load_3D_beam}.
The vertical responses of the monitor point are recorded. 

\begin{figure}
\noindent \centering{}\subfloat[Displacement]{\centering{}\includegraphics[scale=0.8]{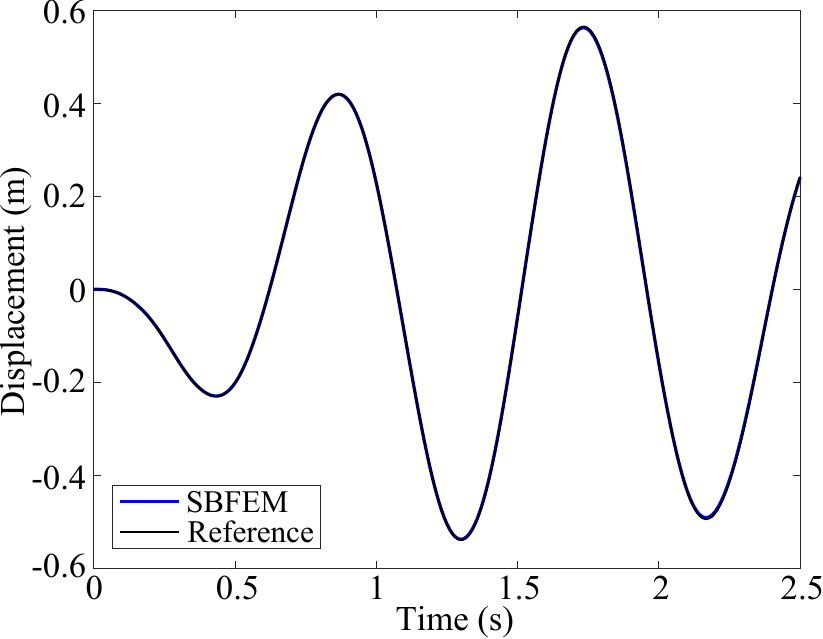}}\subfloat[Velocity]{\centering{}\includegraphics[scale=0.8]{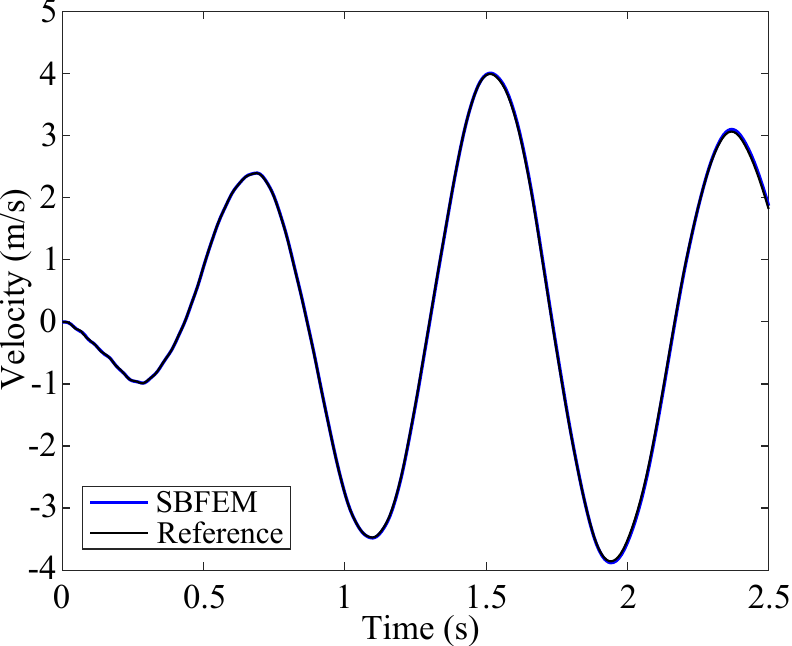}}\subfloat[Acceleration]{\begin{centering}
\includegraphics[scale=0.8]{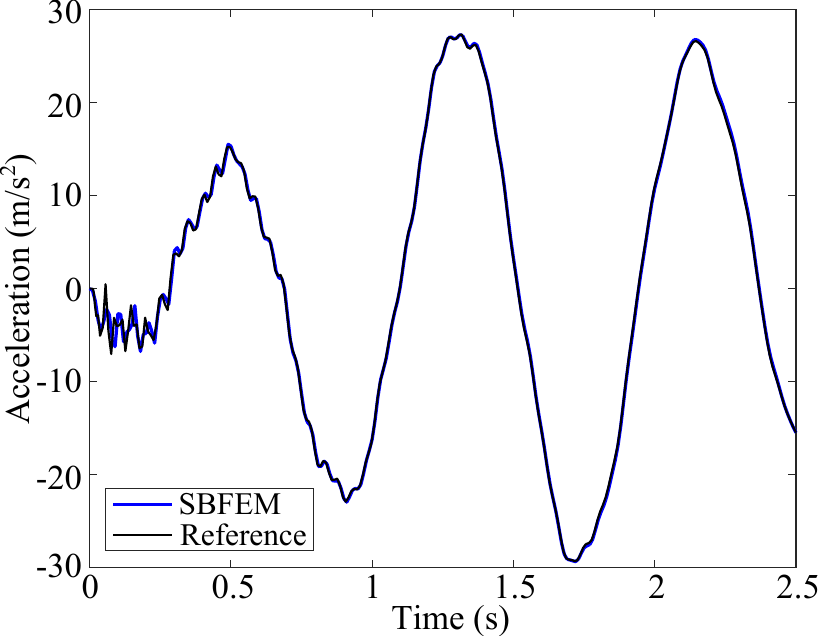}
\par\end{centering}
}\caption{Response history of a 3D cantilever beam: vertical response at monitor
point\label{fig:Response-history-of-3D-beam}}
\end{figure}

The responses of the SBFEM and ABAQUS reference models are compared.
The mesh size of the SBFEM model is chosen as $0.025\,\mathrm{m}$
which is related to the modal analysis results. It has been shown
that the first ten eigenvalues are accurately captured by this model
(error less than $1\%$). The reference model is identical to the
one used in the previous section. The time stepping method is the
Hilber-Hughes-Taylor (HHT) method~\citep{hilber1977} implemented
in ABAQUS/Standard, which is an unconditionally stable method for
linear problems. Again, the default settings for the transient (implicit
dynamic) analysis~\citep{ABA2016} have been employed. A maximum
time step size of $0.01\,\mathrm{s}$ has been chosen and the automatic
time stepping of ABAQUS is activated. The vertical responses at the
monitor point during the $2.5\,\mathrm{s}$ simulation time are obtained
from the two models and plotted in Fig.~\ref{fig:Response-history-of-3D-beam}.
An excellent agreement is found as the difference between the results
obtained from the SBFEM and ABAQUS reference models is negligible.
Only in the acceleration response during the first $0.5\,\mathrm{s}$
small deviations are visible. 

\subsection{Examples for non-matching meshes\label{subsec:Examples-for-non-matching}}

Non-matching meshes often occur in finite element analyses involving
domain decomposition~\citep{Zhang2019,Garg2020} and contact mechanics~\citep{xing2019}
problems. Typically, special techniques are required to enforce the
compatibility and equilibrium at the interfaces of non-matching meshes,
which often introduce additional unknowns. Non-matching meshes in
the finite element sense can be easily converted to matching meshes
owing to flexibility provided by polyhedral SBFEM domains, which allows
for elements with an arbitrary number of faces.

This section presents three examples to demonstrate the advantages
of the matching meshes. The first example investigates the performance
of the proposed elements in contact patch tests having curved interfaces.
In the second example, a patch test including cohesive elements in
conjunction with contact is presented. This example highlights the
convenience of inserting cohesive elements at matching interfaces
and verifies the behavior of cohesive elements combined with SBFEM
user elements. The third example presents a cube with a spherical
inclusion. Here, interaction states containing both contact and traction
are examined and compared with the results obtained from non-matching
meshes.

\subsubsection{Patch test for contact analysis}

ABAQUS/Standard provides comprehensive approaches for defining contact
interactions: general contact, contact pairs and contact elements.
For these approaches, there are different formulations available such
as node-to-surface (NTS), or surface-to-surface (STS) techniques which
are available for contact pairs~\citep{ABA2016}. Although many approaches
can be utilized, the non-uniformity of the contact interface discretization
may reduce the accuracy of results in ABAQUS~\citep{xing2019,Xing2018}.
The performance of contact models in ABAQUS is expected to be enhanced
when matching contact interfaces can be easily generated.

\begin{figure}
\noindent \begin{centering}
\hfill{}\subfloat[Model\label{fig:Model_patch_contact}]{\noindent \centering{}\includegraphics[scale=0.8]{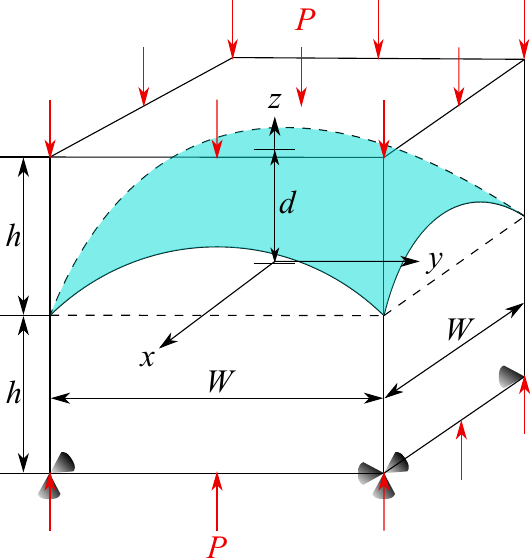}}\hfill{}\subfloat[Mesh scheme\label{fig:Mesh-scheme_patch_contact}]{\noindent \centering{}\includegraphics[bb=0bp 0bp 152bp 162bp,scale=0.8]{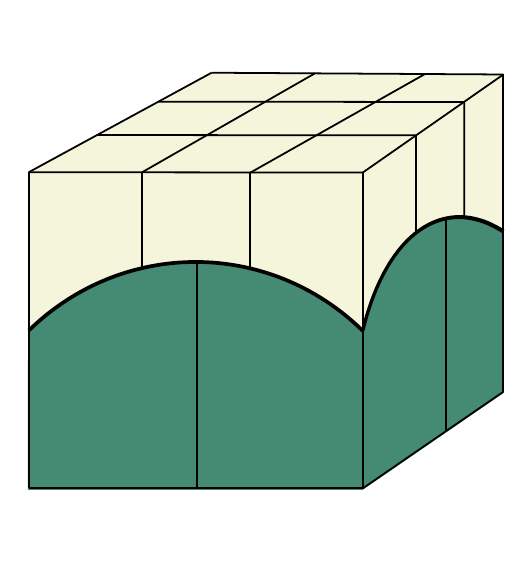}}\hfill{}
\par\end{centering}
\caption{Patch test model for contact analysis\label{fig:Patch-test-model_contact}}
\end{figure}

Three-dimensional contact patch tests are studied for the most commonly
used NTS and STS schemes. The model is illustrated in Fig.~\ref{fig:Patch-test-model_contact}
with dimensions defined as $W=6\,\mathrm{m}$ and $h=3\,\mathrm{m}$.
It features a spherical contact interface between two blocks with
a maximum deviation distance $d$. The shape of the interface is described
as 
\begin{equation}
z_{\mathrm{i}}(x_{\mathrm{i}},y_{\mathrm{i}})=d\left[1-\frac{2\left(x_{\mathrm{i}}^{2}+y_{\mathrm{i}}^{2}\right)}{W^{2}}\right],
\end{equation}
where $x_{\mathrm{i}}$, $y_{\mathrm{i}}$, $z_{\mathrm{i}}$ are
the coordinates of a point on the interface. The parameter $d$ varies
from $0$ to $1.5\,\mathrm{m}$ to examine the contact performance
depending on the actual shape of the curved contact interface. The
model has a Young's modulus of $E=10\,\mathrm{MPa}$, and a Poisson's
ratio of $\nu=0.3$. The contact has a friction coefficient of $\mu=0.5$.
The lower interface is selected as the master surface of the contact.
Six degrees of freedom of the lower block (as indicated in Fig.~\ref{fig:Patch-test-model_contact})
are fixed to prevent rigid body motions. A uniform pressure $P=1\,\mathrm{MPa}$
is applied at the bottom and top surfaces and consequently, a constant
stress state $\sigma_{z}=1\,\mathrm{MPa}$ is expected in the contact
patch test. 

\begin{figure}
\noindent \begin{centering}
\hfill{}\subfloat[Non-matching mesh \label{fig:C3D8-non-matching-mesh}]{\noindent \centering{}\includegraphics[scale=0.8]{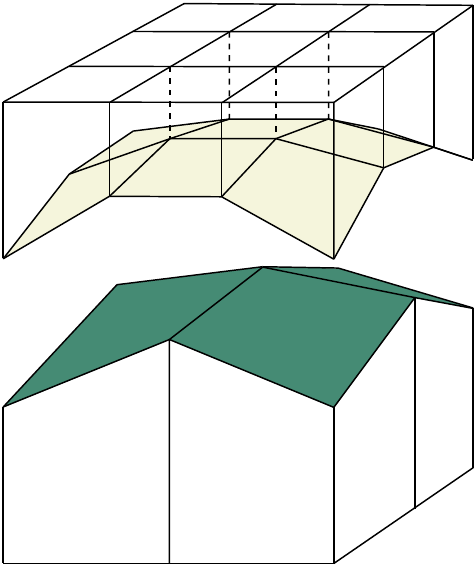}}\hfill{}\subfloat[Matching mesh\label{fig:SBFEM-matching-mesh}]{\noindent \centering{}\includegraphics[scale=0.8]{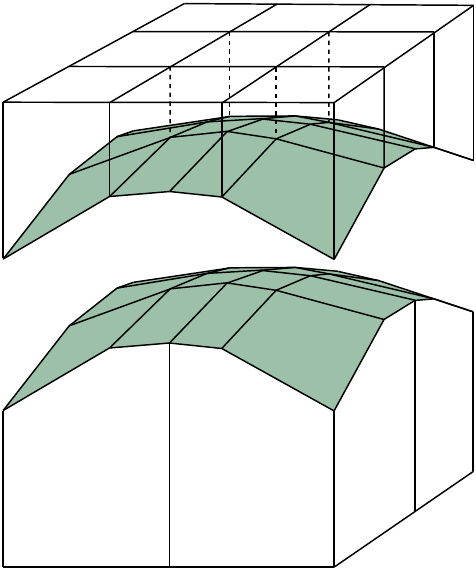}}\hfill{}
\par\end{centering}
\caption{Meshes of contact patch test\label{fig:Meshes-of-contact-patch-test}}
\end{figure}

The basic set-up of the discretization for the model is shown in Fig.~\ref{fig:Mesh-scheme_patch_contact},
where it is observed that the top and bottom contact bodies are divided
into nine and four elements, respectively. C3D8 elements are assigned
to the non-matching meshes directly, as shown in Fig.~\ref{fig:C3D8-non-matching-mesh}.
The special treatment 'slave tolerance' in ABAQUS is used to detect
the effective contact pairs for the ABAQUS models due to initial gaps
or penetrations. Through re-meshing the interface, the hexahedral
non-matching meshes are converted to polyhedral matching meshes as
illustrated in Fig.~\ref{fig:SBFEM-matching-mesh}. The SBFEM user
elements can be directly assigned to the matching meshes (see Fig.~\ref{fig:SBFEM-matching-mesh})
as they allow polyhedral shapes. Element-based surfaces can be created
for the SBFEM models through the methodology proposed in Section~\ref{subsec:Element-based-surface-definition},
which are used to establish the contact pairs. 

\begin{figure}
\begin{centering}
\hfill{}\subfloat[$d$=0m]{\centering{}\includegraphics[width=0.35\textwidth]{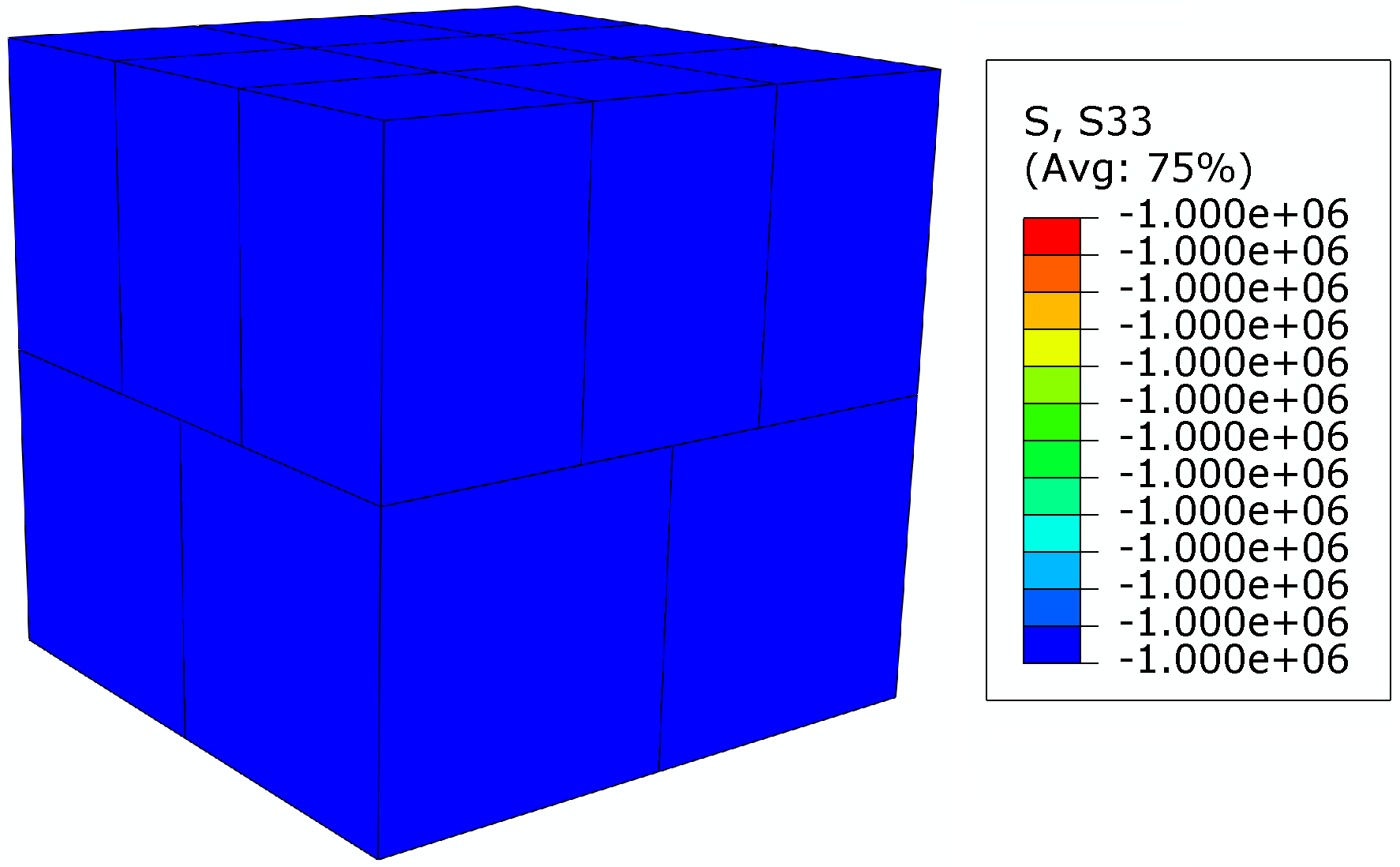}}\hfill{}\subfloat[$d$=0.5m]{\centering{}\includegraphics[width=0.35\textwidth]{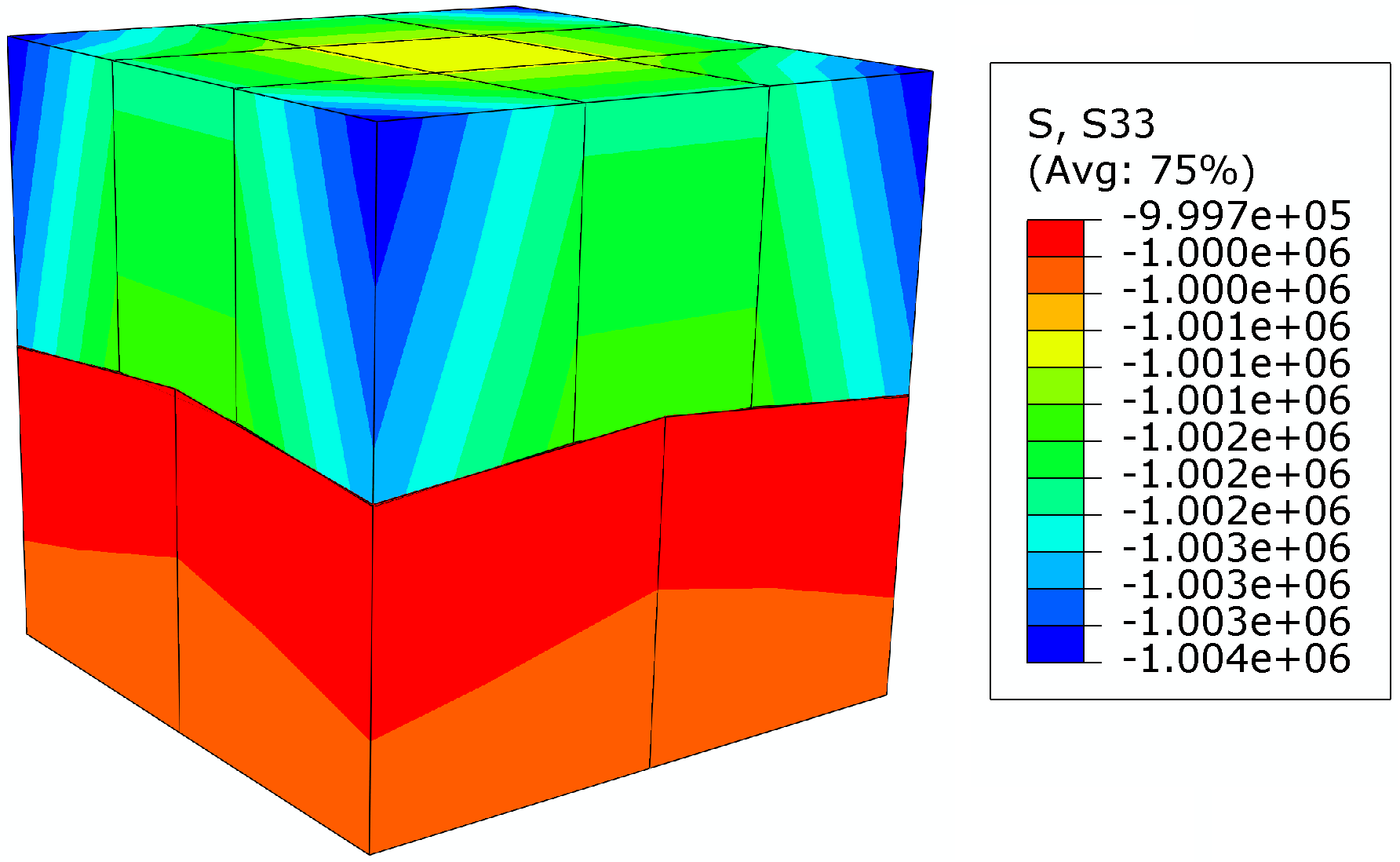}}\hfill{}
\par\end{centering}
\begin{centering}
\hfill{}\subfloat[$d$=1m]{\centering{}\includegraphics[width=0.35\textwidth]{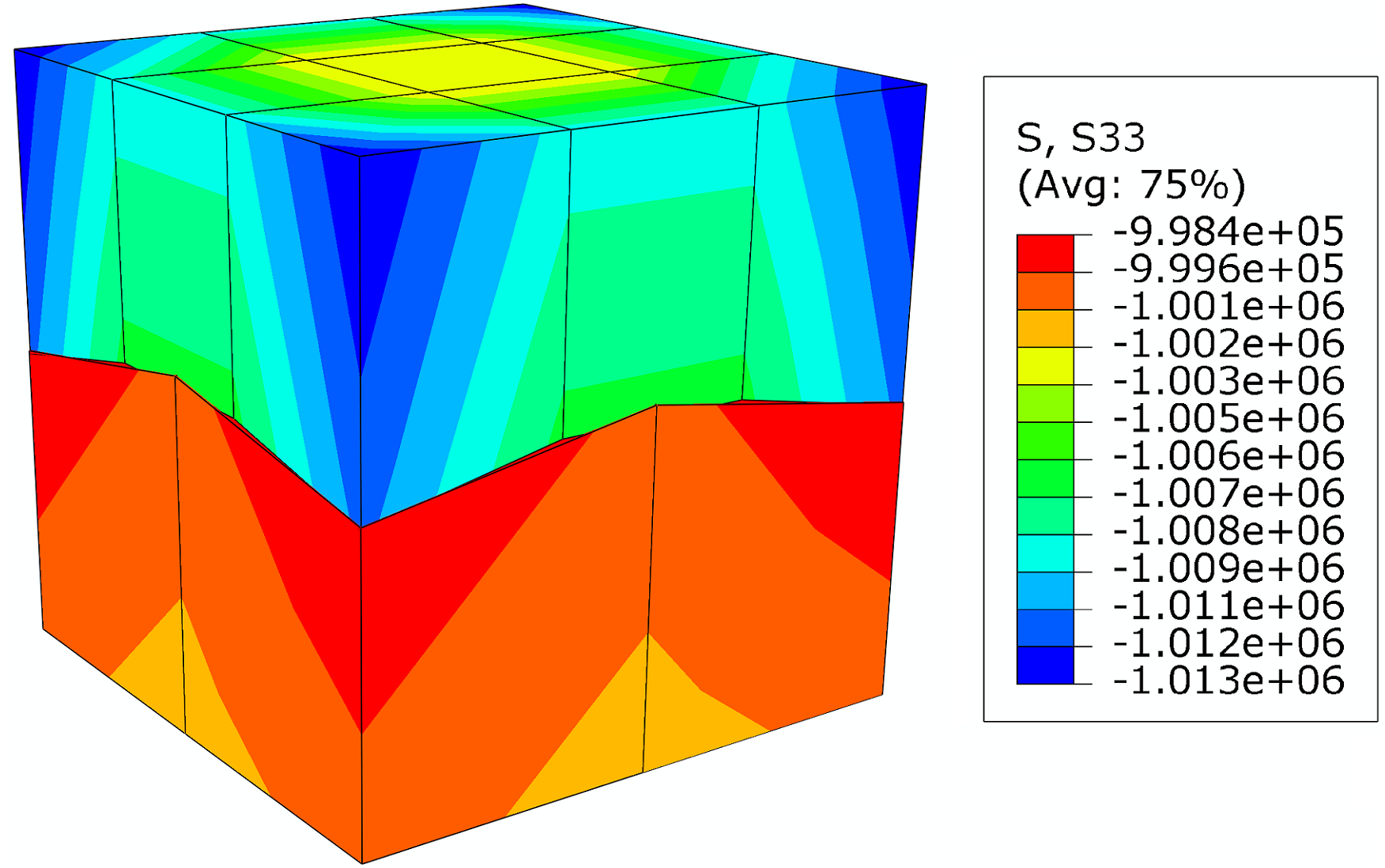}}\hfill{}\subfloat[$d$=1.5m]{\centering{}\includegraphics[width=0.35\textwidth]{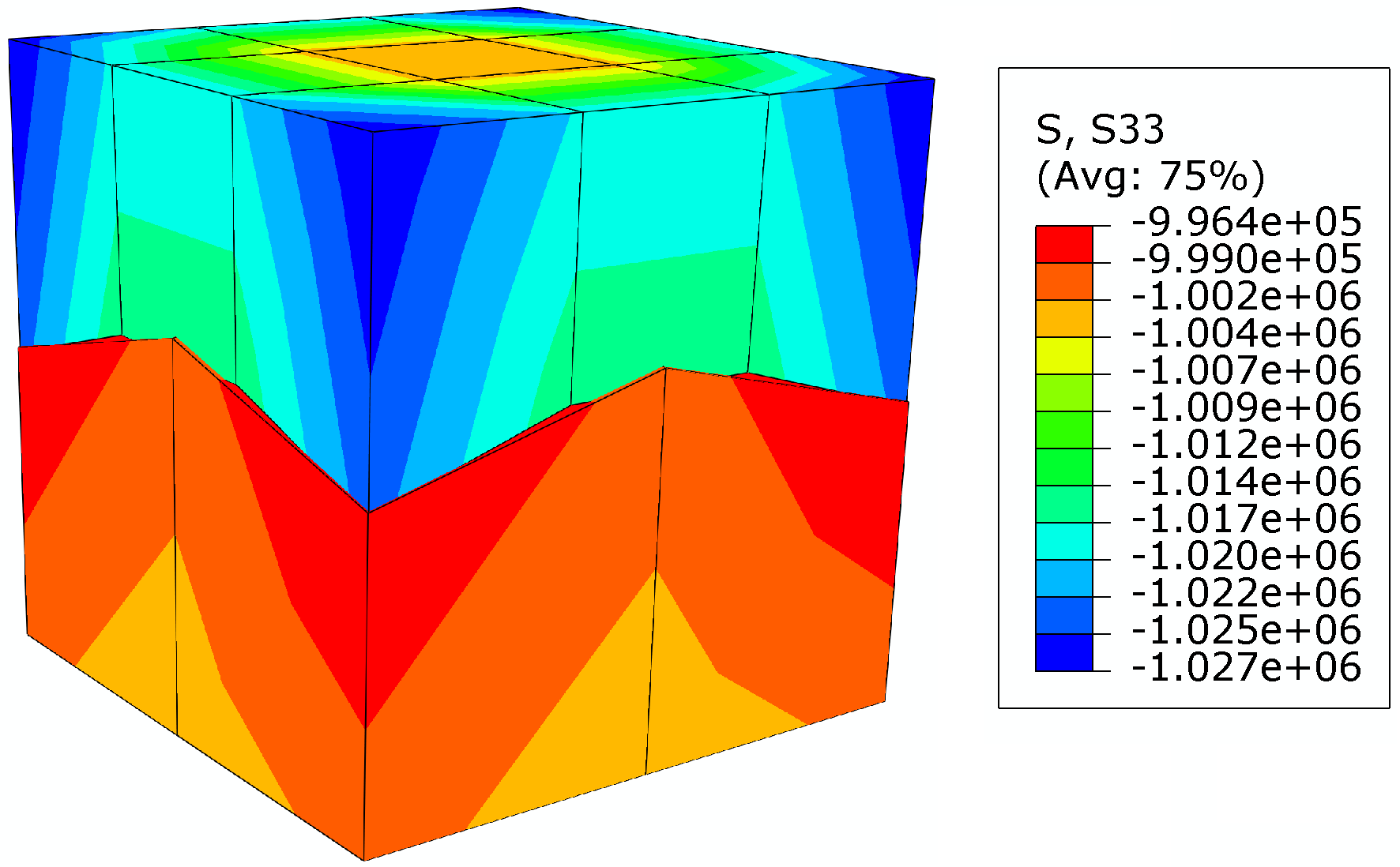}}\hfill{}
\par\end{centering}
\centering{}\caption{Stress contours ($\sigma_{z}$) of ABAQUS models (Unit: Pa)\label{fig:Stress-contours-nonmatching}}
\end{figure}

\begin{figure}
\begin{centering}
\hfill{}\subfloat[$d$=0m]{\centering{}\includegraphics[width=0.33\textwidth]{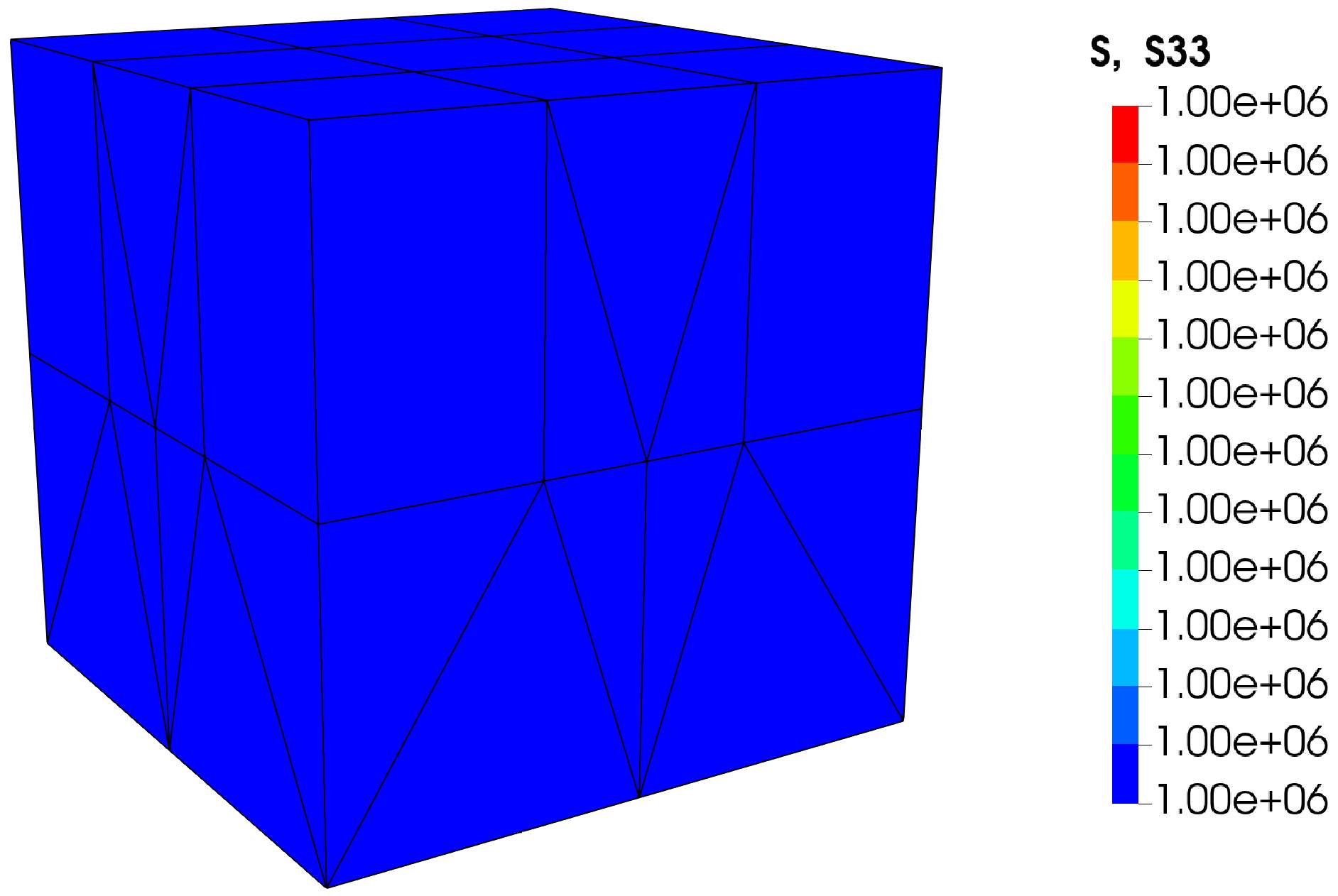}}\hfill{}\subfloat[$d$=0.5m]{\centering{}\includegraphics[width=0.33\textwidth]{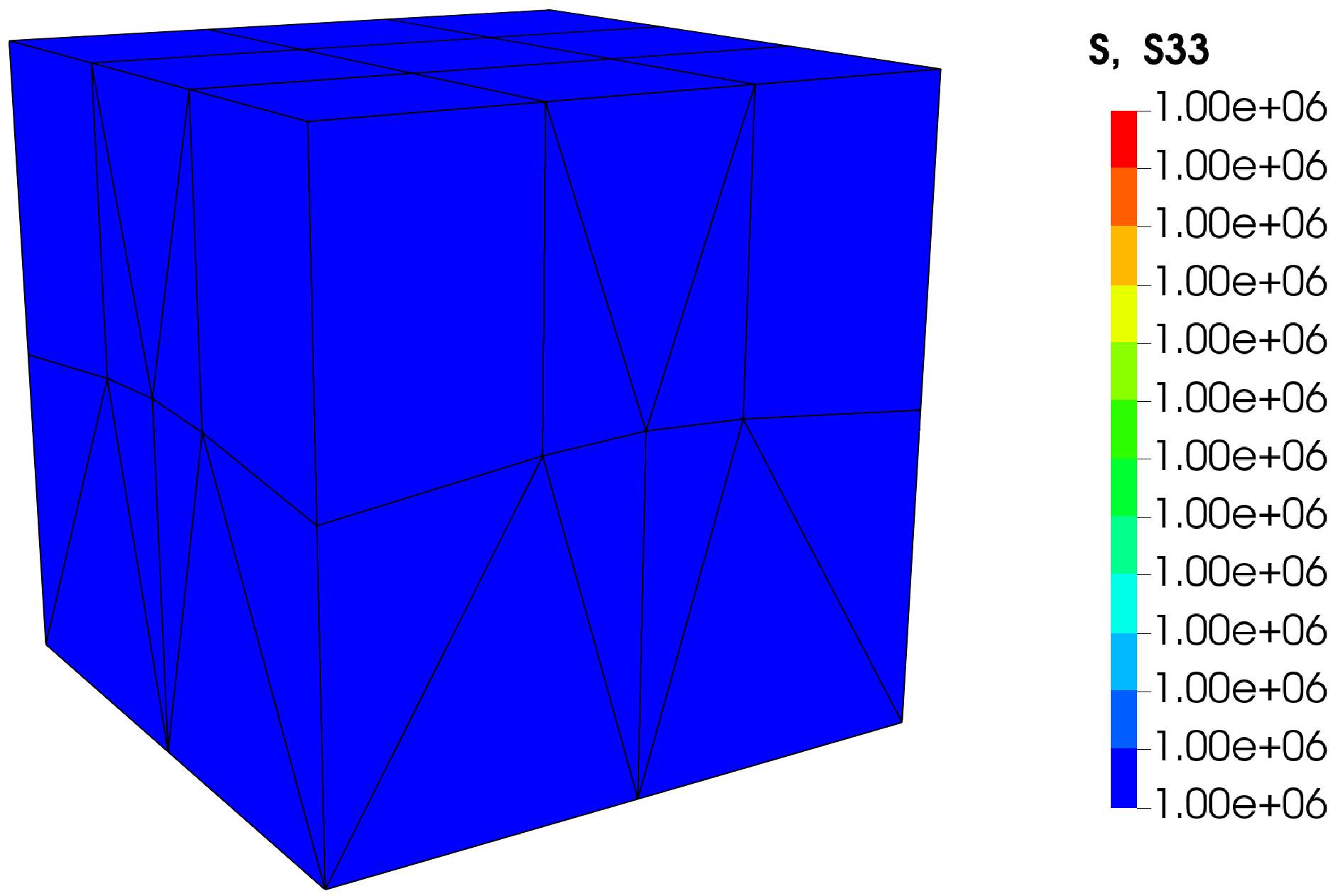}}\hfill{}
\par\end{centering}
\begin{centering}
\hfill{}\subfloat[$d$=1m]{\centering{}\includegraphics[width=0.33\textwidth]{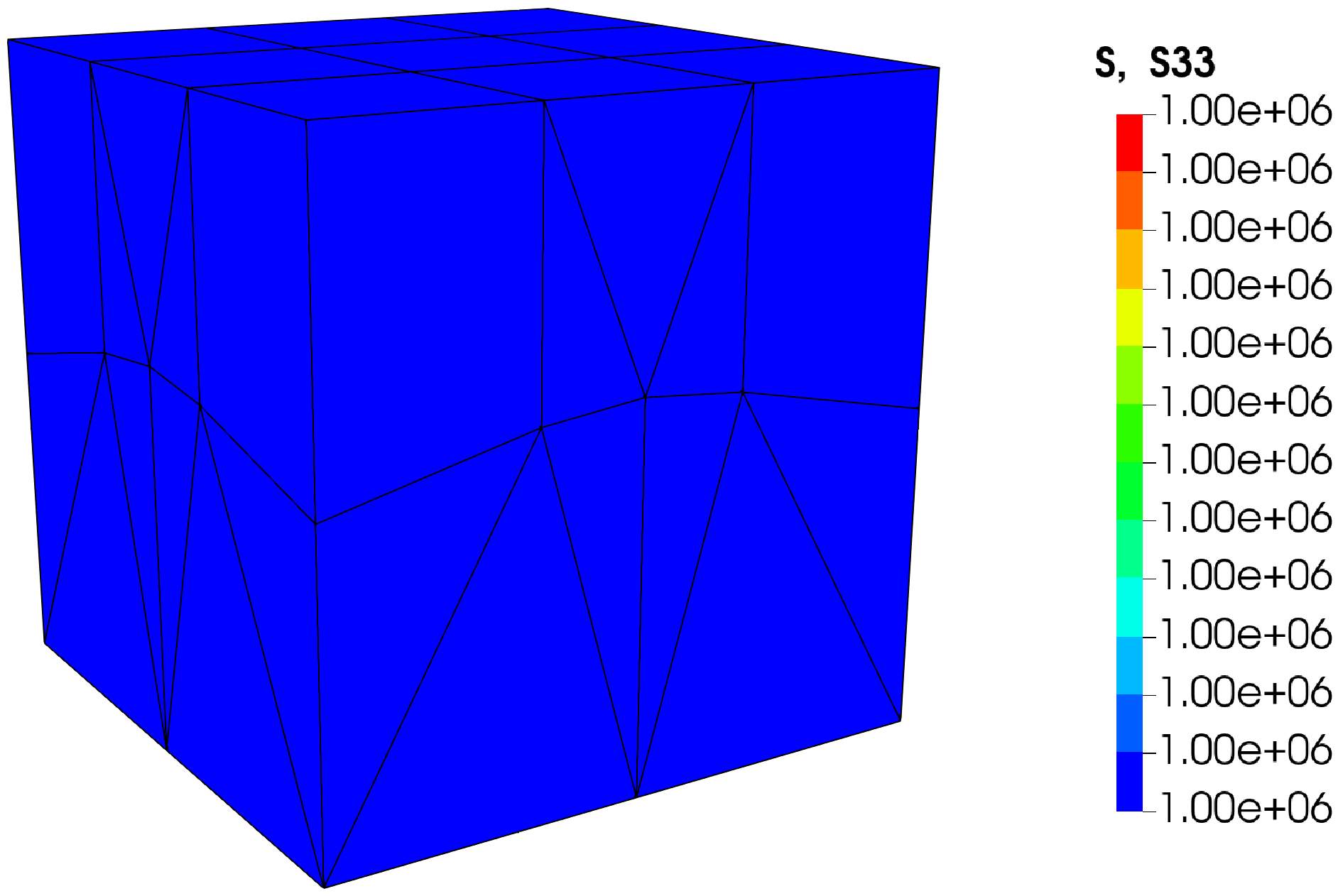}}\hfill{}\subfloat[$d$=1.5m]{\centering{}\includegraphics[width=0.33\textwidth]{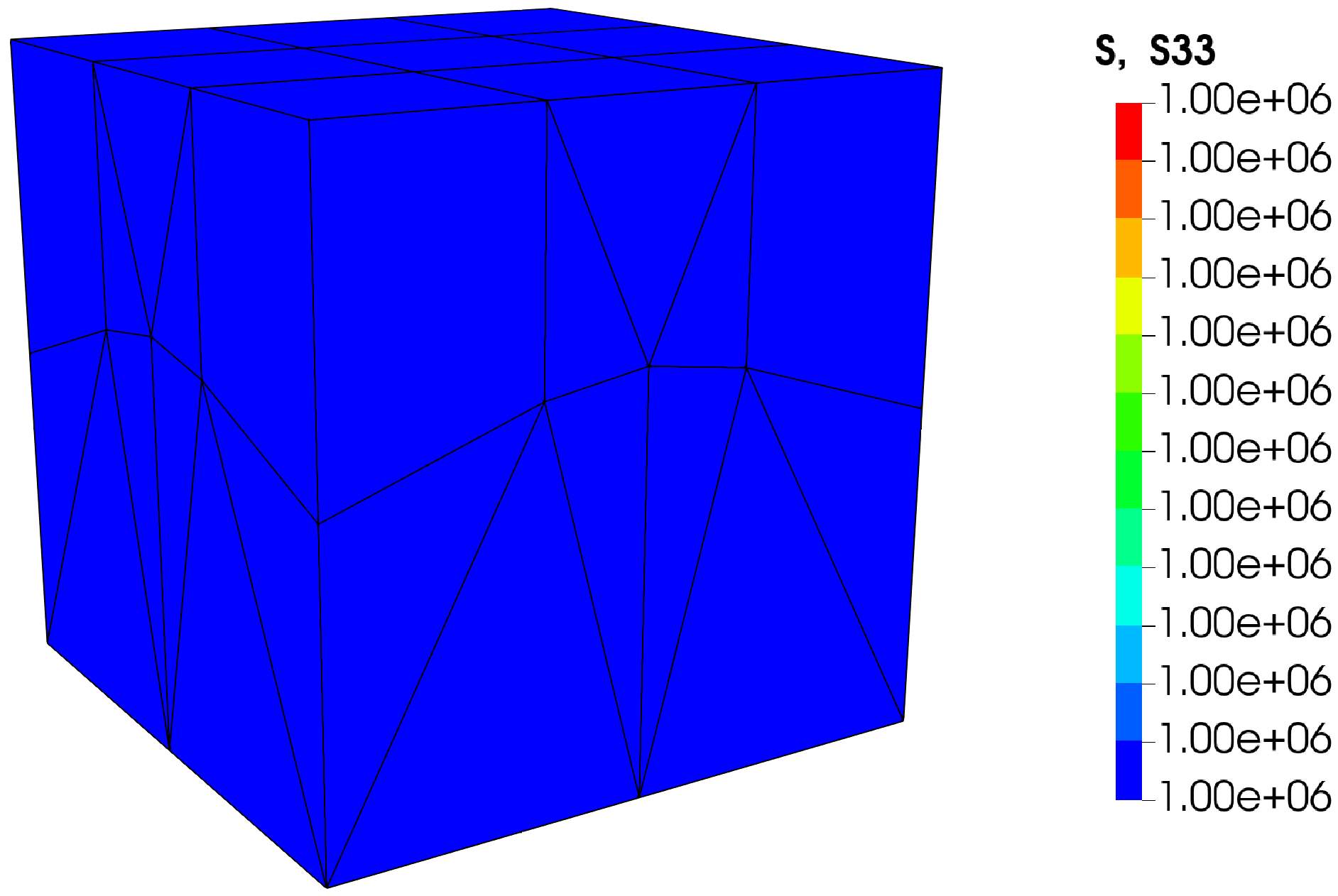}}\hfill{}
\par\end{centering}
\centering{}\caption{Stress contours ($\sigma_{z}$) of SBFEM models (Unit: Pa)\label{fig:Stress-contours-matching}}
\end{figure}

The stress contours of $\sigma_{z}$ (S33), obtained from the ABAQUS
and SBFEM models (STS) are plotted in Figs.~\ref{fig:Stress-contours-nonmatching}~and~\ref{fig:Stress-contours-matching},
respectively. The relative $L^{2}$ error norms for $\sigma_{z}$
are listed in Table~\ref{tab:Error-norm-patch_contact}, in which
the maximum error norms of nodal stress are given. As shown in Fig.~\ref{fig:Stress-contours-nonmatching}
and Table~\ref{tab:Error-norm-patch_contact}, for the ABAQUS models
applying the STS scheme, only the model with flat contact interface
obtains accurate result and passes the patch test. By increasing the
curvature of the contact surfaces, the accuracy of the ABAQUS models
deteriorates. This effect is attributed to initial gaps or penetrations
in non-matching interface meshes which reduce the accuracy of the
contact simulation. All ABAQUS models applying the NTS scheme fail
the patch tests with substantial errors. In contrast, the SBFEM models
provide accurate results (up to machine precision) and pass the patch
tests for various curvatures using both STS and NTS schemes, as depicted
in Fig.~\ref{fig:Stress-contours-matching} and Table~\ref{tab:Error-norm-patch_contact}.
\begin{table}
\centering{}\caption{$L^{2}$ Error norm of $\sigma_{z}$ of different models with different
curvatures\label{tab:Error-norm-patch_contact}}
\begin{tabular}{cccccc}
\toprule 
 &  & $d=0$ & $d=0.5\,\mathrm{m}$ & $d=1\,\mathrm{m}$ & $d=1.5\,\mathrm{m}$\tabularnewline
\midrule
\multirow{2}{*}{STS} & ABAQUS model & <1e-9 & 3.72e-3 & 1.31e-2 & 2.72e-1\tabularnewline
 & SBFEM model & 4.77e-14 & 4.10e-14 & 6.48e-14 & 5.02e-14\tabularnewline
\midrule
\multirow{2}{*}{NTS} & ABAQUS model & 1.24e-1 & 1.16e-1 & 1.12e-1 & 1.11e-1\tabularnewline
 & SBFEM model & 5.13e-14 & 5.24e-14 & 6.76e-14 & 6.17e-14\tabularnewline
\bottomrule
\end{tabular}
\end{table}
 
\subsubsection{Patch test of cohesive element in conjunction with contact}

In composite materials, such as concrete \citep{lopez2008} and fiber-reinforced
plastics~\citep{Garg2020}, interfacial stresses between the bulk
materials develop when they separate. To investigate the behavior
of the interfacial transition zone (ITZ), cohesive elements are widely
used in literature~\citep{huzni2013,Sarrado2016,Trawinski2018}.
The geometry of the microstructure and the interface conditions have
a significant influence on the macroscopic behavior of composite materials~\citep{Spring2014}.
Therefore, surface meshes located at the material interface should
provide a highly accurate representation of the actual geometry being
suitable for simulating the interactions. Matching meshes, which are
easily obtained from polyhedral elements, eliminate the gaps and penetrations
at the interface and two matching surfaces can directly form a cohesive
element. The performance of the combination of existing cohesive elements
in ABAQUS and the proposed user elements is investigated in the following.

\begin{figure}
\noindent \begin{centering}
\includegraphics[scale=0.8]{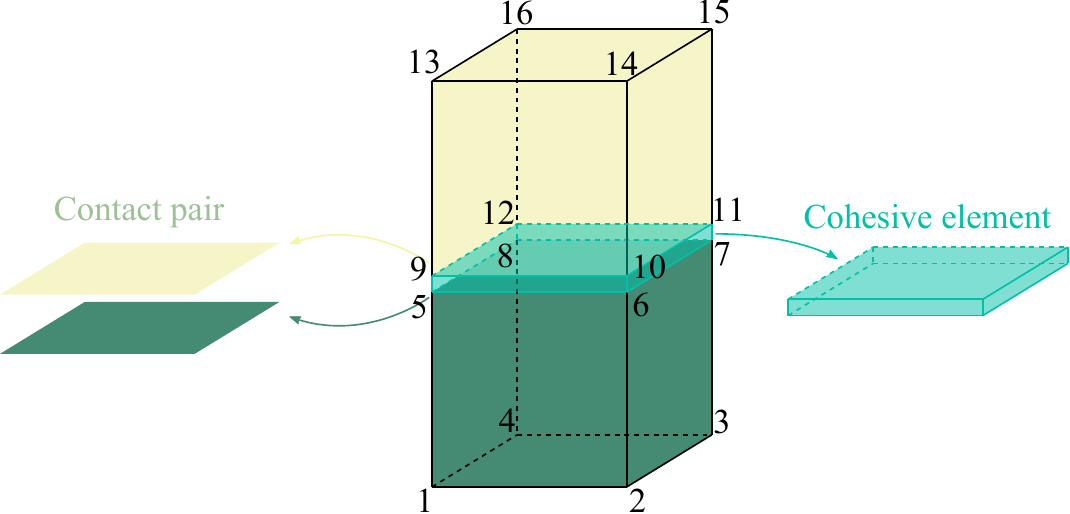}
\par\end{centering}
\noindent \centering{}\caption{Numerical modeling of ITZ\label{fig:ITZ_modeling}}
\end{figure}

The scheme of the numerical modeling for the ITZ based on the matching
meshes is illustrated in Fig.~\ref{fig:ITZ_modeling}. There are
two hexahedral domains representing the bulk materials, and they exhibit
a matching discretization at the interface. For the sake of clarity,
the matching meshes in Fig.~\ref{fig:ITZ_modeling} have been separated,
while essentially they are exactly matching with no gap, i.e., Nodes
5 and 9 have the same coordinates, which is also true for the other
matching nodes. These two surfaces can straightforwardly form a zero
thickness cohesive element. An 8-node three-dimensional cohesive element
COH3D8 can be defined with node ordering '5, 6, 7, 8, 9, 10, 11, 12'.
In this way, no constraints are required to connect the cohesive element
to the other components. To prevent the interpenetration, a contact
pair is established based on the two surfaces of the two blocks. The
cohesive element generates traction applied to the surrounding bulk
materials when they separate. The contact pair will be activated when
interpenetration between bulk materials might occur. For triangular
matching interfaces, a 6-node three-dimensional cohesive element COH3D6
must be used. 

\begin{figure}
\noindent \begin{centering}
\includegraphics[scale=0.8]{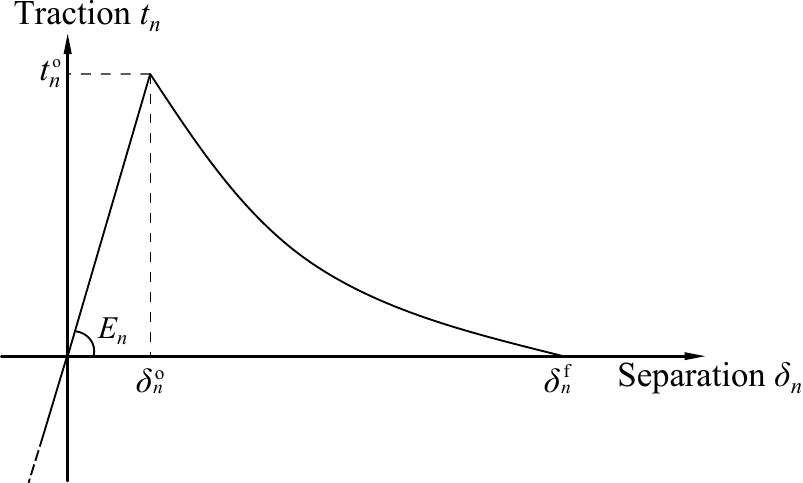}
\par\end{centering}
\noindent \centering{}\caption{Typical traction-separation response with exponential damage evolution\label{fig:Exponential-damage-evolution}}
\end{figure}

For a better understanding, the basic description of the the constitutive
response of cohesive elements using a traction-separation law is provided.
Readers interested in a comprehensive discussion should refer to the
ABAQUS manual~\citep{ABA2016}. Cohesive elements typically employ
a traction-separation response in the normal direction based on an
exponential damage evolution law which is depicted in Fig.~\ref{fig:Exponential-damage-evolution}.
The traction-separation response in tension contains a damage evolution
described by a scalar damage value $D$, and can be expressed as 
\begin{equation}
t_{n}=\begin{cases}
E_{n}\delta_{n} & \delta_{n}^{\mathrm{max}}\leq\delta_{n}^{\mathrm{o}}\\
\left(1-D\right)E_{n}\delta_{n} & \delta_{n}^{\mathrm{o}}<\delta_{n}^{\mathrm{max}}\leq\delta_{n}^{\mathrm{f}}\\
0 & \delta_{n}^{\mathrm{max}}>\delta_{n}^{\mathrm{f}}
\end{cases},
\end{equation}
where $t_{n}$ and $\delta_{n}$ represent the traction and separation,
and $E_{n}$ is the initial elastic modulus of the cohesive element
in the normal direction. The parameters $\delta_{n}^{\mathrm{o}}$,
$\delta_{n}^{\mathrm{f}}$, $\delta_{n}^{\mathrm{max}}$ represent
the separation at different states, they indicate damage initiation,
complete failure, and the maximum value attained during loading, respectively.
The damage variable, $D$, represents the overall damage of the cohesive
element. It has an initial value of 0 before damage initiation and
then monotonically evolves from 0 to 1 until total damage occurs.
For an exponential softening law, when $\delta_{n}^{\mathrm{o}}<\delta_{n}^{\mathrm{max}}<\delta_{n}^{\mathrm{f}}$,
the damage variable $D$ is expressed as
\begin{equation}
D=1-\left(\frac{\delta_{n}^{\mathrm{o}}}{\delta_{n}^{\mathrm{max}}}\right)\left\{ 1-\frac{1-\exp\left[-\alpha\left(\frac{\delta_{n}^{\mathrm{max}}-\delta_{n}^{\mathrm{o}}}{\delta_{n}^{\mathrm{f}}-\delta_{n}^{\mathrm{o}}}\right)\right]}{1-\exp(-\alpha)}\right\} ,\label{eq:damage}
\end{equation}
where $\alpha$ is the exponential coefficient. The cohesive element
will not be damaged if it is in compression, thus the response in
Fig.~\ref{fig:Exponential-damage-evolution} is straight when $\delta_{n}<0$.

\begin{figure}
\noindent \begin{centering}
\includegraphics[scale=0.8]{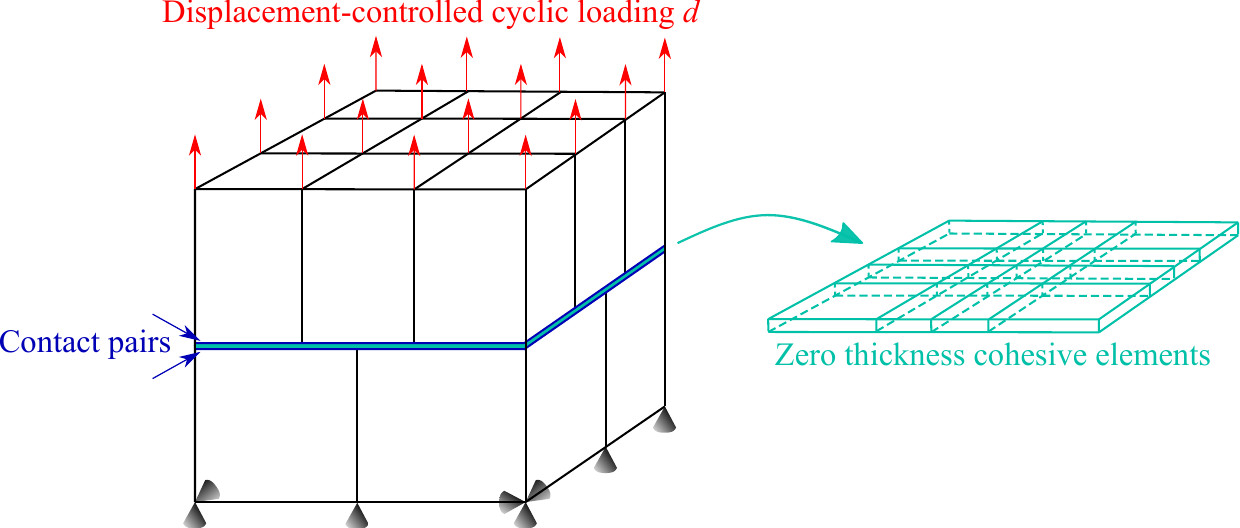}
\par\end{centering}
\noindent \centering{}\caption{Patch test model with cohesive element\label{fig:Patch-test-cohesion}}
\end{figure}

To assess the compatibility of SBFEM user elements and cohesive elements,
we use the same mesh that was already used for the contact investigations,
i.e., a simple hexahedral domains with a flat interface, and insert
cohesive elements at the interface. As shown in Fig.~\ref{fig:Patch-test-cohesion},
zero thickness cohesive elements are generated at the matching interface.
In addition, the contact pairs are established on the matching surfaces
for preventing the interpenetration of the two parts. The bottom of
the lower block is constrained in the vertical direction, and three
DOFs are fixed in the horizontal directions to prevent rigid body
motions. 

The material properties used in this example are listed in Table~\ref{tab:Material-properties-of-cohe}.
The material properties of the cohesive elements are chosen for the
purpose of the verification, they do not represent any specific mechanical
behavior. The tangential properties of the cohesive elements are not
listed because the patch test only generates normal stresses. The
SBFEM elements are relatively rigid, such that the displacement load
$d$ will be mainly transferred to $\delta_{n}$ when it is positive.
When the cohesive elements have no damage ($\delta_{n}^{\mathrm{max}}\leq\delta_{n}^{\mathrm{o}}$),
the ratio between $\delta_{n}$ and $d$ is 0.99997 for $d>0$ based
on elasticity theory, and it will increase if the damage initiation
occurs in the cohesive elements. 
\begin{table}
\centering{}\caption{Material properties of the patch test inserting cohesive element\label{tab:Material-properties-of-cohe}}
\begin{tabular}{cccccc}
\toprule 
 & Young's modulus $E\left[\mathrm{MPa}\right]$ & Poisson's ratio $\nu$ & $\delta_{m}^{\mathrm{o}}\left[\mathrm{m}\right]$ & $\delta_{m}^{\mathrm{f}}\left[\mathrm{m}\right]$ & $\alpha$\tabularnewline
\midrule
Cohesive & $1$ & - & $0.001$ & $0.005$ & 2\tabularnewline
\midrule
Bulk & $200000$ & 0.3 & - & - & -\tabularnewline
\bottomrule
\end{tabular}
\end{table}

A static analysis is performed to observe the behavior of the cohesive
elements. Three harmonically varying cyclic displacement-controlled
loads $d$ are applied on the top surface of the upper block, as plotted
in Fig. \ref{fig:Displacement-history}. The expression of $d$ (Unit:
$\mathrm{mm}$) in terms of time $t(\mathrm{s})$ is 
\begin{equation}
d=\begin{cases}
0.8\sin2\pi t & 0\leq t\leq1\\
2\sin2\pi(t-1) & 1<t\leq2\\
6\sin2\pi(t-2) & 2<t\leq3
\end{cases}.
\end{equation}
The magnitudes of $d$ are chosen to generate different $\delta_{n}^{\mathrm{max}}$
representing different damage states in the three cycles. In the first
cycle ($\ensuremath{0\leq t\leq1\,\mathrm{s}}$), the harmonic loading
generates a $\delta_{n}^{\mathrm{max}}=0.08\,\mathrm{mm}<\delta_{n}^{\mathrm{o}}$,
which indicates there is no damage in the cohesive elements. The second
cycle ($\ensuremath{1\,\mathrm{s}<t\leq2\,\mathrm{s}}$) has a $\delta_{n}^{\mathrm{max}}=2\,\mathrm{mm}$
which satisfies $\delta_{n}^{\mathrm{o}}<\delta_{n}^{\mathrm{max}}<\delta_{n}^{\mathrm{f}}$,
and the cohesive elements are partially damaged in this stage. The
third cycle ($\ensuremath{2\,\mathrm{s}<t\leq3\,\mathrm{s}}$) will
cause complete failure of the cohesive elements, since it results
in $\delta_{n}^{\mathrm{max}}=6\,\mathrm{mm}>\delta_{n}^{\mathrm{f}}$. 

The numerical result for the history of $\delta_{n}$ in the three
cycles is plotted in Fig.~\ref{fig:Displacement-history}. During
the loading history, the normal separation $\delta_{n}$ has the following
results:
\begin{equation}
\delta_{n}\begin{cases}
\approx d & d\geqslant0\\
=0 & d<0
\end{cases}.
\end{equation}
When $d\geqslant0$, $\delta_{n}$ is quite similar to $d$ with negligible
distinction as explained before. If $d<0$, $\delta_{n}$ equals to
0 because the surfaces of cohesive elements share the same DOFs with
the contact surfaces of the two blocks, while the interpenetration
between them is prevented by the contact pairs.

The evolution of the normal traction $t_{n}$ versus $\delta_{n}$
is depicted in Fig.~\ref{fig:Normal-stress-history}. Some feature
points are plotted in Fig.~\ref{fig:Cohesive-element-behavior_contact}
to clarify the behavior of cohesive elements during different loading
cycles. During unloading, the normal traction $t_{n}$ always varies
linearly and goes back to the origin. The response of the normal traction
$t_{n}$ during loading in the three cycles is described as follows:
\begin{figure}
\noindent \begin{centering}
\hfill{}\subfloat[Displacement history\label{fig:Displacement-history}]{\noindent \centering{}\includegraphics[scale=0.8]{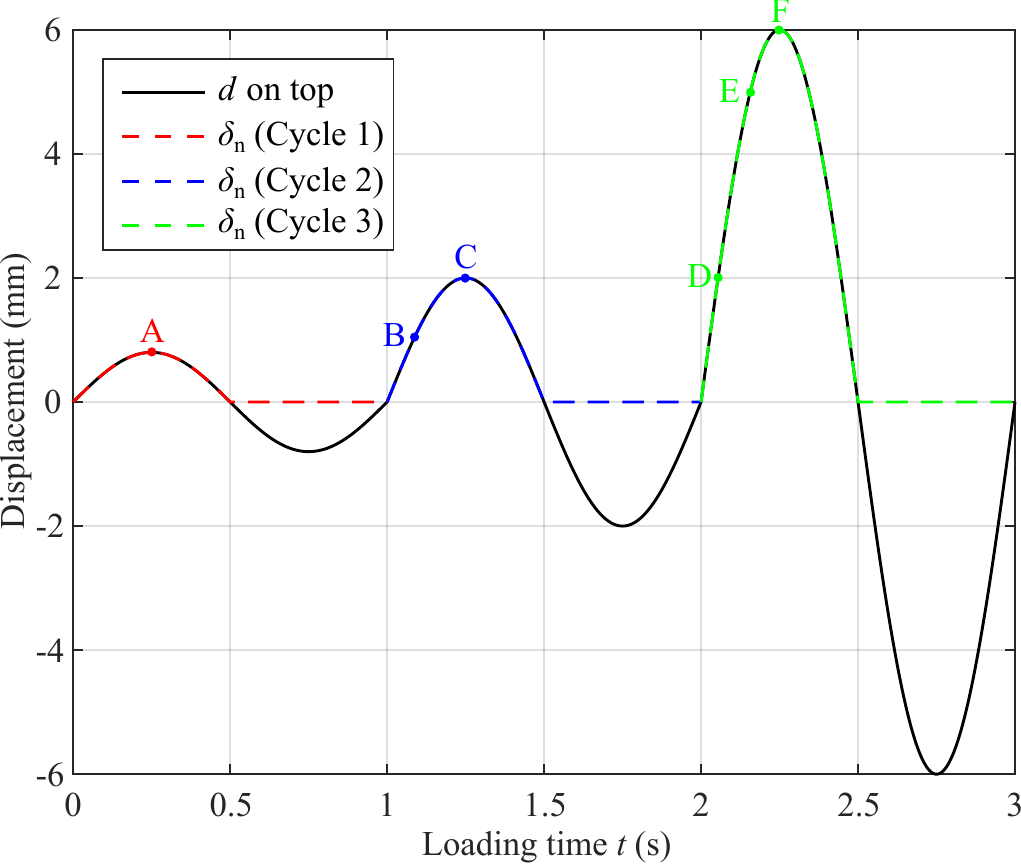}}\hfill{}\subfloat[Normal stress evolution\label{fig:Normal-stress-history}]{\noindent \centering{}\includegraphics[scale=0.8]{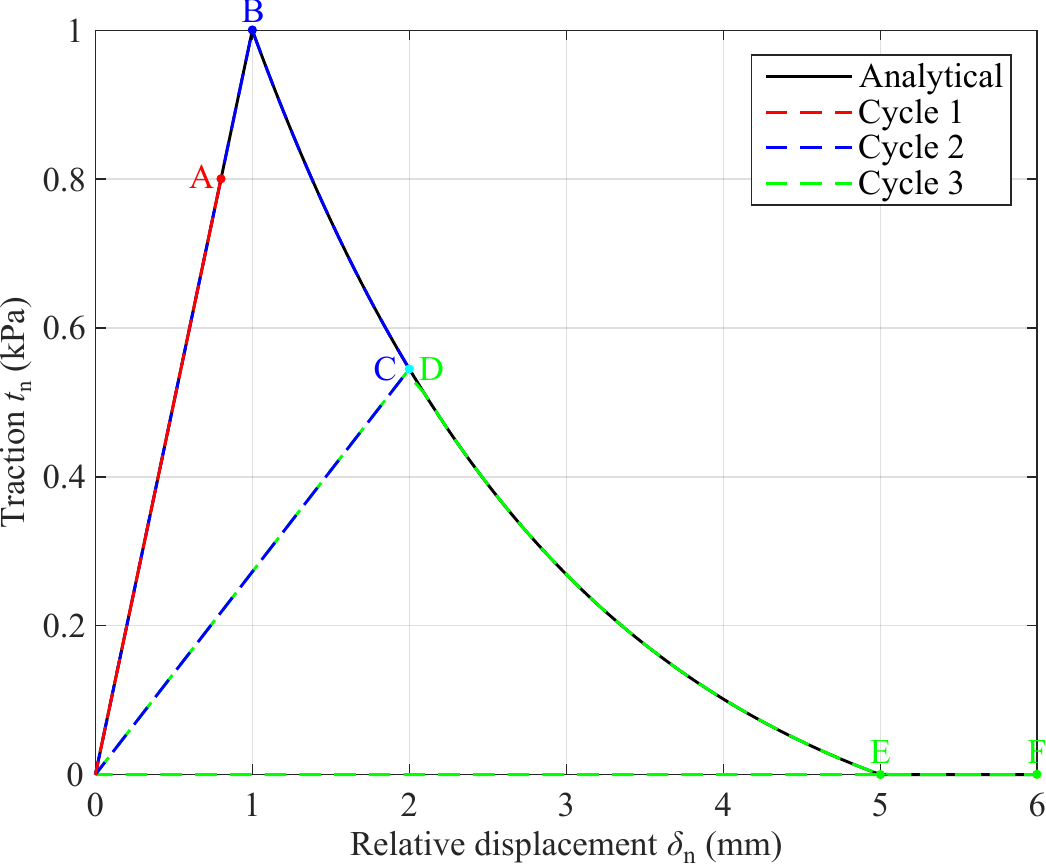}}\hfill{}
\par\end{centering}
\noindent \centering{}\caption{Cohesive element behavior under cyclic loading (with contact)\label{fig:Cohesive-element-behavior_contact}}
\end{figure}

\begin{enumerate}
\item In Cycle 1, $t_{n}$ varies linearly according to $\delta_{n}$ following
$t_{n}=E_{n}\delta_{n}$. It reaches its peak value $0.8\,\mathrm{kPa}$
at Point A and then decreases to 0, after which it remains 0.
\item In Cycle 2, the initial relationship between $t_{n}$ and $\delta_{n}$
is $t_{n}=E_{n}\delta_{n}$ until $\delta_{n}$ increases to $\delta_{n}^{\mathrm{o}}$
(Point B), where the damage initiation occurs and $t_{n}$ reaches
its peak value $1\,\mathrm{kPa}$. After Point B, $t_{n}$ decreases
with the increase of $\delta_{n}$ to $0.54\,\mathrm{kPa}$ (Point
C), following the analytical damage evolution law. At Point C, $\delta_{n}$
reaches its current $\delta_{n}^{\mathrm{max}}$ which is between
$\delta_{n}^{\mathrm{o}}$ and $\delta_{n}^{\mathrm{f}}$, thus the
cohesive element is partially damaged with damage variable $D=0.27$.
The numerical result of $D$ has a good agreement with the analytical
calculation based on Eq.~(\ref{eq:damage}). 
\item In Cycle 3, before $\delta_{n}$ reaches $2\mathrm{mm}$ (Point D),
$t_{n}$ varies linearly following $t_{n}=(1-D)E_{n}\delta_{n}$.
After Point D with the increase of $\delta_{n}$, $t_{n}$ is reduced
to 0 at Point E, where $\delta_{n}^{\mathrm{max}}=\delta_{n}^{\mathrm{f}}$
and the complete failure of the cohesive elements occurs ($D=1$).
Once the complete failure occurs, $t_{n}$ always equals 0, although
$\delta_{n}$ has increased to $6\,\mathrm{mm}$ at Point F.
\end{enumerate}
To conclude, the numerical solutions coincide exactly with the theoretical
solutions. Inserting cohesive elements in matching meshes is convenient
and performs excellently. By defining contact pairs at the interface,
the penetration of bulk materials can be prevented as $\delta_{n}$
has never been negative during the entire loading history. The simulation
of the interaction that involves both traction and contact is easily
achieved by the proposed modeling approach on the matching meshes.

\subsubsection{A cube with a spherical inclusion\label{subsec:Cube_inclusion}}

The inclusion of particles in composite materials has been proved
to increase the stiffness greatly~\citep{Segurado2002}, and the
behavior of the ITZs has a significant effect on the performance of
those composite materials~\citep{Spring2014}. The interfaces in
particle reinforced materials are always complex and most often curved
shapes. The polytree based method proposed by Zhang et al.~\citep{Zhang2019}
is capable of generating complex matching interface meshes, which
offers a convenient way of inserting cohesive elements at the interfaces
to study the behavior of ITZs with complex shapes.

\begin{figure}
\noindent \begin{centering}
\includegraphics[scale=0.8]{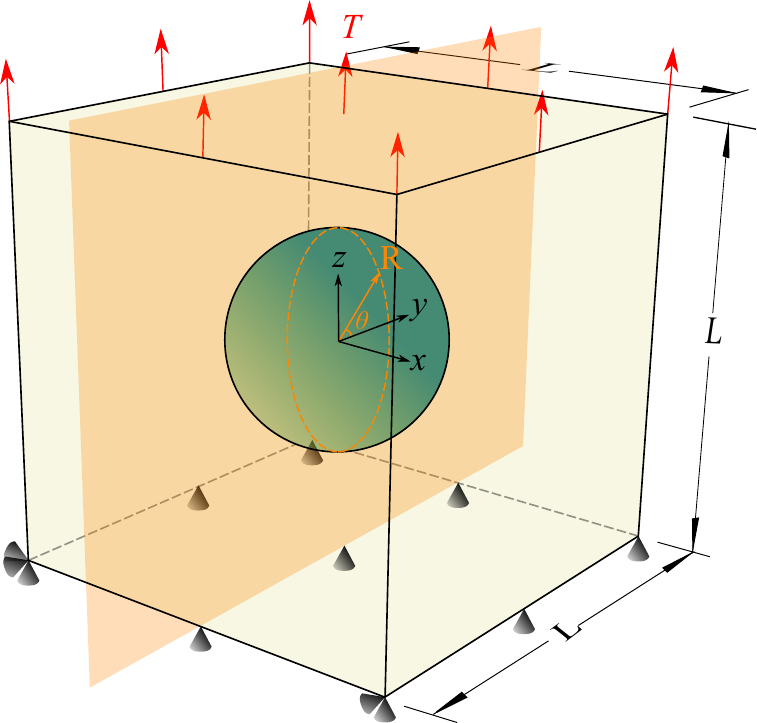}
\par\end{centering}
\noindent \centering{}\caption{A cube with a sphere inclusion\label{fig:A-cube-with-sphere}}
\end{figure}

A cube with a spherical inclusion, as shown in Fig.~\ref{fig:A-cube-with-sphere},
is considered as an example for analyzing ITZs with curved faces.
To this end, we inserted a cohesive layer and a contact pair between
the cube and the embedded sphere to model the separation and prevent
interpenetration, respectively. The length of the cube is $L=4\,\mathrm{m}$
and the radius of the sphere is $R=1\,\mathrm{m}$. Table~\ref{tab:Material-properties-of-cube}
lists the material properties of each part in this example. The cohesive
element has the same initial elastic Young's modulus in the normal
and tangential directions. The damage evolution law of the cohesive
element is of linear type, which actually has the same expression
as the exponential type with $\alpha=0$. The STS approach is used
for the contact modeling and a Coulomb friction coefficient of $\mu=0.2$
is chosen for the sliding. At the bottom of the cube, the vertical
displacement and the rigid movement in horizontal plane are constrained.
A traction of $T=5\,\mathrm{kPa}$ is applied at the top surface of
the cube. The interaction state at the interface will be examined
along a path in the $y-z$ plane ($x=0$), which is shown in Fig.
\ref{fig:A-cube-with-sphere}. 
 
\begin{table}
\centering{}\caption{Material properties of a cube with a spherical inclusion\label{tab:Material-properties-of-cube}}
\begin{tabular}{ccccc}
\toprule 
 & Young's modulus $E\left[\mathrm{MPa}\right]$ & Poisson's ratio $\nu$ & $\delta_{n}^{\mathrm{o}}/\delta_{s}^{\mathrm{o}}/\delta_{t}^{\mathrm{o}}\left[\mathrm{m}\right]$ & $\delta_{m}^{\mathrm{f}}\left[\mathrm{m}\right]$\tabularnewline
\midrule
Cube & $5$ & 0.25 & - & -\tabularnewline
Sphere & $1$ & 0.25 & - & -\tabularnewline
Cohesive layer & $2$ & - & $0.00005$ & $0.01$\tabularnewline
\bottomrule
\end{tabular}
\end{table}

\begin{figure}
\noindent \begin{centering}
\hfill{}\subfloat[Non-matching meshes\label{fig:Non-matching-meshes-(C3D8)}]{\noindent \begin{centering}
\includegraphics[width=0.3\textwidth]{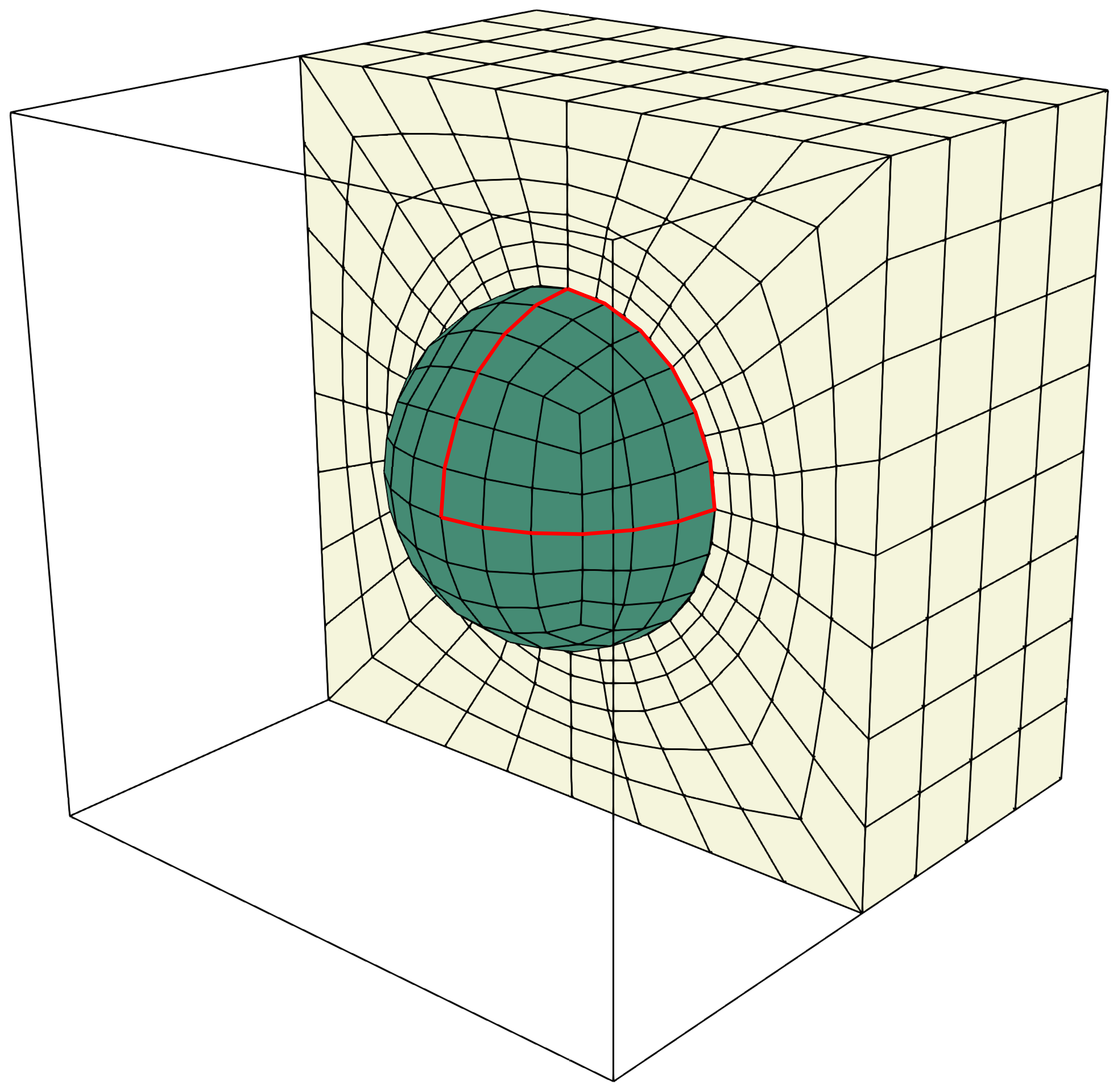}
\par\end{centering}
}\hfill{}\subfloat[Matching meshes\label{fig:Matching-meshes-(SBFEM)}]{\noindent \centering{}\includegraphics[width=0.3\textwidth]{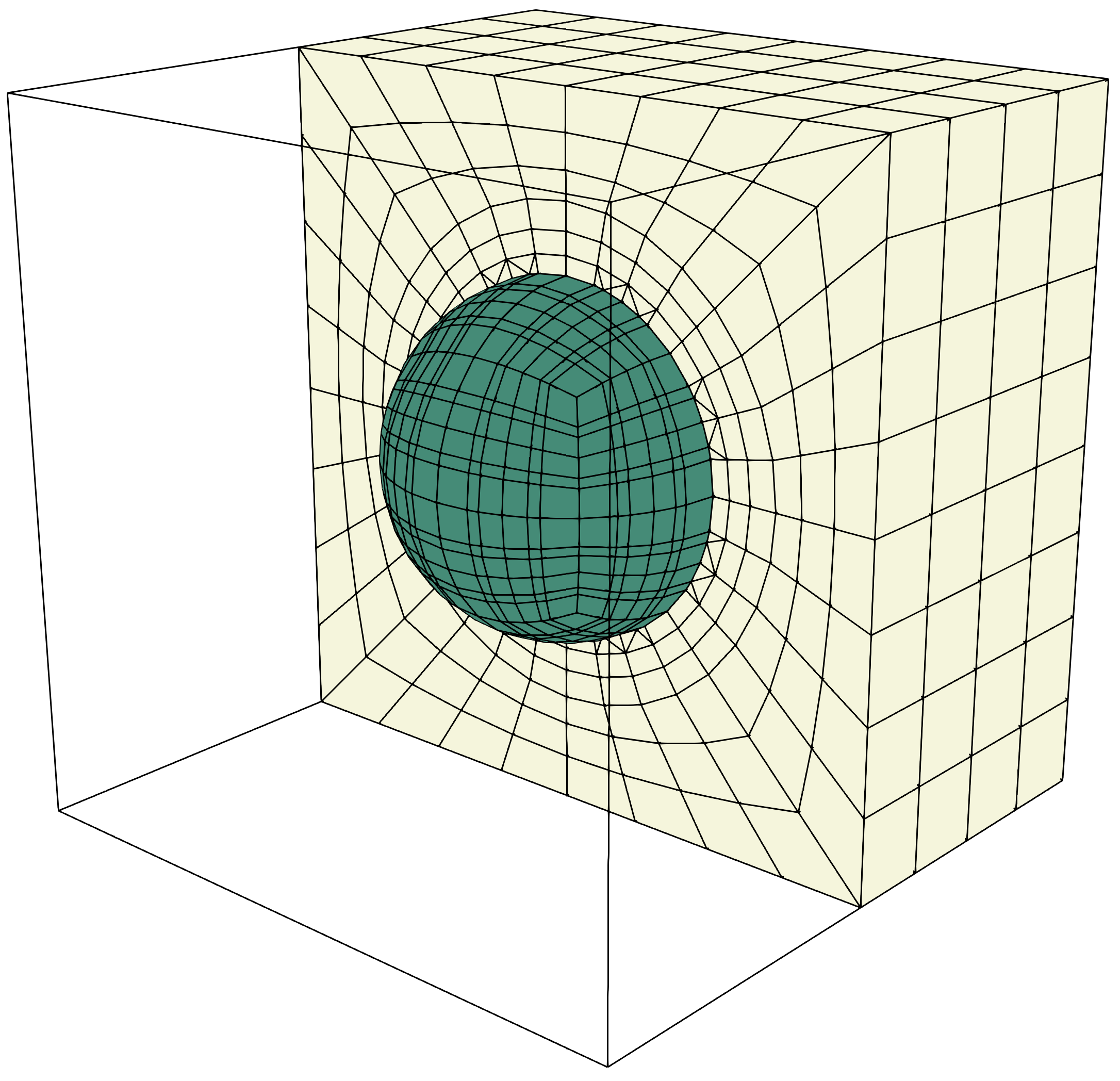}}\hfill{}
\par\end{centering}
\noindent \centering{}\caption{Meshes of a cube with spherical inclusion}
\end{figure}

\begin{figure}
\noindent \begin{centering}
\hfill{}\subfloat[Interface mesh on sphere]{\noindent \centering{}\hspace{0.02\textwidth}\includegraphics[width=0.175\textwidth]{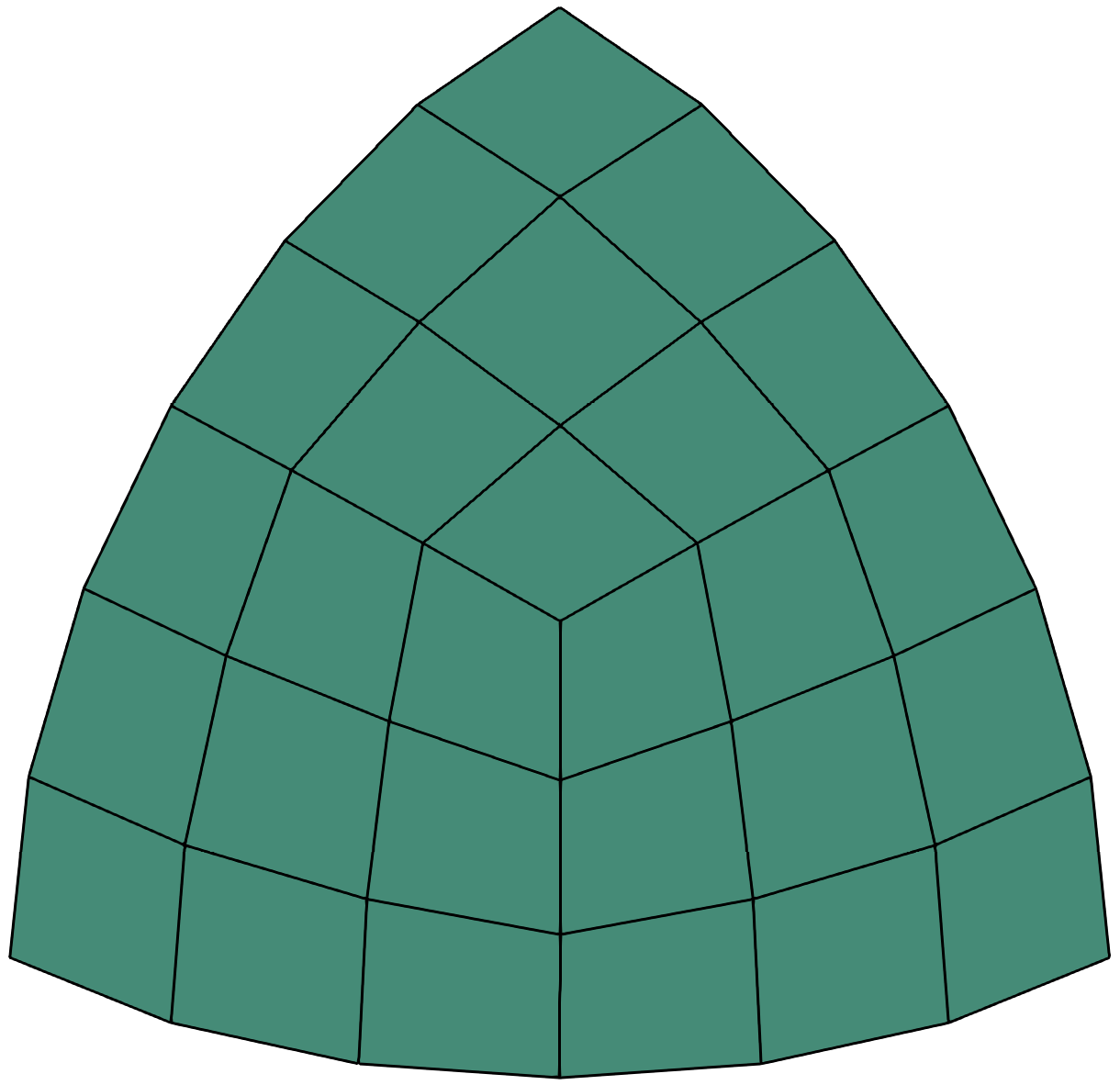}\hspace{0.02\textwidth}}\hfill{}\subfloat[Interface mesh on cube]{\noindent \centering{}\hspace{0.02\textwidth}\includegraphics[width=0.175\textwidth]{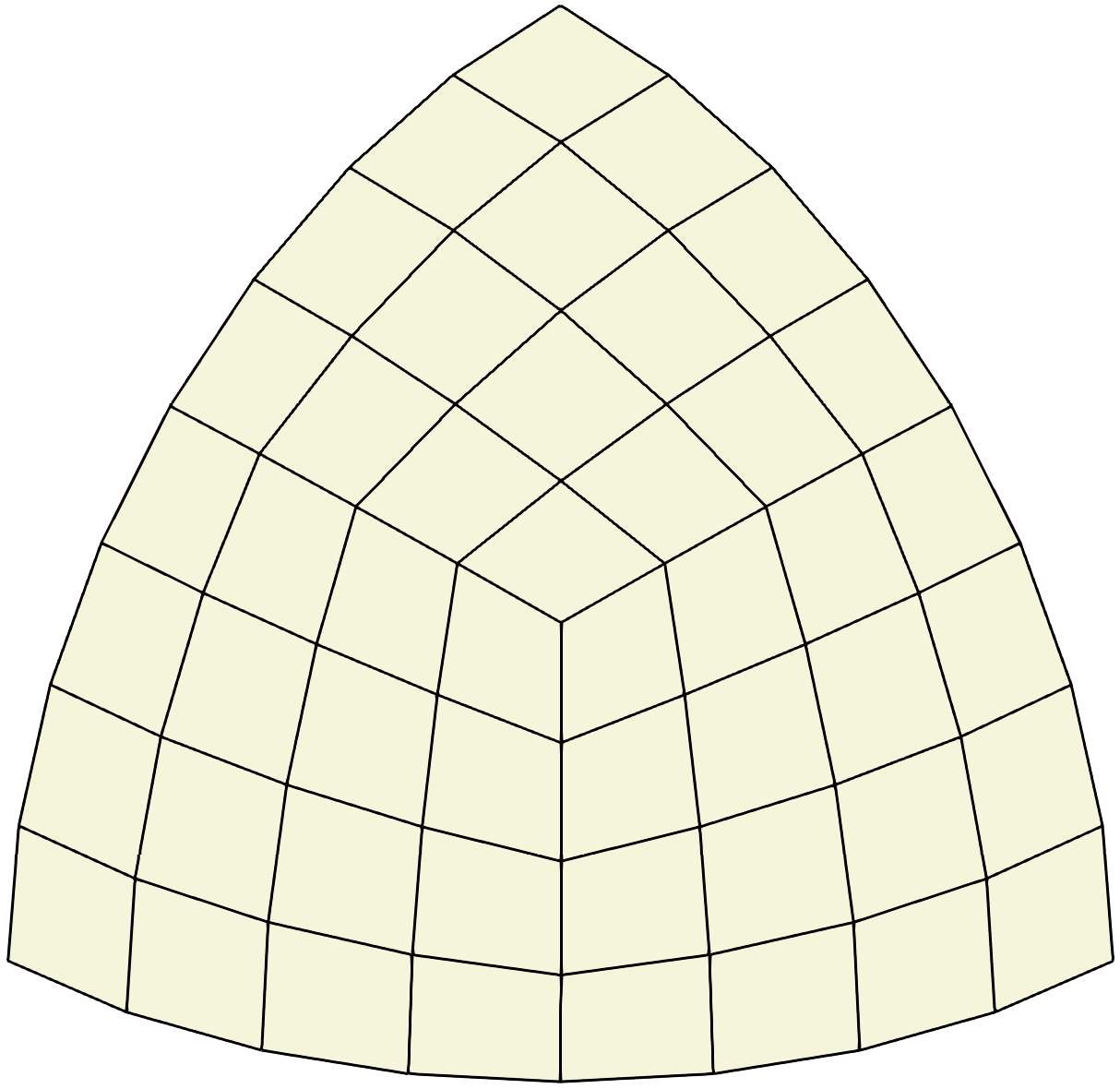}\hspace{0.02\textwidth}}\hfill{}\subfloat[Matching interface mesh]{\noindent \centering{}\hspace{0.02\textwidth}\includegraphics[width=0.175\textwidth]{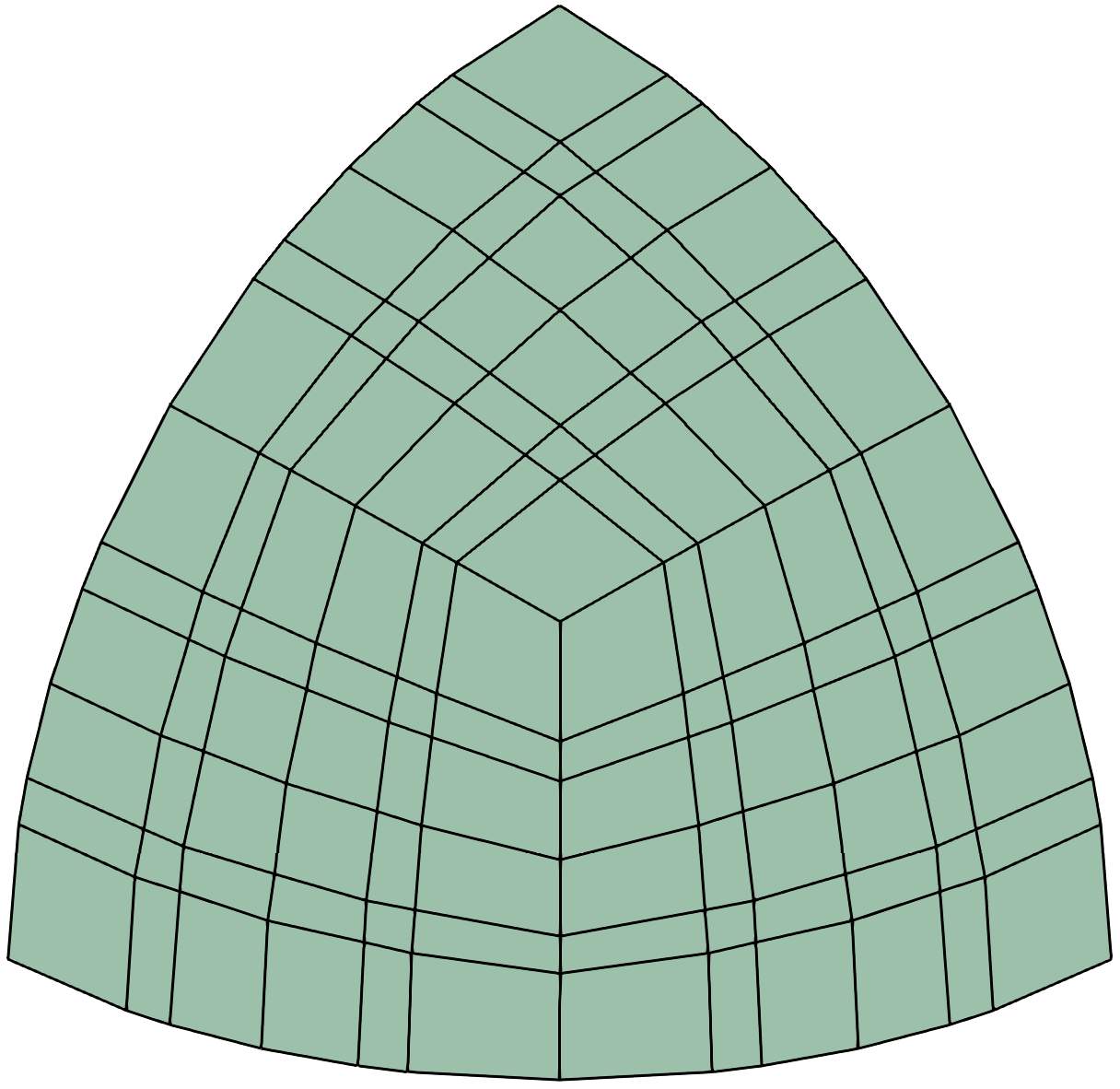}\hspace{0.02\textwidth}}\hfill{}
\par\end{centering}
\caption{Details of matching meshes of an octant\label{fig:Details-of-matching_octant}}
\end{figure}

Two analyses are performed. In one analysis, the model is created
with built-in elements of ABAQUS only (referred to as ABAQUS model
below). In the other analysis, the model includes SBFEM user elements
(referred to as SBFEM model). The ABAQUS model is shown in Fig. \ref{fig:Non-matching-meshes-(C3D8)}.
It features a coarser mesh for the sphere, which makes the interface
meshes non-conforming. In the SBFEM model, matching meshes are generated
by re-meshing the interfacial surfaces, as shown in Fig.~\ref{fig:Matching-meshes-(SBFEM)}.
The re-meshing operation yields polyhedral elements, which can be
easily handled deploying the SBFEM-based user elements. The interface
meshes before and after the re-meshing of an octant (highlighted in
Fig.~\ref{fig:Non-matching-meshes-(C3D8)}) are depicted in Fig.~\ref{fig:Details-of-matching_octant}.
 
In the ABAQUS model, there are 3,168 C3D8 elements and 4,471 nodes.
The cohesive layer is represented by 384 COH3D8 elements, which are
conforming to the inner surface of the cube. The cohesive layer in
this model has a thickness of 1$\mathrm{\mu m}$ and it is connected
to the two solid blocks by tie constraints. In the SBFEM model, only
the surface meshes are adjusted in the re-meshing, thus the number
of solid elements does not change. The interfaces are discretized
into 864 matching quadrilaterals, which form 864 COH3D8 elements and
1728 overlaid elements. The total number of nodes of the matching
mesh is 4,827. As discussed before, the cohesive elements formed by
matching meshes have zero thickness and no constraint is required
to connect them with adjacent elements. A reference solution is also
obtained using ABAQUS built-in elements and a matching mesh. The reference
model consists of 201,800 C3D8 elements and 6,144 COH3D8 elements.

\begin{figure}
\noindent \begin{centering}
\hfill{}\subfloat[ABAQUS model]{\noindent \begin{centering}
\includegraphics[width=0.35\textwidth]{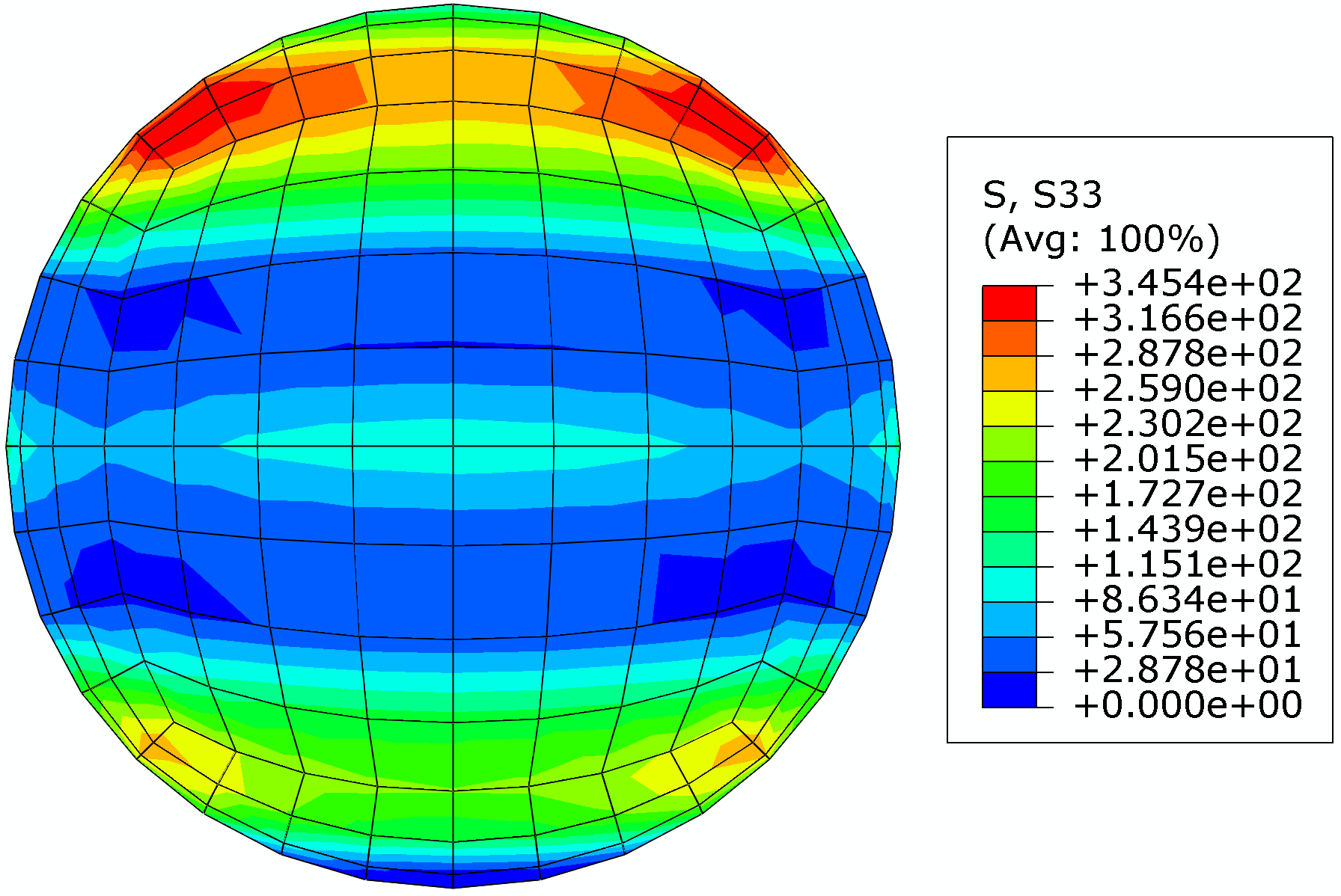}
\par\end{centering}
}\hfill{}\subfloat[SBFEM model]{\noindent \centering{}\includegraphics[width=0.35\textwidth]{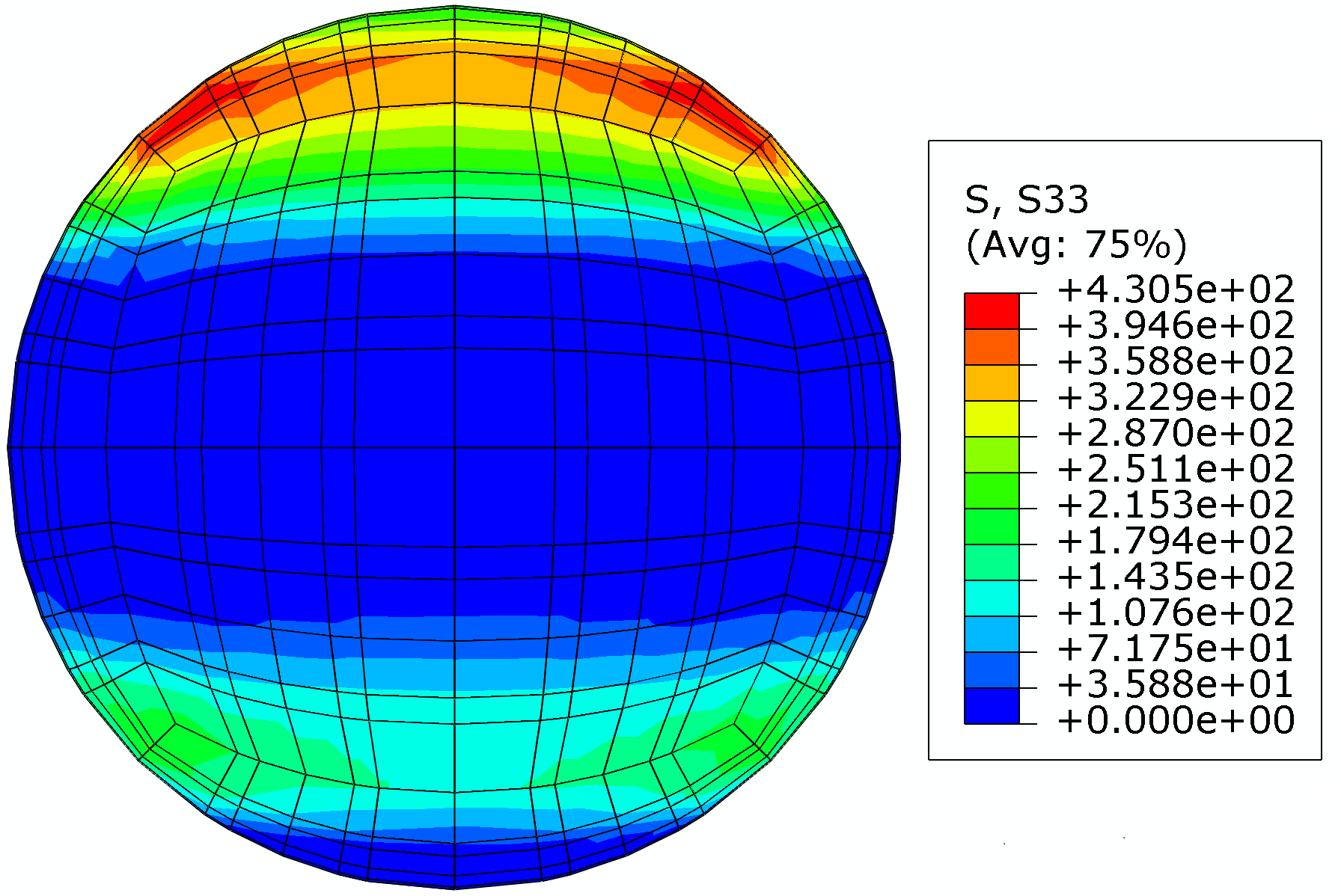}}\hfill{}
\par\end{centering}
\noindent \begin{centering}
\subfloat[Reference model]{\noindent \centering{}\includegraphics[width=0.35\textwidth]{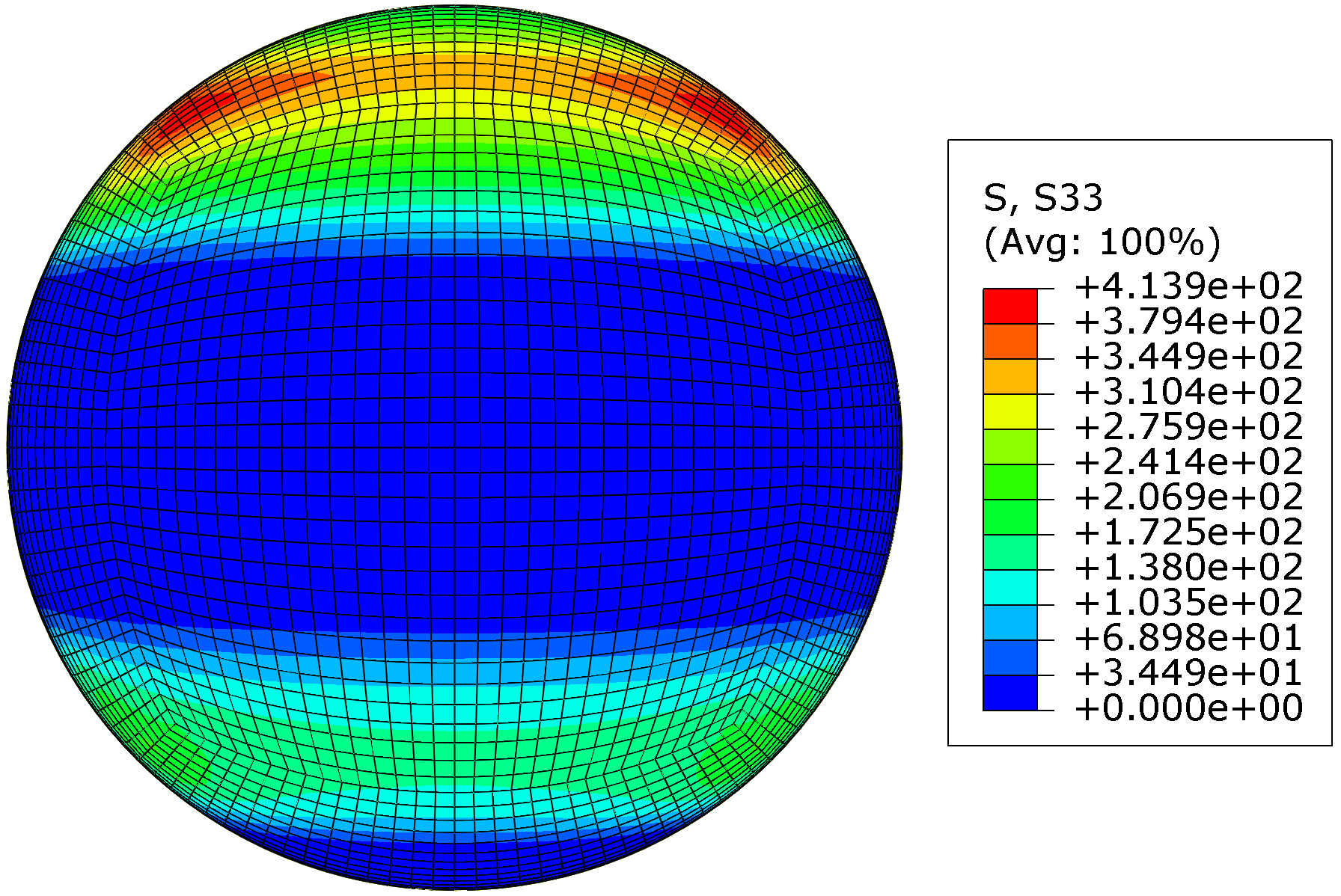}}
\par\end{centering}
\noindent \centering{}\caption{Comparison of the traction contours of cohesive elements (Unit: Pa)
\label{fig:S33-contour-comparison-nonmatching}}
\end{figure}

The values for the normal traction $t_{n}$ of the cohesive elements
obtained from the three models are compared in Fig.~\ref{fig:S33-contour-comparison-nonmatching}.
Note that the parameter S33 in ABAQUS denotes the normal traction
$t_{n}$ of three-dimensional cohesive element. The contour of the
SBFEM model is smoother compared to the ABAQUS model, and it is closer
to the reference solution. 

\begin{figure}
\noindent \begin{centering}
\includegraphics[scale=0.8]{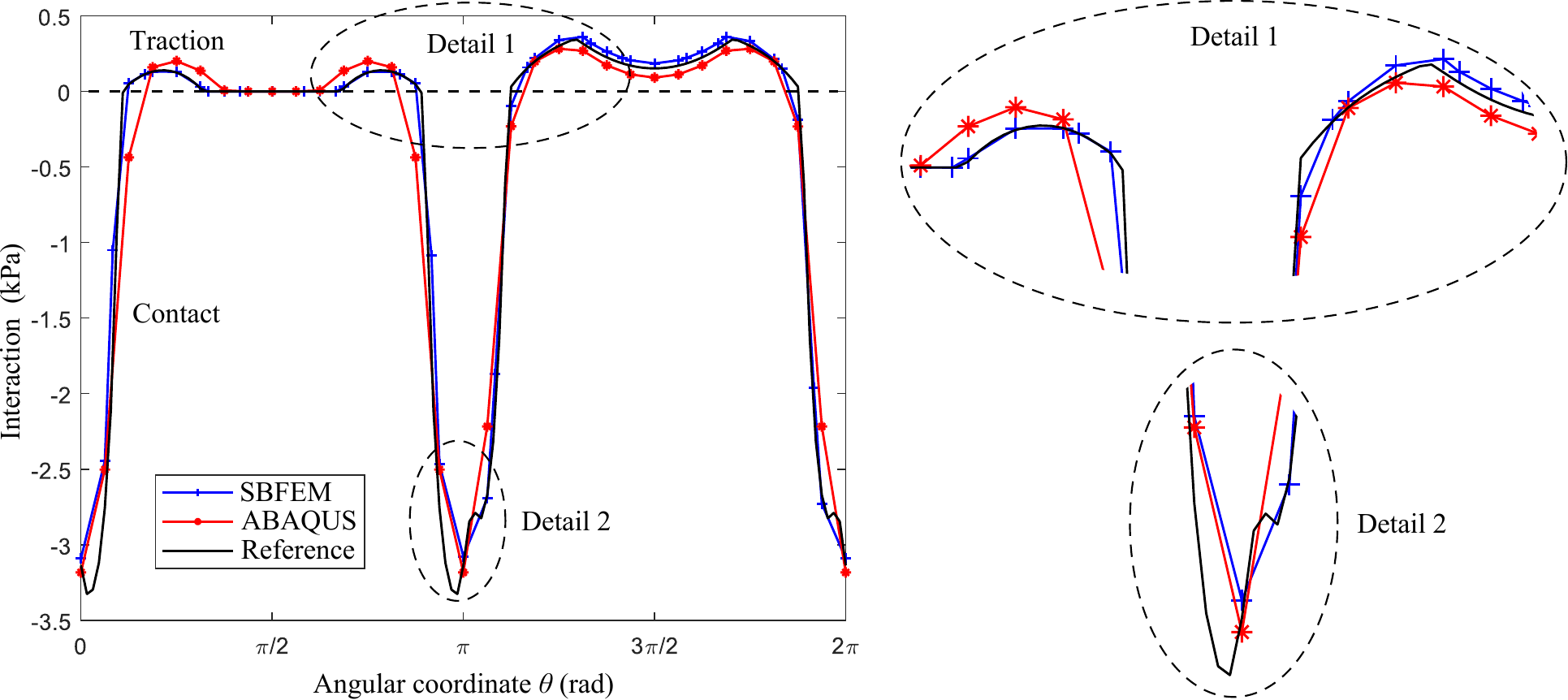}
\par\end{centering}
\noindent \centering{}\caption{Interaction stress comparison in Y-Z plane\label{fig:Interaction-stress-YZ-linear}}
\end{figure}

The interaction occurring at the interface along the designated path
is plotted in Fig.~\ref{fig:Interaction-stress-YZ-linear}. A positive
value indicates the interaction is traction, whereas a negative value
denotes the interaction is contact. The horizontal axis is the angle
$\theta$ denoted in Fig.~\ref{fig:A-cube-with-sphere}. As we can
see in the figure, the result of the SBFEM model is more accurate
than that of the ABAQUS model for both traction and contact zones.
At the region $\text{\ensuremath{\pi}}/3<\theta<2\pi/3$, the interaction
value is 0, which indicates the cohesive elements around this region
have been damaged completely. 

The computational times for the ABAQUS and SBFEM models are compared
in Table \ref{tab:Comparison-of-nonlinear}.  Because the increment
size setting affects the computational time, the default automatic
setting of ABAQUS~\citep{ABA2016} is applied for both models. For
the SBFEM model, a pre-calculation step is created before the actual
analysis step to obtain stiffness and mass matrices, as discussed
in Section \ref{subsec:General-workflow}. The time used for the element
analysis of UEL comprises of both the time consumed in the pre-calculation
and the time in the actual step of analysis. For this problem, the
 $\mathrm{UEL}$ time accounts for only 18.2\% of the total CPU time.
The SBFEM model, which has matching interfacial meshes, converges
faster than the ABAQUS model. As a result, the SBFEM model requires
15 increments to complete the analysis while the ABAQUS model requires
36 increments. The amount of CPU time saved from the iterative solution
is much larger than the extra time spent on the element analysis of
UEL.\textbf{ }Using the SBFEM model leads to a 70.8\% saving of the
total CPU time from the ABAQUS model.\textbf{ }
\begin{table}
\centering{}\caption{Comparison of computational times (s) for non-linear analysis\label{tab:Comparison-of-nonlinear}}
\begin{tabular}{ccccc}
\toprule 
Model & No. of elements & UEL time & ABAQUS solution time & Total CPU time\tabularnewline
\midrule 
ABAQUS & 3552 & - & - & 373.70\tabularnewline
\midrule 
SBFEM & 5824 & 19.90 & 89.4 & 109.30\tabularnewline
\bottomrule
\end{tabular}
\end{table}

Apart from the convenience of inserting cohesive elements, the matching
meshes have obvious higher accuracy and efficiency compared to the
non-matching ones, especially for curved surfaces. The matching meshes
are efficient for the numerical simulation of curved ITZs, which commonly
exist in composite materials.

\subsection{Examples for automatic mesh generation}

In previous studies, the octree algorithm has been proven to be a
robust and fully automatic mesh generation approach~\citep{Yerry1984,Shephard1991}.
This technique generates octree cells by recursively subdividing one
cell into eight smaller ones of equal size until only one material
exists in one cell or a certain threshold is reached. Despite its
robustness, the application of octree algorithms has been hindered
in conventional FEM due to the issue of hanging nodes. 

SBFEM user elements, which allow highly flexible shapes, are complementary
to the meshes generated by the octree algorithm. Their combination
provides a promising technique for integrating geometric models and
numerical analysis in a fully automatic manner. In recent years, octree
algorithms have been successfully employed in the SBFEM to generate
polyhedral meshes based on digital images~\citep{saputra2017,ASaputra2020}
and STL models~\citep{Liu2017}. They are extremely simple and efficient
in meshing complex geometries such as concrete and sculptures. Especially
the STL-based automatic mesh generation approach is capable of handling
curved boundaries by trimming the octree cells~\citep{Liu2017}. 

As commercial software, ABAQUS has powerful non-linear solver and
it allows parallel equation solution, which benefits the solutions
of interfacial problems, especially for geometrically complex problems.
Besides, ABAQUS provides comprehensive contact approaches and offers
enriched element library that we can combine SBFEM user element with.
Especially the contact cohesive behavior and cohesive element, those
robust algorithms can be utilized for simulating the traction-separation
response occurring in interfacial problems.

In this section, two numerical examples are presented to show the
possibilities of using octree cells in ABAQUS in conjunction with
our user elements. The first example performs a quasi-static analysis
of a porous asphalt concrete specimen meshed by the image-based octree
algorithm, in which the interface stripping is simulated by cohesive
elements. In the second example, contact analysis is performed on
a knee joint meshed by the STL-based octree algorithm with trimming.

\subsubsection{Analysis of a porous asphalt concrete specimen\label{subsec:Analysis-of-asphalt}}

\begin{figure}
\noindent \begin{centering}
\hfill{}\subfloat[Digital image \citep{hu2015}\label{fig:Digital-image-concrete}]{\noindent \centering{}\includegraphics[width=0.35\textwidth]{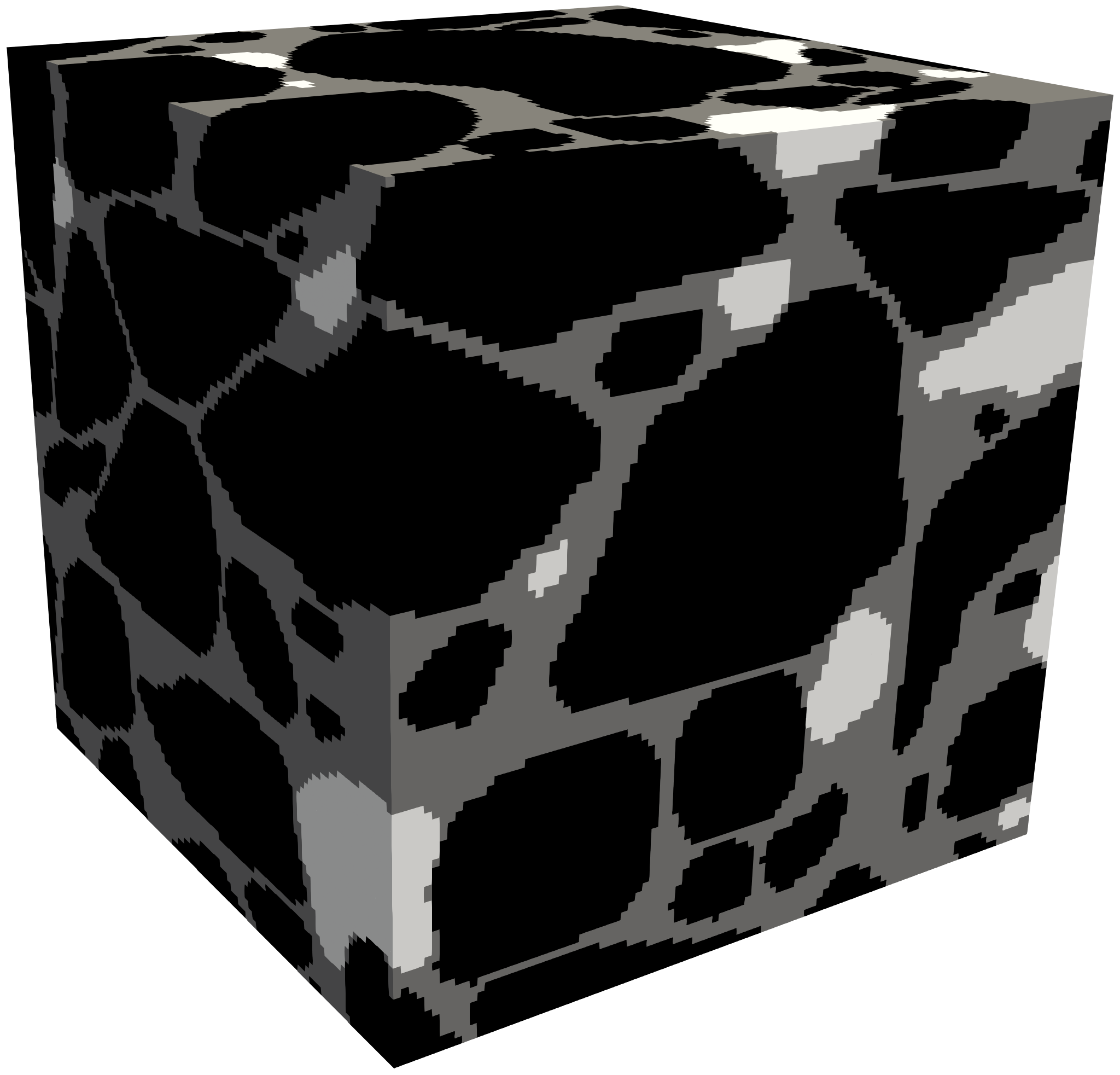}}\hfill{}\subfloat[Mesh\label{fig:Mesh-concrete}]{\noindent \centering{}\includegraphics[width=0.35\textwidth]{../Figures/concrete_mesh}}\hfill{}
\par\end{centering}
\caption{A three-dimensional asphalt concrete specimen\label{fig:A-three-dimensional-concrete}}
\end{figure}

Interface stripping is a common distress appeared in asphalt mixture
due to the adhesive failure, which is commonly simulated through cohesive
elements in ABAQUS~\citep{Hu2019}. The internal states of the asphalt
mixtures can be obtained from high-resolution X-CT scans and processed
by digital image processing techniques~\citep{hu2015,Hu2019}. The
microstructure can be reconstructed by octree meshes based on the
digital images in a fully automatic manner~\citep{ASaputra2020,saputra2017}.
A porous asphalt concrete specimen is illustrated in Fig.~\ref{fig:Digital-image-concrete},
which is represented by digital images consisting of $200\times200\times200$
voxels. The image is a part of the X-CT scans of an asphalt concrete
specimen from Ref.~\citep{hu2015}. In the figure, the black, gray,
and white phases represent the aggregates, asphalt mastic, and air-voids,
respectively. 

The images can be decomposed automatically using the octree structure,
as explained in Ref.~\citep{saputra2017}. The discretization of
the asphalt concrete specimen is shown in Fig.~\ref{fig:Mesh-concrete}.
The elements representing air-voids are removed. At the interface,
cohesive elements are inserted for the simulation of the stripping
damage, and STS contact with a friction coefficient 0.5~\citep{Hu2019}
is assigned to prevent interpenetration. The minimum and maximum element
sizes are $2^{3}$ and $16^{3}$ voxels, respectively, and the size
of one pixel is $0.1\,\mathrm{mm}$. There are 539,170 polyhedral
user elements, 208,880 cohesive elements, and 909,402 nodes in the
model. The number of nodes in one polyhedral element varies from 8
to 26. The main feature of such octree meshes is the meshes at the
interfaces between material phases are fine. Besides, the interfacial
meshes are conforming naturally. The meshes having such features can
be constructed by the SBFEM user elements, and they are exactly the
meshes required for the simulation of the ITZs as discussed in the
previous section. 

The mechanical properties used in this examples are listed in Table~\ref{tab:Material-properties-of-pac},
in which the viscoelasticity of the asphalt mastic is not taken into
consideration. The interface has the same damage behaviors in the
normal and two tangential directions, and allows a maximum traction
$4\,\mathrm{MPa}$ in each direction. Because the dynamic modulus
is often adapted for the asphalt mastic~\citep{hu2015}, a quasi-static
analysis is performed on the model. The bottom of the concrete specimen
is fixed, and the vertical planes are constrained for the simulation
of confinement. A displacement controlled load up to $-0.5\,\mathrm{mm}$
is applied on the top surface with a loading rate $0.02\,\mathrm{mm}/\mathrm{s}$.
\begin{table}
\caption{Material properties of a porous asphalt concrete specimen\label{tab:Material-properties-of-pac}}
\begin{tabular}{cccccc}
\toprule 
 & Young's modulus $E\left[\mathrm{MPa}\right]$ & Poisson's ratio $\nu$ & Density $\rho\left[\mathrm{kg/m^{3}}\right]$ & $\delta_{n}^{\mathrm{o}}/\delta_{s}^{\mathrm{o}}/\delta_{t}^{\mathrm{o}}\left[\mathrm{mm}\right]$ & $\delta_{m}^{\mathrm{f}}\left[\mathrm{mm}\right]$\tabularnewline
\midrule
Aggregate~\citep{Hu2019} & $55000$ & 0.25 & 2400 & - & -\tabularnewline
Asphalt mastic~\citep{buttlar1999} & $500$ & 0.40 & 2000 & - & -\tabularnewline
Interface~\citep{Hu2019} & $20000$ & - &  & $0.0002$ & $0.0125$\tabularnewline
\bottomrule
\end{tabular}
\end{table}

\begin{figure}
\noindent \begin{centering}
\hfill{}\subfloat[Mises stress contour (Unit: MPa)\label{fig:Mises-stress-contour-concrete}]{\centering{}\includegraphics[width=0.45\textwidth]{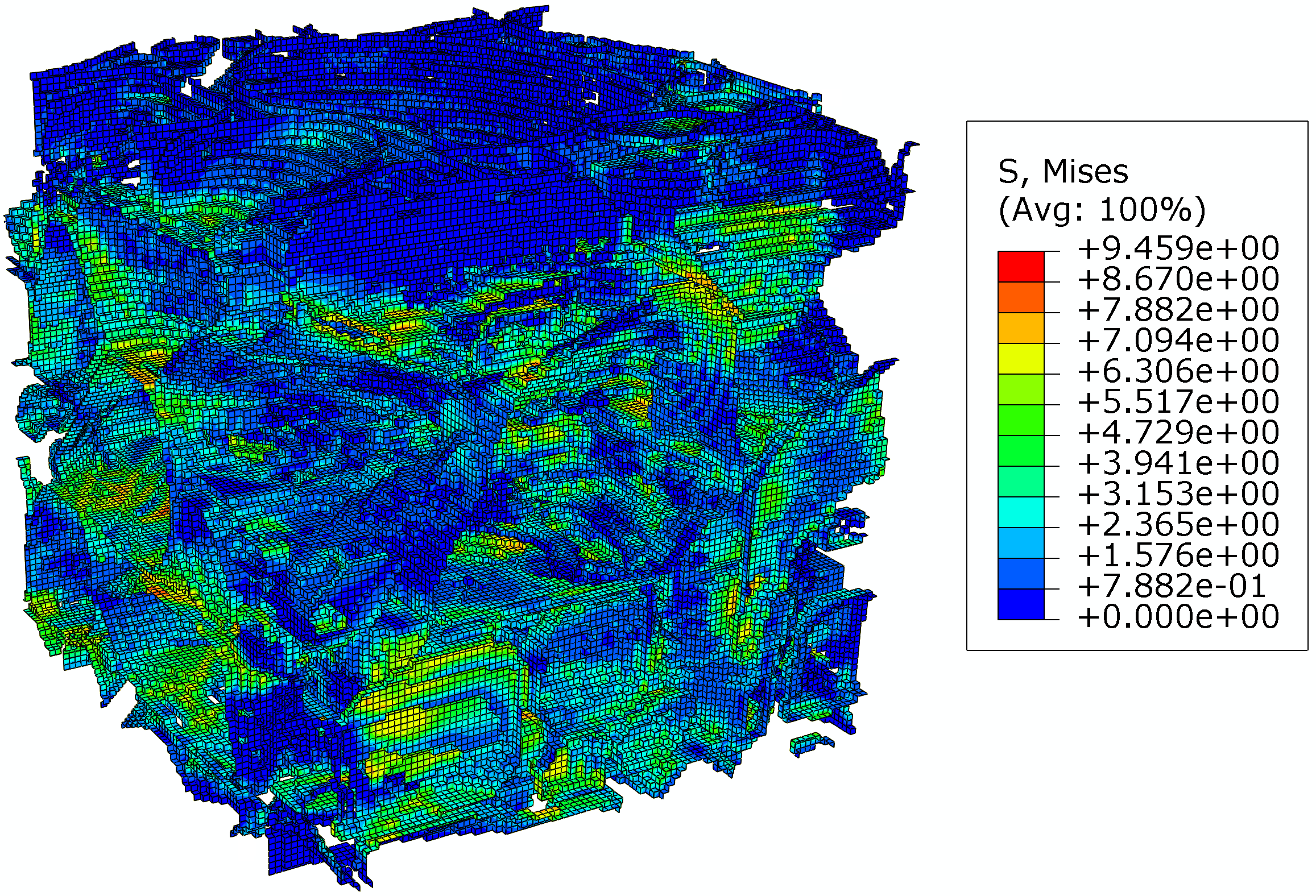}}\hfill{}\subfloat[Damage contour\label{fig:Damage-contour-concrete}]{\centering{}\includegraphics[width=0.45\textwidth]{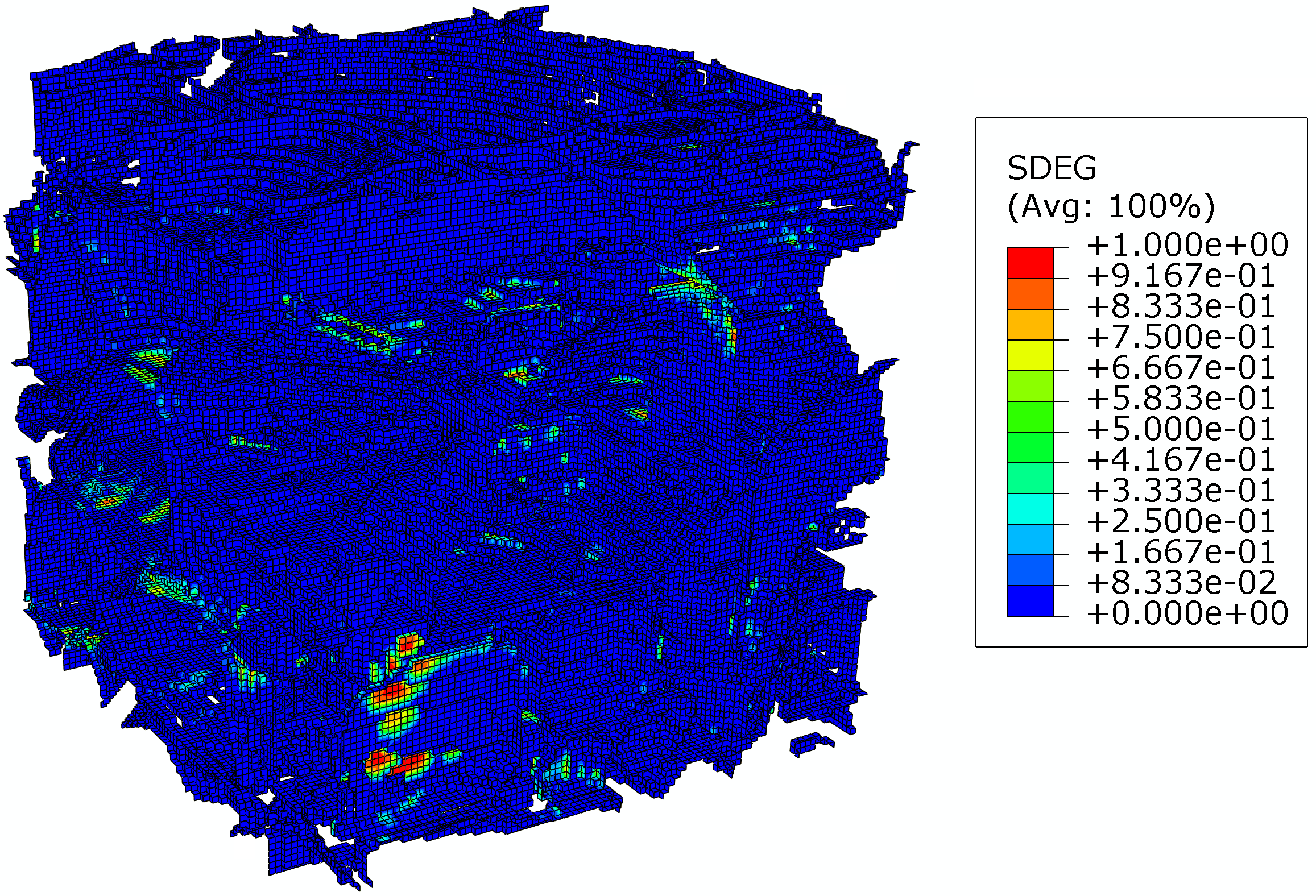}}\hfill{}
\par\end{centering}
\noindent \centering{}\caption{Interface behaviors of a porous asphalt concrete specimen\label{fig:Interface-behavior-concrete}}
\end{figure}

The stress and damage states of the cohesive elements at the end are
depicted in Fig.~\ref{fig:Interface-behavior-concrete}. From Fig.~\ref{fig:Mises-stress-contour-concrete}
we can see, the Mises stress has reached up to $9.459\,\mathrm{MPa}$
in some area, which may cause damage initiation. The damage contour
in Fig.~\ref{fig:Damage-contour-concrete} (the parameter SDEG in
ABAQUS represents the scalar damage value $D$) shows clearly that
the damage has occurs at the interface. In some areas, the damage
value reaches to 1, where the interface has complete failure. Those
areas are concentrated at the edges of interface or smooth interfaces.
This phenomenon is reasonable, because the damage in the normal and
tangential directions tends to initiate at the edges of interface
and smooth interfaces, respectively. 

The total CPU time of this example is 233.51 hours, in which 11.48
hours are spent on the element analysis of UEL. The UEL time comprises
of 1.96 hours on the pre-calculation and 9.52 hours on the step of
nonlinear analysis. It accounts for only 4.9\% of the total CPU time.

\subsubsection{Analysis of a knee joint\label{subsec:Analysis-of-knee}}

\begin{figure}
\noindent \begin{centering}
\includegraphics[scale=0.135]{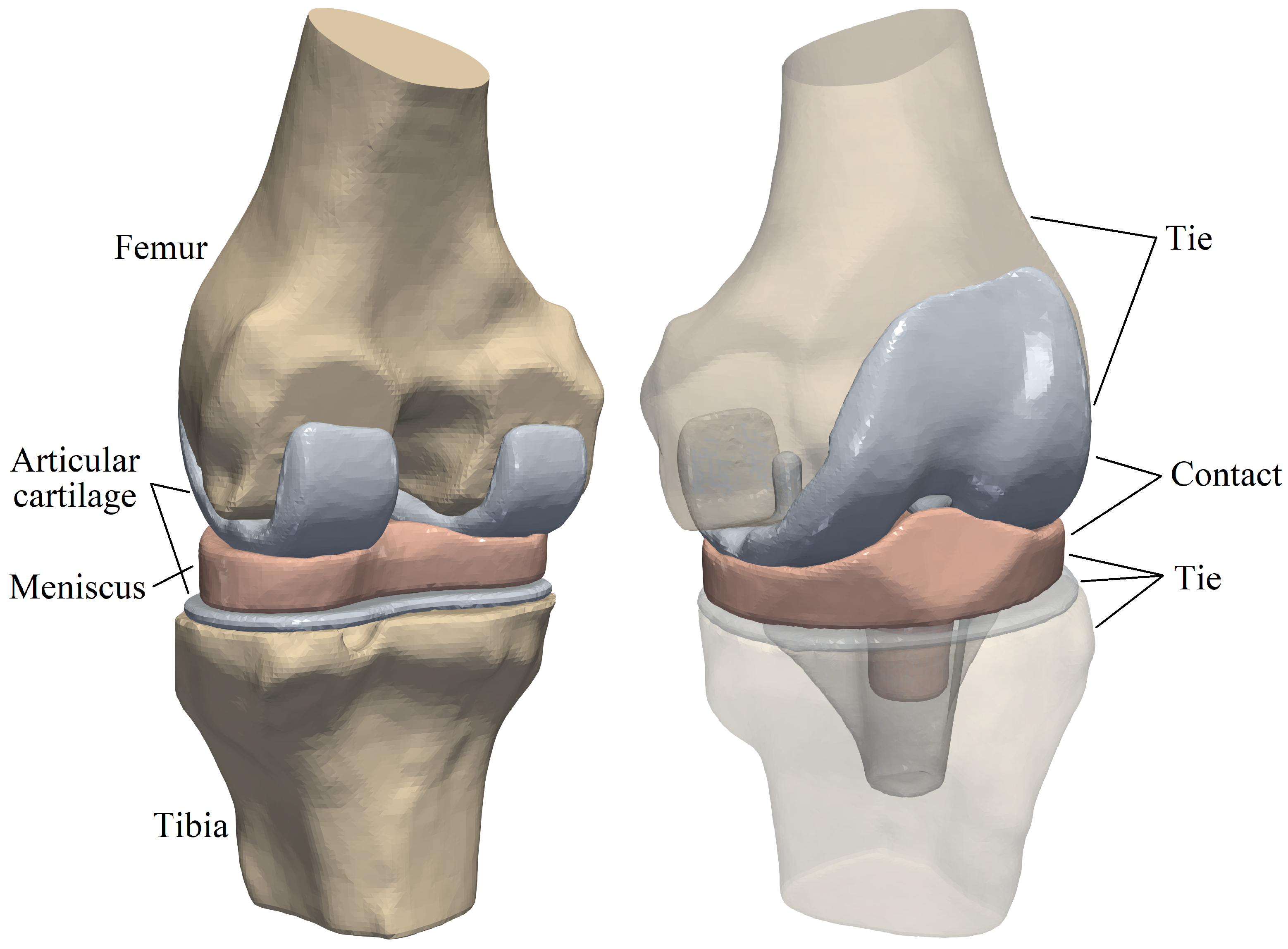}
\par\end{centering}
\noindent \centering{}\caption{Geometry of a knee model\label{fig:Geometry-of-a-knee-model}}
\end{figure}

A knee joint given in STL format \citep{ISIFC2014} is considered
for performing a contact analysis. The knee joint is composed of femur
and tibia, a meniscus and two articular cartilages as illustrated
in Fig.~\ref{fig:Geometry-of-a-knee-model}. The femur and the tibia
are covered with articular cartilages, thus the interactions between
bones and articular cartilages can be treated as perfect bond (tie
constraint in ABAQUS). The meniscus is inserted into the articular
cartilage covering the tibia, thus the interaction between them can
also be regarded as perfect bond. The interaction between the meniscus
and the articular cartilage covering the femur is treated as contact.

\begin{figure}
\noindent \begin{centering}
\includegraphics[scale=0.14]{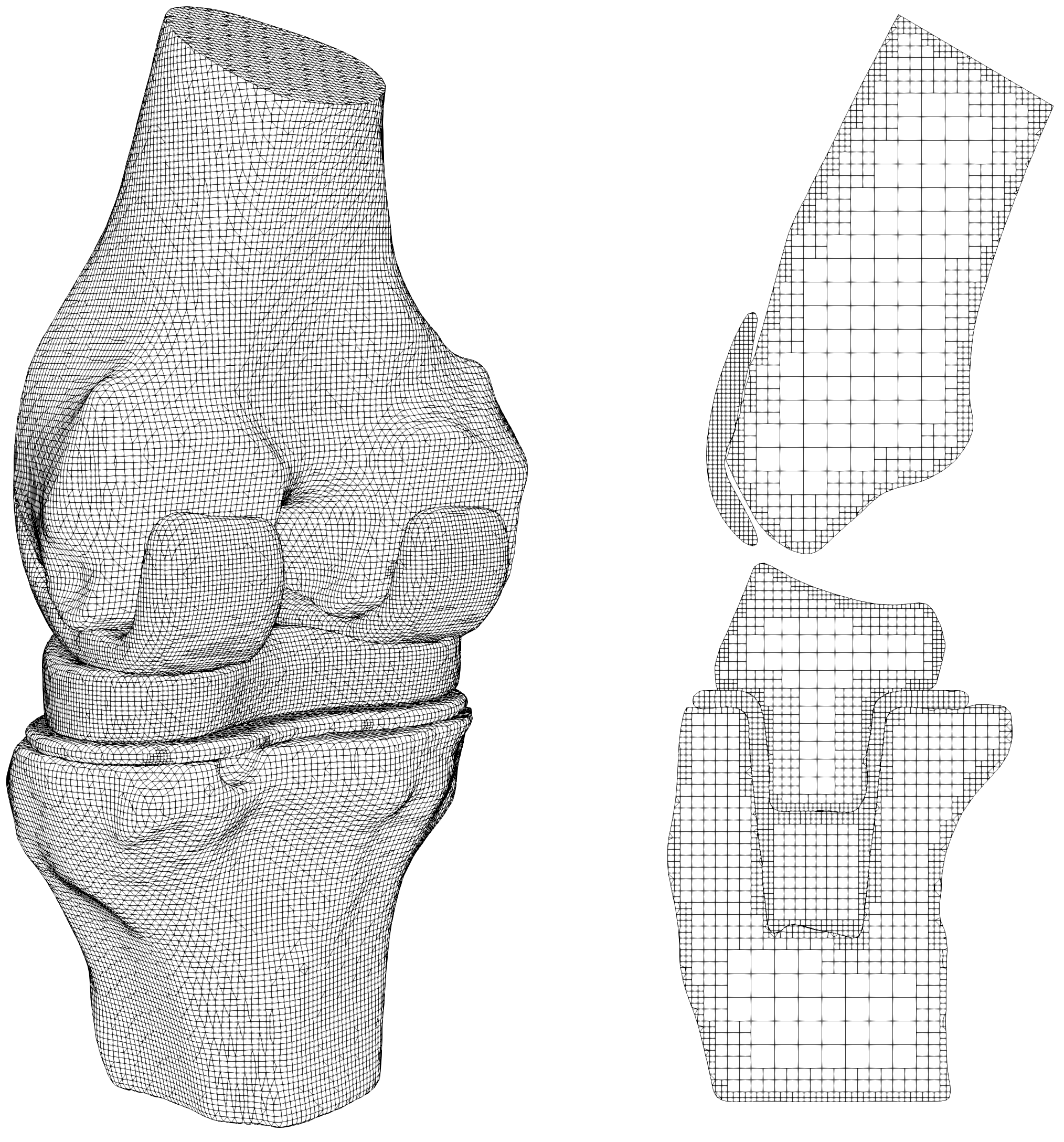}
\par\end{centering}
\noindent \centering{}\caption{Knee mesh generated by octree algorithm with trimming\label{fig:Knee-mesh}}
\end{figure}

The mesh generated by the octree algorithm with trimming~\citep{Liu2017}
is shown in Fig.~\ref{fig:Knee-mesh}. The base sizes are $0.75\,\mathrm{mm}$
and $1\,\mathrm{mm}$ for the contact parts and other parts, respectively.
There are 253,079 elements and 361,296 nodes, and the number of nodes
in one element varies from 4 to 27. A slice in the vertical plane
passing the geometrical center is also shown in Fig. \ref{fig:Knee-mesh}.
The contact interface meshes are non-matching for this example, because
the contact domains are geometrically non-conforming and large sliding
might occurs at the interface during the loading process. The meshes
generated through the STL-based octree algorithm are suitable for
contact analysis. In the standard FEM, local mesh refinement may be
required to generate finer meshes on the contact surfaces. However,
in an automatic manner, octree cells with fast mesh size transition
can be generated through the octree mesh generation technique. Compared
to the interior elements, the elements near the boundaries are smaller,
which will generate more accurate surfaces for tie constraints and
contact pairs. Besides, compared to the image-based technique, the
STL-based octree algorithm allows trimming on the octree cells, which
leads to more accurate meshes representing complex boundaries especially
for curved shapes. 

The tissue material properties in the computational modeling of human
knee joints have been studied in Ref.~\citep{peters2018}, however,
there are considerable variations of the material properties within
the wide body of literature. The mechanical properties used in this
example are mainly taken from Ref.~\citep{wang2014comparison} and
are listed in Table \ref{tab:Material-properties-of-knee}. The contact
is STS contact with a friction coefficient of $\mu=0.05$ \citep{McCann2009}.
\begin{table}
\centering{}\caption{Material properties of the knee joint~\citep{wang2014comparison}\label{tab:Material-properties-of-knee}}
\begin{tabular}{ccccc}
\toprule 
 & Femur & Tibia & Articular cartilage & Meniscus\tabularnewline
\midrule
Young's modulus $E\left[\mathrm{MPa}\right]$ & $20,000$ & $20,000$ & $30$ & $2$\tabularnewline
Poisson's ratio $\nu$ & 0.2 & 0.2 & 0.25 & 0.35\tabularnewline
\bottomrule
\end{tabular}
\end{table}

\begin{figure}
\noindent \begin{centering}
\includegraphics[scale=0.16]{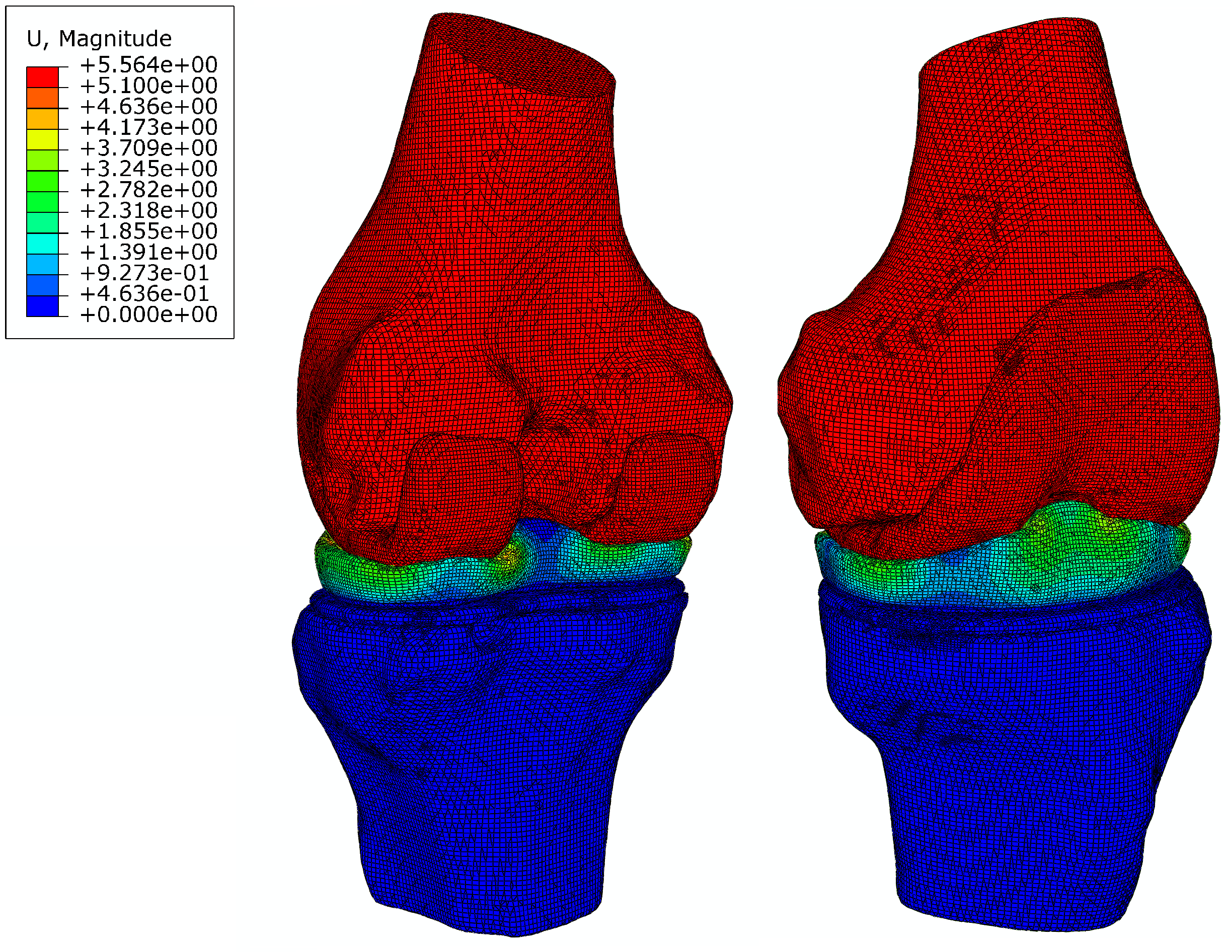}
\par\end{centering}
\noindent \centering{}\caption{Displacement contours of the knee joint (Unit: $\mathrm{mm}$)\label{fig:Displacement-contours-of-knee}}
\end{figure}

The bottom surface of the tibia head is fixed, and two translational
DOFs in the horizontal plane of the top surface of the femur head
are also constrained. On the top surface of the femur head, a vertical
load $F$ of up to $1000\,\mathrm{N}$ \citep{wang2014comparison}
is applied to induce contact in the knee joint. The displacement contours
($F=1000\,\mathrm{N}$) from different views are shown in Fig.~\ref{fig:Displacement-contours-of-knee}.
It is obvious that the deformation occurs mainly in the meniscus.
The femur, tibia and articular cartilages are similar to rigid bodies.
This phenomenon meets the expectation because the Young's modulus
of the meniscus is significantly smaller than those of the bones and
articular cartilages. The meniscus is squeezed into the joint space
and there is lateral extension of the meniscus. The large deformation
of the meniscus increases the contact area which is helpful to reduce
the contact pressure. 

\begin{figure}
\noindent \begin{centering}
\hfill{}\subfloat[Contact pressure on articular cartilage]{\noindent \centering{}\hspace{0.04\textwidth}\includegraphics[scale=0.1]{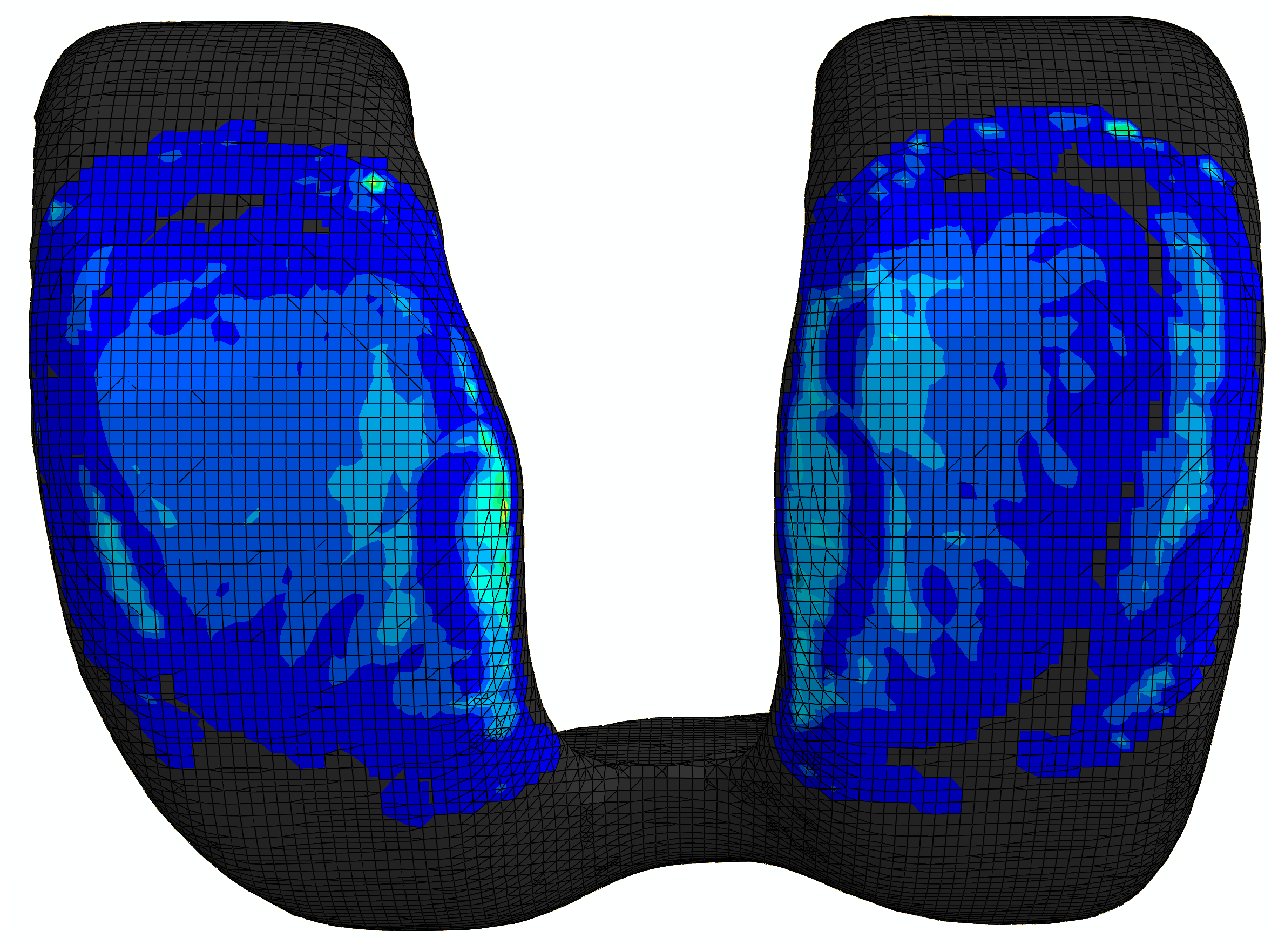}\hspace{0.04\textwidth}}\hfill{}\includegraphics[scale=0.1]{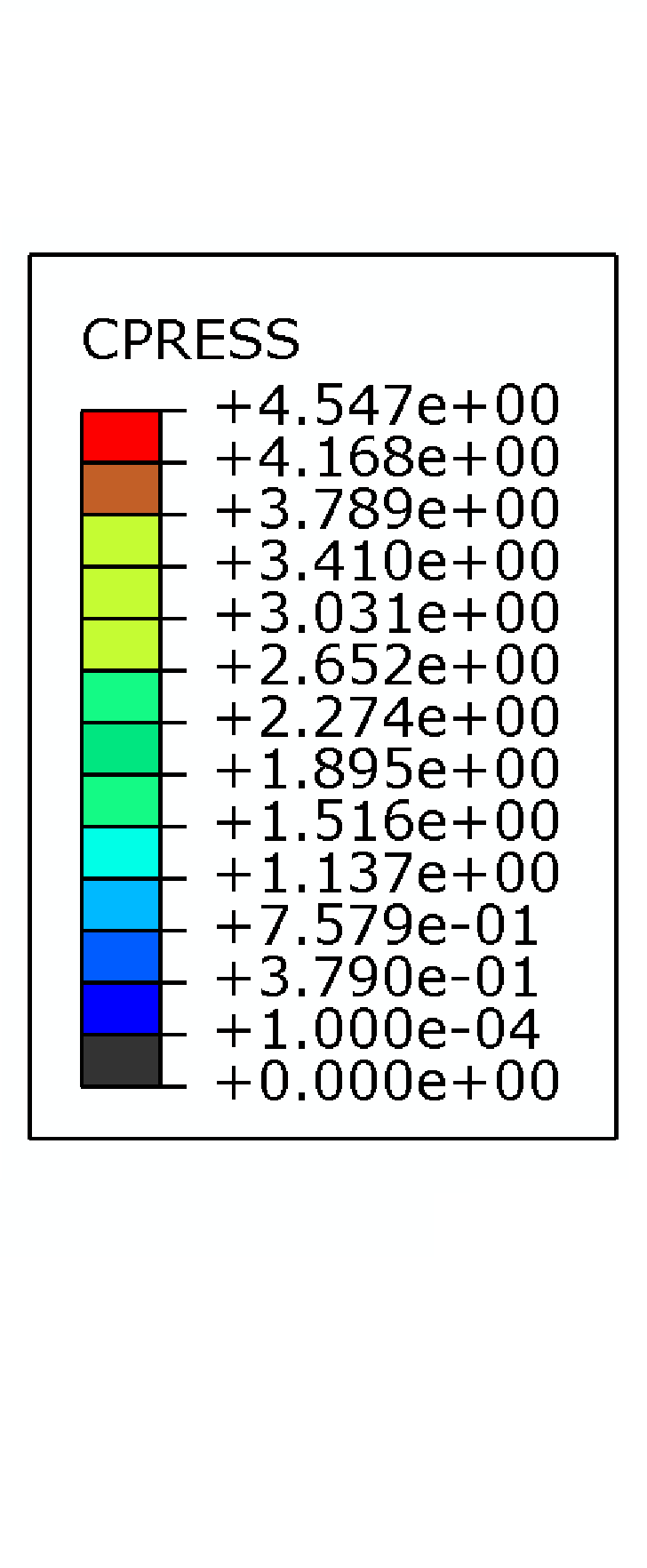}\hfill{}\subfloat[Contact pressure on meniscus]{\noindent \centering{}\hspace{0.04\textwidth}\includegraphics[scale=0.1]{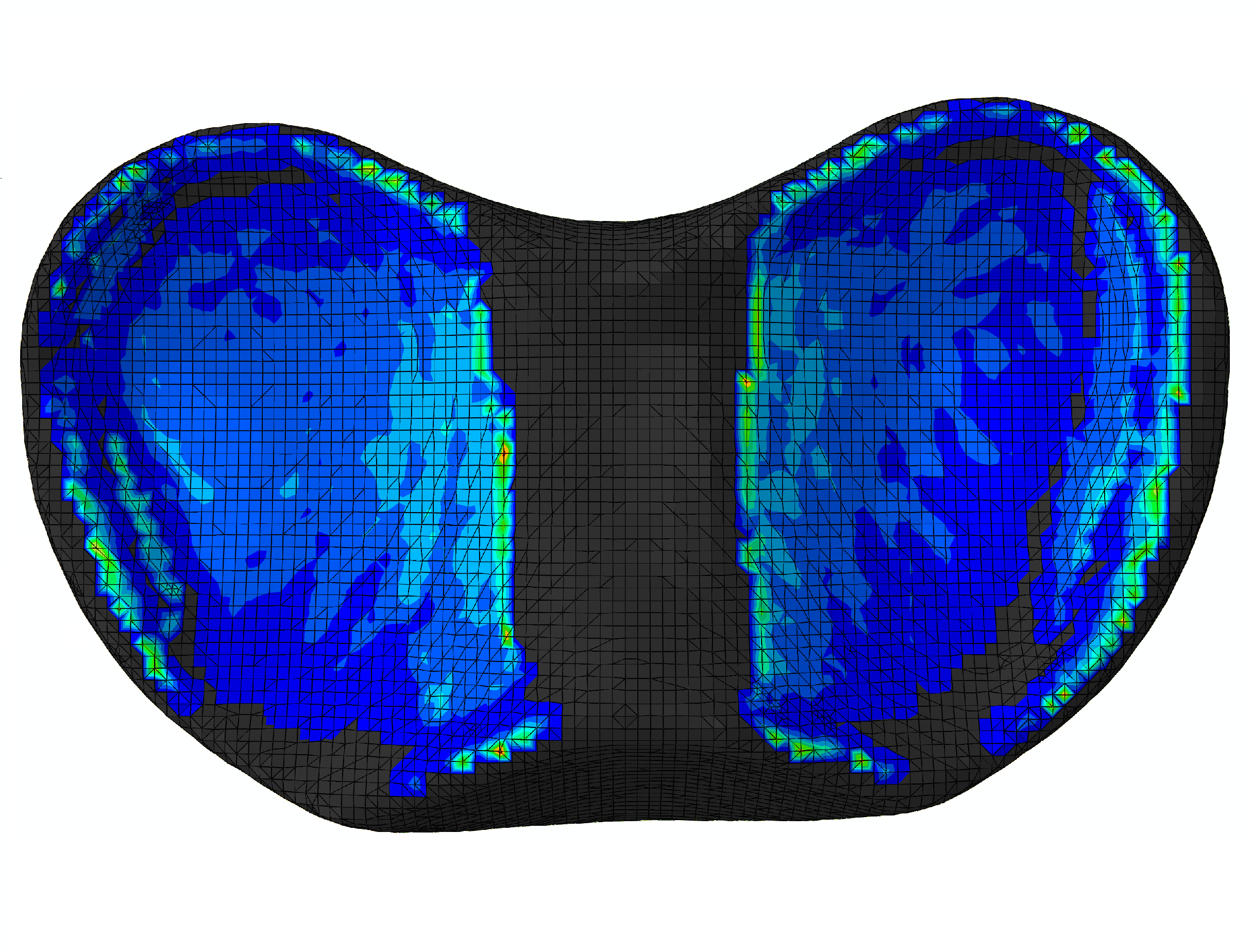}\hspace{0.04\textwidth}}\hfill{}
\par\end{centering}
\noindent \centering{}\caption{Contact pressure of knee joint (Unit: $\mathrm{MPa}$)\label{fig:Contact-pressure-of-knee}}
\end{figure}

The contact pressure distributions ($F=1000\,\mathrm{N}$) are depicted
in Fig.~\ref{fig:Contact-pressure-of-knee}. To illustrate the distribution
clearly, contact pressure values smaller than $10^{-4}\,\mathrm{MPa}$
are not included in the contour. It is obvious that the contact pressure
on the articular cartilage has a similar distribution and magnitude
compared to that on the meniscus. Besides, the contact pressure on
the articular cartilage has a similar distribution from Ref.~\citep{wang2014comparison}.

\begin{figure}
\noindent \begin{centering}
\hfill{}\subfloat[Contact Area]{\noindent \centering{}\includegraphics[scale=0.8]{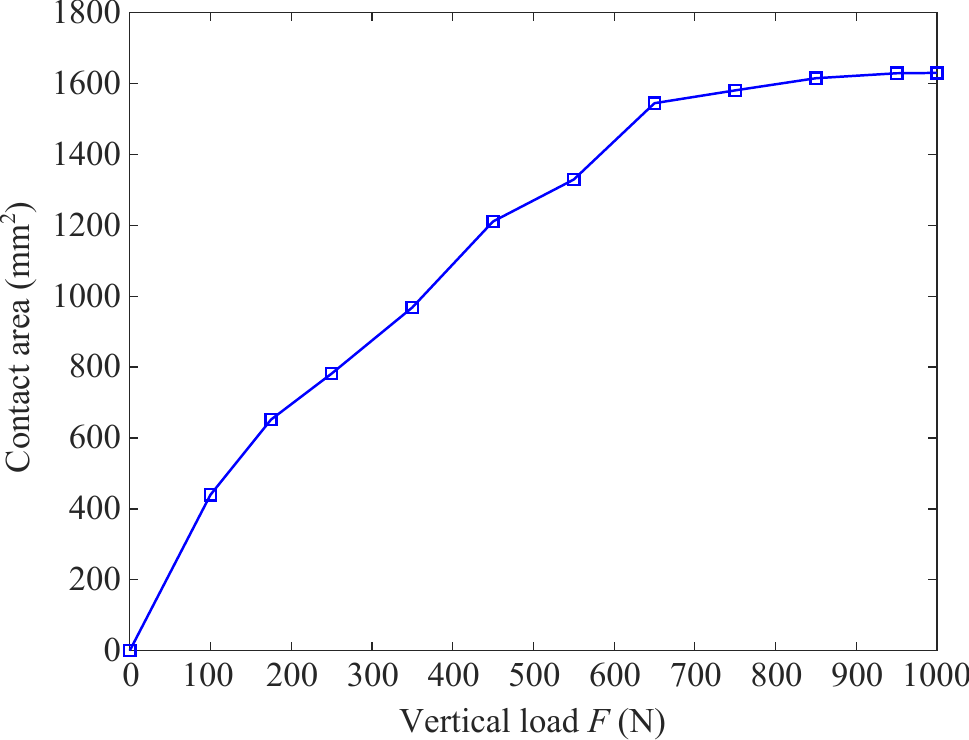}}\hfill{}\subfloat[Average contact pressure]{\noindent \centering{}\includegraphics[scale=0.8]{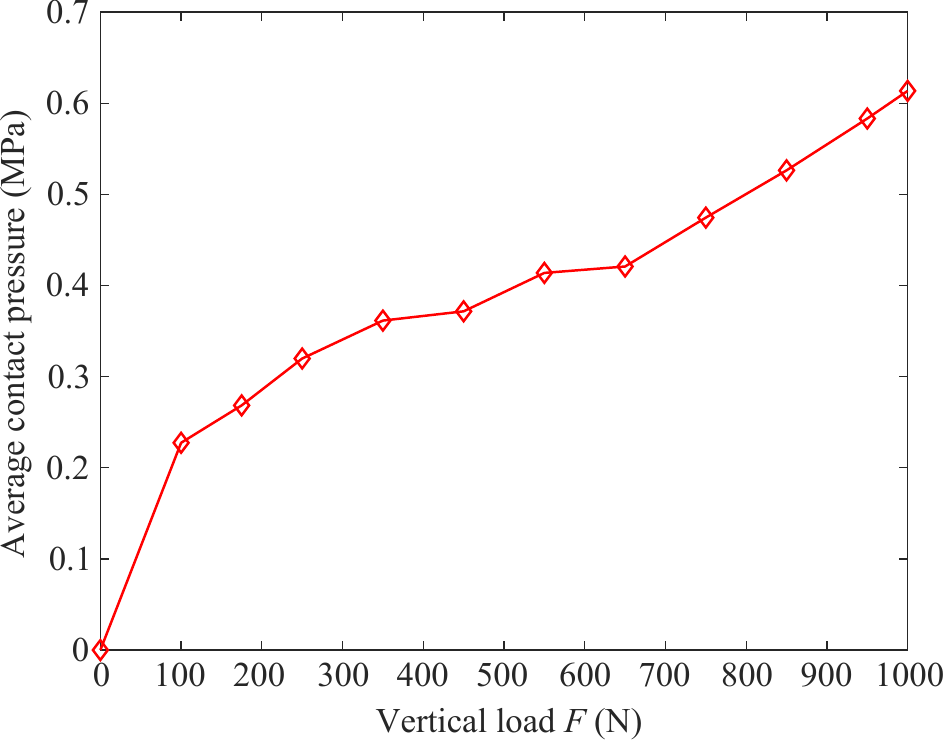}}\hfill{}
\par\end{centering}
\noindent \centering{}\caption{Development of the contact on the articular cartilage\label{fig:Development-of-the-knee-contact}}
\end{figure}

The development of the contact on the articular cartilage is recorded,
as depicted in Fig.~\ref{fig:Development-of-the-knee-contact}. Note
that only the area where the contact pressure is greater than $10^{-4}\,\mathrm{MPa}$
has been taken into account. Generally speaking, with increasing the
vertical load $F$, the contact area is increasing while its increasing
rate is reducing. The average contact pressure increases during the
loading history. Before the vertical load increases up to $650\,\mathrm{N}$,
the increasing rate of the average contact pressure is basically reducing.
However, when $F>650\,\mathrm{N}$ the average contact pressure increases
almost linearly because the contact area increases only slightly.

The total CPU time of this example is 16.68 hours, in which 0.90 hours
are spent on the element analysis of UEL. The UEL time comprises of
0.52 hours on the pre-calculation and 0.38 hours on the step of nonlinear
analysis. It accounts for only 5.4\% of the total CPU time.

\section{Conclusion\label{sec:Conclusion}}

In this paper, a polyhedral element based on the SBFEM is implemented
as a user element within the commercial finite element software ABAQUS.
The implementation involves the user subroutines $\mathrm{UEXTERNALDB}$
and $\mathrm{UEL}$ to store the topology and perform the calculations
of the polyhedral element, respectively. The required data structures
for storing the polyhedral mesh and the elemental topology have been
presented and described in detail. A comprehensive explanation of
the framework of the $\mathrm{UEL}$ has been provided in conjunction
with the theoretical derivations needed for the implementation. Furthermore,
a modified approach of overlaying standard elements on the SBFEM user
elements is proposed to create element-based surfaces, which allows
establishing interactions for interfacial problems.

The availability of polyhedral user elements enhances the performances
of ABAQUS for interfacial problems, and it significantly reduces the
meshing burden encountered in the standard FEM. To verify the implementation,
three benchmark tests (static, modal, and transient) are performed.
It is observed that very accurate results are obtained when compared
to theoretical and numerical reference solutions. The advantages of
polyhedral elements, i.e., the flexibility of dealing with non-matching
meshes and its compatibility with automatic mesh generation techniques,
are illustrated by five numerical examples. By introducing matching
meshes for complex geometries consisting of several parts featuring
different element sizes, the performance of existing contact modeling
approaches in ABAQUS can be significantly enhanced. Matching meshes
provide interface representations without initial gaps or penetrations,
and thus, zero thickness cohesive elements for the simulation of ITZs
in the composite materials can be easily inserted. The SBFEM user
elements are highly complementary to octree-based mesh generation
techniques, which are efficient and robust to mesh complex geometries
and promising for integrating geometric models and numerical analysis
in a fully automatic manner.

In a nutshell, the SBFEM user element will enable ABAQUS users to
benefit from the advantages of this method. The commercial FEM software
ABAQUS offers a user-friendly interface and outstanding nonlinear
solvers, which will facilitate the use of the SBFEM to scholars. The
source code of the implementation is chosen to be published open-source
governed by the GPLv3 license, it can be downloaded from https://github.com/ShukaiYa/SBFEM-UEL
with several input files of numerical examples presented in the paper.

\section*{Acknowledgment}

The work presented in this paper is partially supported by the Australian
Research Council (Grant Nos. DP180101538 and DP200103577), the National
Major Scientific Research Program of China (Grant No. 2016YFB0201003),
and the National Natural Science Foundation of China (Grant No. 51779222).
The authors would like to thank Professor~Jing~Hu of Southeast University,
China, for providing the X-CT scans of the three-dimensional asphalt
concrete specimen in Section~\ref{subsec:Analysis-of-asphalt} and
Dr.~Junqi~Zhang for generating the matching meshes used in Section~\ref{subsec:Examples-for-non-matching}.

\newpage{}

\bibliographystyle{unsrt}
\nocite{*}
\bibliography{./CMA_113766}

\end{document}